\documentclass[aps,prl,10pt,twocolumn,showpacs,preprintnumbers,amsmath,amssymb,superscriptaddress]{revtex4-1}

\usepackage[caption=false]{subfig}
\usepackage{lineno}
\usepackage{graphicx}  
\usepackage{dcolumn}   
\usepackage{bm}        
\usepackage{amssymb}   
\usepackage{amsmath}
\usepackage{blkarray, multirow, graphicx, diagbox, color, xcolor, colortbl}
\usepackage{bbm, bbold}
\usepackage{ifthen}
\usepackage[colorlinks, linkcolor = blue, citecolor = blue, filecolor = black, urlcolor = blue]{hyperref}
\usepackage{xkeyval}
\usepackage{moreverb}
\usepackage{rotating}
\usepackage{slashbox}
\usepackage{xspace}
\usepackage{nicefrac}
\usepackage[]{units}
\usepackage[]{natbib}
\usepackage{physics}
\usepackage{braket}
\usepackage{hyphenat}

\usepackage[customcolors]{hf-tikz} 
\usepackage{tikz}
\usepackage{calc}
\usetikzlibrary{external}
\usepackage{pgffor}
\usepackage{pgfplots}
\pgfplotsset{compat=1.13}
\usepackage{pgfplotstable}
\usepgfplotslibrary{groupplots}
\usepgfplotslibrary{fillbetween}

\usetikzlibrary
{
	calc,
	decorations,
	plotmarks,
	patterns,
	positioning,
	petri,
	arrows,
	decorations.markings,
	backgrounds,
	fit,
	graphs,
	shapes.geometric,
	decorations.pathmorphing,
	shapes.misc,
	shapes,
	tikzmark,
	pgfplots.colorbrewer
}

\pgfplotsset{
        cycle from colormap manual style/.style={
            x=3cm,y=10pt,ytick=\empty,
            stack plots=y,
            every axis plot/.style={line width=2pt},
        },
}

\definecolor{colorA}{rgb} {0.58,0,0.8275}
\definecolor{colorB}{rgb} {0.11,0.663,0.51}
\definecolor{colorC}{rgb} {0.3373,0.7059,0.9137}
\definecolor{colorD}{rgb} {0.902,0.6235,0}
\definecolor{colorE}{rgb} {0.9451,0.902,0.3255}
\definecolor{colorF}{rgb} {0.3373,0.3255,0.902}
\definecolor{colorG}{rgb} {0.9451,0.3255,0.3373}

\newcommand{\nodagger}[0]{\phantom{\dagger}}
\newcommand{\noprime}[0]{\phantom{\prime}}
\newcommand{\overbar}[1]{\mkern 1.5mu\overline{\mkern-1.5mu#1\mkern-1.5mu}\mkern 1.5mu}

\usepackage[acronym,nohypertypes={acronym,notation}]{glossaries}
\newacronym{1D}{1D}{one\hyp dimensional}
\newacronym[shortplural={MPS}]{MPS}{MPS}{matrix\hyp product state}
\newacronym{MPO}{MPO}{matrix\hyp product operator}
\newacronym{SVD}{SVD}{singular\hyp value decomposition}
\newacronym{QCS}{QCS}{quantum\hyp computer simulator}
\newacronym{QC}{QC}{quantum computer}
\newacronym{FSM}{FSM}{finite\hyp state machine}
\newacronym{ACA}{ACA}{adaptive cross\hyp approximation}
\newacronym{CDW}{CDW}{charge\hyp density wave}
\newacronym{SDW}{SDW}{spin\hyp density wave}
\newacronym{ARPES}{ARPES}{angle-resolved photoemission spectroscopy}
\newacronym{OBC}{OBC}{open-boundary conditions}
\newacronym{PBC}{PBC}{periodic-boundary conditions}
\newacronym{TEBD}{TEBD}{time-evolution block-decimation}
\newacronym{TDVP}{TDVP}{time\hyp dependent variational principle}
\newacronym{iff}{iff}{if and only if}
\newacronym{DFT}{DFT}{density\hyp functional theory}
\newacronym{DMFT}{DMFT}{dynamical mean\hyp field theory}
\newacronym{DMRG}{DMRG}{density\hyp matrix renormalization group}
\newacronym{QMC}{QMC}{quantum Monte Carlo}
\newacronym{AIM}{AIM}{Anderson impurity model}
\newacronym{SIAM}{SIAM}{single impurity Anderson model}
\newacronym{LDA}{LDA}{local\hyp density approximation}
\newacronym{LBNL}{LBNL}{Lawrence Berkeley National Laboratory}
\newacronym{VQE}{VQE}{variational\hyp quantum eigensolver}
\newacronym{ED}{ED}{exact diagonalization}
\newacronym{QPT}{QPT}{quantum phase transition}
\newacronym{QCP}{QCP}{quantum critical point}
\newacronym{ETH}{ETH}{eigenstate thermalization hypothesis}
\newacronym{EHM}{EHM}{extended Hubbard model}
\newacronym{AKLT}{AKLT}{Affleck\hyp Lieb\hyp Kennedy\hyp Tasaki}
\newglossaryentry{QR}{name={QR},description={QR decomposition}}
\newacronym{TNS}{TNS}{tensor\hyp network state}
\newacronym{SM}{SM}{supplemental material}
\newacronym{NOO}{NOO}{natural orbital occupation}
\newacronym{NO}{NO}{natural orbital}
\newacronym{LRO}{LRO}{long\hyp range order}
\newacronym{qLRO}{qLRO}{quasi\hyp long\hyp range order}
\newacronym{SC}{SC}{Superconductivity}
\newacronym{tr-ARPES}{tr-ARPES}{time- and angle-resolved photoemission spectroscopy}

\newboolean{buildtikzpics}
\setboolean{buildtikzpics}{true}
\newif\ifrebuildtikz
\newif\ifChangeMode
\ChangeModetrue
\ChangeModefalse
\ifthenelse{\boolean{buildtikzpics}}
{
	\rebuildtikztrue
	\usetikzlibrary{external}
	\tikzexternalize[optimize=false,prefix=figures/autogen/]
}
{
	\rebuildtikzfalse
}

\usepackage{ulem}
\usepackage{cleveref}
\Crefname{appendix}{Appendix}{Appendices}
\Crefname{equation}{Equation}{Equations}
\Crefname{figure}{Figure}{Figures}
\Crefname{section}{Section}{Sections}
\Crefname{tabular}{Tabular}{Tabulars}
\crefname{appendix}{App.}{Apps.}
\crefname{equation}{Eq.}{Eqs.}
\crefname{figure}{Fig.}{Figs.}
\crefname{section}{Sec.}{Secs.}
\crefname{tabular}{Tab.}{Tabs.}

\def\EqExp{0.501253707483214}
\def\EqExpShort{-0.50}
\def\EqScale{0.439187090842296} 
\def\EqShift{-0.00132342169091444}
\def\NeqExp{0.923280604068559} 
\def\NeqExpShort{-0.92} 
\def\NeqScale{0.837691909688444} 
\def\NeqShift{0.000889348348624644}

\begin{document}

\title{Detecting superconductivity out-of-equilibrium}

\author{S.~Paeckel}
\affiliation{Fakult\"at f\"ur Physik, Ludwig-Maximilians-Universit\"at M\"unchen, D-80333 M\"nchen, Germany}
\affiliation{Institut f\"ur Theoretische Physik, Georg-August-Universit\"at G\"ottingen, D-37077 G\"ottingen, Germany}

\author{B.~Fauseweh}
\affiliation{Max-Planck-Institut f\"ur Festk\"orperforschung, Heisenbergstra\ss{}e 1, D-70569 Stuttgart, Germany}
\affiliation{Theoretical Division, Los Alamos National Laboratory, Los Alamos, New Mexico 87545, USA}

\author{A.~Osterkorn}
\affiliation{Institut f\"ur Theoretische Physik, Georg-August-Universit\"at G\"ottingen, D-37077 G\"ottingen, Germany}

\author{T.~K\"ohler}
\affiliation{Department of Physics and Astronomy, Uppsala University, Box 516, S-751 20 Uppsala, Sweden}
\affiliation{Institut f\"ur Theoretische Physik, Georg-August-Universit\"at G\"ottingen, D-37077 G\"ottingen, Germany}

\author{D.~Manske}
\affiliation{Max-Planck-Institut f\"ur Festk\"orperforschung, Heisenbergstra\ss{}e 1, D-70569 Stuttgart, Germany}

\author{S.R.~Manmana}
\affiliation{Institut f\"ur Theoretische Physik, Georg-August-Universit\"at G\"ottingen, D-37077 G\"ottingen, Germany}

\date{\today}

\begin{abstract}
Recent pump-probe experiments on underdoped cuprates and similar systems suggest the existence of a transient superconducting state above $\mathrm{T}_c$.
This poses the question how to reliably identify the emergence of \acrlong{LRO}, in particular superconductivity, out\hyp of\hyp equilibrium.
We investigate this point by studying a quantum quench in an extended Hubbard model and by computing various observables, which are used to identify (quasi\hyp )long\hyp range order in equilibrium. 
Our findings imply that, in contrast to current experimental studies, it does not suffice to study the time evolution of the optical conductivity to identify superconductivity. 
In turn, we suggest to utilize time-resolved ARPES experiments to probe for the formation of a condensate in the two\hyp particle channel.
\end{abstract}


\maketitle

\paragraph{Introduction}
\Gls{SC} is one of the hallmarks of condensed\hyp matter systems and has inspired researchers since its discovery in 1911, and later by the advent of high-temperature \gls{SC} in cuprate materials~\cite{highTc_original,Wu1987,dagotto,Lee2006}.
While, in particular for the latter class of materials, a lot of questions are subject of ongoing research, many characteristics of the \gls{SC} phase are by now well established as long as the system is in equilibrium.
However, recent experiments (e.g.,~\cite{Foerst2014,Mankowsky2014,Hu2014,Hunt2016,Mitrano2016a} on copper oxides, or on $\mathrm{K}_3\mathrm{C}_{60}$) report the observation of possible photo-induced transient \gls{SC} phases, which can exist at elevated temperatures, even above the equilibrium\hyp critical temperature $\mathrm{T}_c$~\cite{Foerst2011,Fausti189,Kaiser2014,buzzi2020photomolecular}.
In these investigations, ultrashort THz pulses excite single phonon modes, which decay very slowly compared to the typical time scale of the electron dynamics and thereby offer the possibility to control the interaction parameters of the electronic system~\cite{Singla2015}.
Subsequently, the $\omega$-dependent optical conductivity is determined as a function of time via reflectivity measurements using a probe pulse, and \gls{SC} correlations are identified by the emergence or enhancement of a signal at $\omega \rightarrow 0$.
This is by now a standard experimental procedure, which, however, leaves many questions open, in particular concerning the characterization of the state induced by the pump excitation (see, e.g., Refs.~\onlinecite{Eckstein2010,PhysRevB.96.054506,PhysRevB.96.235142,Bittner2019,Fauseweh2017,schwarz2020momentumresolved}).
In this Letter, we address this issue regarding further experimental measures to probe \gls{SC} in such non\hyp equilibrium setups.
For the sake of simplicity, we focus on \gls{1D} systems, for which powerful numerical techniques are available in terms of \glspl{MPS}~\cite{White1992,Schollwoeck2010,Paeckel2019}.
We argue that it does not suffice to study only the optical conductivity, since the pump as well as the probe pulse can induce currents, which can modify the low-frequency behavior, without being a direct proof for \gls{SC}.
Nevertheless, we are able to provide evidence for the emergence of \gls{SC} in the course of time by studying the time evolution of spectral functions, which are accessible to variants of \gls{tr-ARPES} experiments~\cite{Damascelli2003,Damascelli2004,Lynch2005,Eckstein2008,Freericks2009,Ligges2018,wang2018theoretical}.
We propose to study the usual single-particle and a pairing spectral function, which we introduce below.
We find particularly in the latter quantity clear signatures for the accumulation of weight at $k = 0$.  
This raises the question, whether a condensation or quasicondensation of pairs is achieved, which would realize a transient non\hyp equilibrium \gls{SC} state.
It is interesting to ask, whether non\hyp equilibrium situations can be beyond the realm of validity of the Mermin-Wagner-Hohenberg theorem \cite{merminwagner_orig,merminwagner_erratum,hohenberg_orig}, which inhibits the formation of true \gls{LRO} in 1D systems.
The scope of this Letter is, therefore, three-fold: To demonstrate that the time evolution of the optical conductivity does not suffice to unambiguously establish transient \gls{SC} order, to present spectral functions as a more reliable probe, and to 
further characterize the observed transient state by investigating correlation matrices.
The general validity of our findings is supported by comparing the extended Hubbard model~\cite{voit,PRBVoit2,Jeckelmann2002,Tsuchiizu2002,Sandvik2004,Ejima2007,PhysRevB.85.205127} and a variant of the \gls{1D} $t$-$J$ model~\cite{tJoriginal0,tJoriginal1,tJoriginal2,dagotto,PhysRevB.83.205113NO,PhysRevLett.107.115301,PhysRevA.96.043618}.
\paragraph{Model and Methods}
We study the time evolution of Hubbard chains~\cite{Hubbard1963,Gutzwiller1963,Kanamori1963,HubbBook} following a quantum quench~\cite{Eisert2015a}. 
Recent experiments~\cite{Fausti189,Hu2014,Hunt2016,Mitrano2016a} on high-$\mathrm{T}_c$ superconductors suggest that if there exists preformed pairing, e.g., in the normal state slightly above $\mathrm{T}_c$, pumping particular phonon modes induces charge coherences, which drive the system into a transient superconducting state.
Therefore, our starting point is to assume that lattice distortions modify the strength of the couplings~\cite{Foerst2011,Singla2015} and thereby alter the nearest\hyp neighbor interaction between the electrons. 
Thus, we consider a quench in the \gls{1D} extended Hubbard model, 
\begin{align}
	\hat{H} 
	&= 
	\hat{T} 
	+
	U \sum_j \hat{n}^{\nodagger}_{j,\uparrow}\hat{n}^{\nodagger}_{j,\downarrow}
	+ 
	V \sum_j \hat{n}^{\nodagger}_{j}\hat{n}^{\nodagger}_{j+1}
\end{align}
with $\hat{T} = -t_{\rm hop} \sum_{j,\sigma} \left( \hat{c}^{\dagger}_{j,\sigma}\hat{c}^{\nodagger}_{j+1,\sigma} + \mathrm{h.c.} \right)$ being the kinetic energy.
Therein, $\hat{c}^{(\dagger)}_{j,\sigma}$ are $S-\nicefrac{1}{2}$ fermionic ladder operators, which obey the canonical anticommutation relations $\left\{ \hat{c}^{\nodagger}_{i,\sigma}, \hat{c}^{\dagger}_{j,\sigma^{\prime}} \right\} = \delta_{i,j}\delta_{\sigma,\sigma^{\prime}}$, $\left\{\hat{c}^{\nodagger}_{i,\sigma}, \hat{c}^{\nodagger}_{j,\sigma^{\prime}} \right\} = \left\{\hat{c}^{\nodagger}_{i,\sigma}, \hat{c}^{\nodagger}_{j,\sigma^{\prime}} \right\} = 0$, and we denote by $\hat{n}_{j}=\hat{n}_{j,\uparrow}+\hat{n}_{j,\downarrow}$ the total electron occupation at site $j$.
For later convenience, we also define doublon ladder operators $\hat{d}^{\nodagger}_j \equiv \hat{c}^{\nodagger}_{j,\uparrow}\hat{c}^{\nodagger}_{j,\downarrow}$.
As motivated above, we start in a \gls{CDW}, which favors double occupancies ($U/t_{\rm hop}=-4$, $V/t_{\rm hop}=\nicefrac{1}{4}$)~\cite{Bittner2019,BittnerPhD}. 
We perform a sudden quench in the nearest\hyp neighbor interaction $V/t_{\rm hop}=\nicefrac{1}{4}\rightarrow -\nicefrac{1}{4}$ into the s-wave superconducting phase at zero temperature~\cite{supmat}, keeping the local Hubbard interaction fixed at $U/t_{\rm hop} = -4$.
Note that this value of the attractive interaction is comparable to the electronic bandwidth. Although this is out of the scope for cuprate high-$T_c$ superconductors, it is, however, similar to the electronic structure in alkali-doped fullerides, where the effective phonon-mediated intraorbtial interaction is comparable to the electronic bandwidth coming from $t_{1u}$ bands \cite{Nomurae1500568}.
We then calculate the real-time evolution using a combined single- and two-site \gls{TDVP} scheme in the \gls{MPS} formulation of the \gls{DMRG} for lattices~\cite{White1992,Schollwoeck2010,PhysRevB.94.165116,Paeckel2019} with up to $L=80$ sites, open boundary conditions, and a maximal bond dimension of $m_{\mathrm{max}} = 1000$ states.
By comparing calculations with different values of $m_{\mathrm{max}}$ we estimate the accuracy of the presented results to be of a few percent~\cite{supmat}.
To investigate the formation and stability of transient \gls{SC}, we studied the differential optical conductivity after a probe pulse~\cite{Lenar2014,Shao2016}, spectral functions, and the correlation matrices~\cite{supersolid_penrose,rigol:031603} of single\hyp{} and two\hyp particle excitations \cite{Damascelli2003,Kouzakov2003,Damascelli2004,Lynch2005,Eckstein2008,Freericks2009,Truetzschler2017,Ligges2018,Stahl2019}.
We complement our studies by considering a similar quench in the \gls{1D} $t$-$J_\perp$ model~\cite{PhysRevLett.107.115301,PhysRevA.96.043618},
$
	\hat{H}_\text{$t$-$J_\perp$} 
	=
	\hat{T}
	+ 
	\nicefrac{J_\perp}{2} \sum_j \left( \hat S_j^+ \hat S_{j+1}^- + \hat S_j^- \hat S_{j+1}^+ \right),
$ at filling $n=0.2$ by quenching $J_{\perp}/t_{\rm hop}=2 \rightarrow 6$, i.e., from a Luttinger liquid~\cite{giamarchi} to a singlet \gls{SC} phase \footnote{The operators $S_j^\pm$ are the usual $S-\nicefrac{1}{2}$ ladder operators on site $j$. Note that double occupancies are forbidden in the $t$-$J$ model.}. 
We choose our energy and time units by setting $t_{\rm hop} \equiv 1$ and $\hbar \equiv 1$.
\paragraph{Time-dependent Optical Conductivity}
\begin{figure} 
	\subfloat[\label{fig:optical-conductivity:1}]
	{
		\centering
		\tikzsetnextfilename{optical_conductivity_real_CDW_SC_quench_U_m4p0_V_0p25_dV_m0p5}
		\begin{tikzpicture}
			\begin{axis}
			[
				axis lines=center,
				xlabel={$\omega$},
				xlabel style = {at=(xticklabel cs:1.15),yshift=10},
				ylabel={$\Delta t$},
				ylabel style = {at=(yticklabel cs:-0.0),yshift=5pt},
				zlabel={$\sigma_{1}$}, 
				zlabel style = {at=(zticklabel cs:1.125), xshift=5pt},
				yticklabels = {,,},
				xmin=1e-6,
				xmax=3.5,
				ymin=-1,
				ymax=26,
				zmax=2.5,
				zmin=-0.5,
				grid,
			]
			\addplot3
				[
					fill opacity = 0.75,
					draw opacity = 0.0,
					fill=colorD!60!white,
					restrict expr to domain = {x}{0.1:3.3},
					restrict expr to domain = {z}{-2:2},
				]
				table
				[
					x expr 	= \thisrowno{0},
					y expr	= 22,
					z expr	= -1.0*\thisrowno{2},
				]
				{../spectral_functions/eq/optical_conductivity/U_m4p0_V_m0p25/results/optical_conductivity_merged_A0_0p5_t0_0p0_tau_0p05_omega_2p38} \closedcycle;
			\addplot3
				[
					draw=colorD!80!black,
					thick,
					restrict expr to domain = {x}{0.1:3.3},
					restrict expr to domain = {z}{-2:2},
				]
				table
				[
					x expr 	= \thisrowno{0},
					y expr	= 22,
					z expr	= -1.0*\thisrowno{2},
				]
				{../spectral_functions/eq/optical_conductivity/U_m4p0_V_m0p25/results/optical_conductivity_merged_A0_0p5_t0_0p0_tau_0p05_omega_2p38};
				\node[text=colorD!80] at (axis cs:2.5,2*12+2,0) {\scriptsize equilibrium SC};
			\foreach \dt [count=\x from 1] in {{10p0},{9p0},{8p0},{7p0},{6p0},{5p0},{4p0},{3p0},{2p0},{1p0}}
			{
				\addplot3
					[
						fill opacity = 0.75,
						draw opacity = 0.0,
						fill=colorC!60!white,
						restrict expr to domain = {x}{0.1:3.3},
						restrict expr to domain = {z}{-2:2},
					]
					table
					[
						x expr 	= \thisrowno{0},
						y expr	= {2*(11-\x)},
						z expr	= -1.0*\thisrowno{2},
					]
					{../spectral_functions/global_quench/1TDVP/probe_pulse/U_m4p0_V_0p25_dV_m0p5/results/optical_conductivity_merged_A0_0p5_t0_\dt_tau_0p05_omega_2p38} \closedcycle;
				\addplot3
					[
						draw=colorC!80!black,
						thick,
						restrict expr to domain = {x}{0.1:3.3},
						restrict expr to domain = {z}{-2:2},
					]
					table
					[
						x expr 	= \thisrowno{0},
						y expr	= {2*(11-\x)},
						z expr	= -1.0*\thisrowno{2},
					]
					{../spectral_functions/global_quench/1TDVP/probe_pulse/U_m4p0_V_0p25_dV_m0p5/results/optical_conductivity_merged_A0_0p5_t0_\dt_tau_0p05_omega_2p38};
				\pgfmathparse{int(round(11 - \x))}
				\edef\PrintLabel{
				    \noexpand\node at (axis cs:3.5,2*\pgfmathresult,0.0) {\noexpand\tiny$\pgfmathresult$};
				}
				\
				\PrintLabel
			}
			\begin{scope}
				\clip (axis cs:1e-6,0.01,-0.5) -- (axis cs:3.5,0.01,-0.5) -- (axis cs:3.5,0.01,2.5) -- (axis cs:1e-6,0.01,2.5) -- (axis cs:1e-6,0.01,-0.5);
				\addplot3
					[
						fill opacity = 0.75,
						draw opacity = 0.0,
						fill=colorF!60!white,
						restrict expr to domain = {x}{0.1:3.3},
						restrict expr to domain = {z}{-6:6},
					]
					table
					[
						x expr 	= \thisrowno{0},
						y expr	= 0.01,
						z expr	= -1.0*\thisrowno{2},
					]
					{../spectral_functions/eq/optical_conductivity/U_m4p0_V_0p25/results/optical_conductivity_merged_A0_0p5_t0_0p0_tau_0p05_omega_2p38} \closedcycle;
				\addplot3
					[
						draw=colorF!80!black,
						thick,
						restrict expr to domain = {x}{0.1:3.3},
						restrict expr to domain = {z}{-6:6},
					]
					table
					[
						x expr 	= \thisrowno{0},
						y expr	= 0.01,
						z expr	= -1.0*\thisrowno{2},
					]
					{../spectral_functions/eq/optical_conductivity/U_m4p0_V_0p25/results/optical_conductivity_merged_A0_0p5_t0_0p0_tau_0p05_omega_2p38};

			\end{scope}
			\end{axis}
			\node[text=colorF!80] at (2.5,0.45) {\scriptsize equilibrium CDW};
		\end{tikzpicture}
	}
	
	\subfloat[\label{fig:optical-conductivity:2}]
	{
		\centering
		\tikzsetnextfilename{optical_conductivity_imag_CDW_SC_quench_U_m4p0_V_0p25_dV_m0p5}
		\begin{tikzpicture}
			\pgfplotsset
			{
				/pgfplots/colormap={moreland}{
					rgb255=(59,76,192)
					rgb255=(77,104,215)
					rgb255=(98,130,234)
					rgb255=(119,154,247)
					rgb255=(141,176,254)
					rgb255=(163,194,255)
					rgb255=(184,208,249)
					rgb255=(204,217,238)
					rgb255=(221,221,221)
					rgb255=(236,211,197)
					rgb255=(245,196,173)
					rgb255=(247,177,148)
					rgb255=(244,154,123)
					rgb255=(236,127,99)
					rgb255=(222,96,77)
					rgb255=(203,62,56)
					rgb255=(180,4,38)
				}
			}
			\begin{axis}
			[
				axis lines=center,
				xlabel={$\omega$},
				xlabel style = {at=(xticklabel cs:1.15),yshift=10},
				ylabel={$\Delta t$},
				ylabel style = {at=(yticklabel cs:-0.0),yshift=5pt},
				zlabel={$\sigma_{2}$}, 
				zlabel style = {at=(zticklabel cs:1.125), xshift=5pt},
				yticklabels = {,,},
				xmin=1e-6,
				xmax=3.5,
				ymin=-1,
				ymax=26,
				zmax=9,
				zmin=-1.75,
				grid,
			]
				\coordinate (insetPosition) at (axis cs: 0.0,38.0,0.0);
				\begin{scope}
					\clip (axis cs:1e-6,22,-0.5) -- (axis cs:3.5,22,-0.5) -- (axis cs:3.5,22,9) -- (axis cs:1e-6,22,9) -- (axis cs:1e-6,22,-0.5);
					\addplot3
						[
							fill opacity = 0.75,
							draw opacity = 0.0,
							fill=colorD!60!white,
							restrict expr to domain = {x}{0.01:3.3},
						]
						table
						[
							x expr 	= \thisrowno{0},
							y expr	= 22,
							z expr	= \thisrowno{3},
						]
						{../spectral_functions/eq/optical_conductivity/U_m4p0_V_m0p25/results/optical_conductivity_merged_A0_0p5_t0_0p0_tau_0p05_omega_2p38} \closedcycle;
					\addplot3
						[
							draw=colorD!80!black,
							thick,
							restrict expr to domain = {x}{0.01:3.3},
						]
						table
						[
							x expr 	= \thisrowno{0},
							y expr	= 22,
							z expr	= \thisrowno{3},
						]
						{../spectral_functions/eq/optical_conductivity/U_m4p0_V_m0p25/results/optical_conductivity_merged_A0_0p5_t0_0p0_tau_0p05_omega_2p38};
				\end{scope}
				\node[text=colorD!80] at (axis cs:2.5,2*12+2,0.5) {\scriptsize equilibrium SC};
			\foreach \dt [count=\x from 1] in {{10p0},{9p0},{8p0},{7p0},{6p0},{5p0},{4p0},{3p0},{2p0},{1p0}}
			{
				\addplot3
					[
						fill opacity = 0.75,
						draw opacity = 0.0,
						fill=colorA!60!white,
						restrict expr to domain = {x}{0.1:3.3},
					]
					table
					[
						x expr 	= \thisrowno{0},
						y expr	= {2*(11-\x)},
						z expr	= \thisrowno{3},
					]
					{../spectral_functions/global_quench/1TDVP/probe_pulse/U_m4p0_V_0p25_dV_m0p5/results/optical_conductivity_merged_A0_0p5_t0_\dt_tau_0p05_omega_2p38} \closedcycle;
				\addplot3
					[
						draw=colorA!80!black,
						thick,
						restrict expr to domain = {x}{0.1:3.3},
					]
					table
					[
						x expr 	= \thisrowno{0},
						y expr	= {2*(11-\x)},
						z expr	= \thisrowno{3},
					]
					{../spectral_functions/global_quench/1TDVP/probe_pulse/U_m4p0_V_0p25_dV_m0p5/results/optical_conductivity_merged_A0_0p5_t0_\dt_tau_0p05_omega_2p38};
				\pgfmathparse{int(round(11 - \x))}
				\edef\PrintLabel{
				    \noexpand\node at (axis cs:3.5,2*\pgfmathresult,0,0.0) {\noexpand\tiny$\pgfmathresult$};
				}
				\
				\PrintLabel
			}
			\addplot3
				[
					fill opacity = 0.75,
					draw opacity = 0.0,
					fill=colorF!60!white,
					restrict expr to domain = {x}{0.1:3.3},
				]
				table
				[
					x expr 	= \thisrowno{0},
					y expr	= 0.01,
					z expr	= \thisrowno{3},
				]
				{../spectral_functions/eq/optical_conductivity/U_m4p0_V_0p25/results/optical_conductivity_merged_A0_0p5_t0_0p0_tau_0p05_omega_2p38} \closedcycle;
			\addplot3
				[
					draw=colorF!80!black,
					thick,
					restrict expr to domain = {x}{0.1:3.3},
				]
				table
				[
					x expr 	= \thisrowno{0},
					y expr	= 0.01,
					z expr	= \thisrowno{3},
				]
				{../spectral_functions/eq/optical_conductivity/U_m4p0_V_0p25/results/optical_conductivity_merged_A0_0p5_t0_0p0_tau_0p05_omega_2p38};
			\end{axis}
			\node[text=colorF!80] at (2.5,0.45) {\scriptsize equilibrium CDW};
		\end{tikzpicture}
	}
	
	\subfloat[\label{fig:optical-conductivity:3}]
	{
		\tikzsetnextfilename{response_currents_CDW_SC_quench_U_m4p0_V_0p25_dV_m0p5}
		\begin{tikzpicture}
			\pgfplotsset
			{
				/pgfplots/colormap={moreland}{
					rgb255=(59,76,192)
					rgb255=(77,104,215)
					rgb255=(98,130,234)
					rgb255=(119,154,247)
					rgb255=(141,176,254)
					rgb255=(163,194,255)
					rgb255=(184,208,249)
					rgb255=(204,217,238)
					rgb255=(221,221,221)
					rgb255=(236,211,197)
					rgb255=(245,196,173)
					rgb255=(247,177,148)
					rgb255=(244,154,123)
					rgb255=(236,127,99)
					rgb255=(222,96,77)
					rgb255=(203,62,56)
					rgb255=(180,4,38)
				}
			}
			\pgfplotsset
			{
				/pgfplots/colormap={temp}{
					rgb255=(36,0,217) 		
					rgb255=(25,29,247) 		
					rgb255=(41,87,255) 		
					rgb255=(61,135,255) 	
					rgb255=(87,176,255) 	
					rgb255=(117,211,255) 	
					rgb255=(153,235,255) 	
					rgb255=(189,249,255) 	
					rgb255=(235,255,255) 	
					rgb255=(255,255,235) 	
					rgb255=(255,242,189) 	
					rgb255=(255,214,153) 	
					rgb255=(255,172,117) 	
					rgb255=(255,120,87) 	
					rgb255=(255,61,61) 		
					rgb255=(247,40,54) 		
					rgb255=(217,22,48) 		
					rgb255=(166,0,33)		
				}
			}
			\begin{axis}
			[
				axis lines=center,
				xlabel={time $t-\Delta t$}, 
				xlabel style = {at=(xticklabel cs:1.35),yshift=5pt,xshift=-10pt},
				ylabel={$\Delta t$},
				ylabel style = {at=(yticklabel cs:-0.0),xshift=1.5pt,yshift=1.5pt},
				zlabel={$j$}, 
				zlabel style = {at=(zticklabel cs:1.185), xshift=16.5pt},
				yticklabels = {,,},
				xmin=0,
				xmax=25,
				ymin=-1,
				ymax=50,
				grid,
			]
				\coordinate (insetPosition) at (axis cs:10.0,0.0,-2.35);
				\addplot3
					[
						fill opacity	= 0.5,
						draw opacity	= 0.0,
						fill		= colorD!60!white,
						restrict expr to domain = {x}{0:25},
					]
				table
					[
						x expr		= \thisrowno{0}-0.2,
						y expr		= {44},
						z expr		= \thisrowno{1},
					]
					{data/L_64/sc_groundstate_U_m4p0_V_m0p25/optical_conductivity/response_currents_A0_0p5_t0_0p2_tau_0p05_omega_2p38} \closedcycle;
				\addplot3
					[
						draw=colorD!80!black,
						thick,
						restrict expr to domain	= {x}{0:25},
					]
				table
					[
						x expr	= \thisrowno{0}-0.2,
						y expr	= {44},
						z expr	= \thisrowno{1},
					]
					{data/L_64/sc_groundstate_U_m4p0_V_m0p25/optical_conductivity/response_currents_A0_0p5_t0_0p2_tau_0p05_omega_2p38};
				\node[text=colorD!80] at (axis cs:2.5/3*24,4*12+1,0.05) {\scriptsize equilibrium SC};
				\foreach \dt [count=\x from 1] in {{10p0},{9p0},{8p0},{7p0},{6p0},{5p0},{4p0},{3p0},{2p0},{1p0}}
				{
					\addplot3
						[
							fill opacity = 0.5,
							draw opacity = 0.0,
							fill=colorB!60!white,
							restrict expr to domain = {x}{0.5:25},
						]
						table
						[
							x expr 	= \thisrowno{0},
							y expr	= {4*(11-\x)},
							z expr	= \thisrowno{1},
						]
						{../spectral_functions/global_quench/1TDVP/probe_pulse/U_m4p0_V_0p25_dV_m0p5/results/response_currents_A0_0p5_t0_\dt_tau_0p05_omega_2p38} \closedcycle;
					\addplot3
						[
							draw=colorB!80!black,
							thick,
							restrict expr to domain = {x}{0.5:25},
						]
						table
						[
							x expr 	= \thisrowno{0},
							y expr	= {4*(11-\x)},
							z expr	= \thisrowno{1},
						]
						{../spectral_functions/global_quench/1TDVP/probe_pulse/U_m4p0_V_0p25_dV_m0p5/results/response_currents_A0_0p5_t0_\dt_tau_0p05_omega_2p38};
					\pgfmathparse{int(round(11 - \x))}
					\edef\PrintLabel{
					    \noexpand\node at (axis cs:26.25,4*\pgfmathresult-1.75,0,-0) {\noexpand\tiny$\pgfmathresult$};
					}
					\PrintLabel
				}
			\end{axis}
			\begin{axis}
				[
					name		= inset,
					width		= 0.225\textwidth,
					height		= 0.125\textheight,
					at		= {(insetPosition)},
					axis on top,
					enlargelimits	= false,
					ticklabel style	= {font=\scriptsize},
					xlabel style	= {font=\scriptsize,yshift=2.5pt},
					ylabel style	= {font=\scriptsize,yshift=-3.5pt},
					ylabel		= {time $t\,[\nicefrac{1}{t_\mathrm{hop}}]$},
					xlabel		= {site $i$},
					colorbar right,
					colormap name	= temp,
					ytick		= {0,50,100},
					xtick		= {0,16,32,48},
					point meta min=-0.06,
					point meta max=0.06,
					every colorbar/.append style =
						{
							width			=	2mm,
							scaled y ticks		= 	false,
							ylabel style		= 	{font=\tiny,yshift=-2.5pt},
							ytick			= 	{-0.06,0.0,0.06},
							yticklabels		=	{$-0.06$, $\phantom{-}0.00$, $\phantom{-}0.06$},
							ylabel shift 		=	-4pt,
						},
					title					=	{\scriptsize $\Delta n_{i}(t)$},
					title style = {yshift=-0.75em, xshift=0.75em,},
				]
				\addplot graphics
				[
					xmin = 0, 
					xmax = 63, 
					ymin = 0, 
					ymax = 100
				]
				{../spectral_functions/precompiled/eq_U_m4p0_V_m0p25_pulse_response_densities_diff.eps}; 
			\end{axis}
		\end{tikzpicture}
	}
	\caption
	{
		Real \protect\subref{fig:optical-conductivity:1} and imaginary \protect\subref{fig:optical-conductivity:2} part of the optical conductivity $\sigma_{1,2}(\omega,\Delta t)$ with probe pulses applied at different time delays $\Delta t$ and the corresponding quantities in the \gls{CDW} ground state (violet) and the  \gls{SC} ground state (orange).
		The Bottom panel \protect\subref{fig:optical-conductivity:3} shows the real-time evolution of the response current $j(t-\Delta t,\Delta t)$ after application of a probe pulse at time delay $\Delta t$ following the quench (green) and in the \gls{SC} ground state (orange).
		The charge flow after applying the probe pulse in the \gls{SC} ground state at $\Delta t = 0$ is shown in the inset.
	}
	\label{fig:optical-conductivity}
\end{figure}
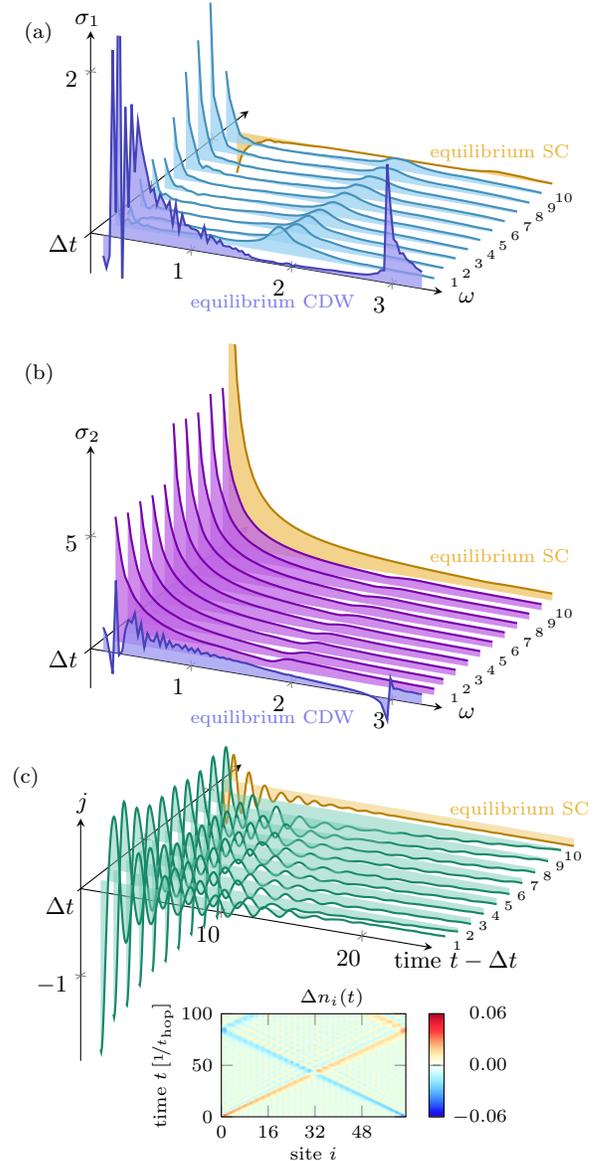
The experimental setups~\cite{Fausti189,Hu2014,Hunt2016,Mitrano2016a} we refer to typically measure the reflectivity after pump\hyp probe excitations, from which the optical conductivity is extracted.
We follow Refs.~\onlinecite{Lenar2014,Shao2016} and compute the differential optical conductivity
\begin{align}
	\sigma(\omega, \Delta t) &= \frac{j(\omega,\Delta t)}{\mathrm i(\omega + \mathrm i\eta) A(\omega,\Delta t) L} \equiv \sigma_1(\omega,\Delta t) + \mathrm i \sigma_2(\omega, \Delta t)
\end{align}
after the quench at $t=0$.
$A(\omega, \Delta t)$ is the vector potential of the Fourier\hyp transformed Gaussian probe pulse\footnote{For the width of the probe pulse we chose $\tau=0.05$ throughout this letter} $A(t,\Delta t)=A_{0}e^{-\frac{(t-\Delta t)^2}{2\tau^2}}\cos(\omega_{0} t)$ in Peierls substitution~\cite{PhysRevB.14.2239} applied after the time delay $\Delta t$.
The overall response current $j(t,\Delta t) = \mathrm i \left\langle \sum_{j,\sigma}\left(e^{-\mathrm i A(t,\Delta t)}\hat{c}^{\dagger}_{j,\sigma}\hat{c}^{\nodagger}_{j+1,\sigma} - \mathrm{h.c.}\right) \right\rangle$ was calculated from a real-time evolution of the perturbed system, and the $\omega$\hyp dependence $j(\omega, \Delta t)$ is obtained by a Fourier transform~\footnote{In~\cite{Shao2016} the non\hyp equilibrium response function is obtained by considering the Fourier transform of the difference $j(t,\Delta t) - j_{0}(t)$ with $j_0(t)$ the current without probe pulse. As in our quench protocol without probe pulse there is always $j_0(t)\equiv 0$ we drop this term for brevity.} including values up to time $t=25/t_{\mathrm{hop}}$ after the probe pulse.
We show the real (\cref{fig:optical-conductivity:1}) and imaginary (\cref{fig:optical-conductivity:2}) parts of $\sigma(\omega,\Delta t)$ for delays up to $\Delta t = 10$ and a system with $L=64$ sites.
For the real part, we find a sudden transfer of spectral weight from the \gls{CDW} signal at around $\omega \approx 3$ towards $\omega \approx 1.7$, which is due to the sudden change of the Hamiltonian.
%
%
At the same time, in the imaginary part a peak forms near $\omega = 0$.
We compare this to the \gls{SC} ground state and realize that, at the first glance, similar behavior is induced.
However, in particular in $\sigma_1$, clear differences appear, and $\sigma_2$ shows additional features.
Also, it is hard to decide whether $\sigma_2(\omega,\Delta t)$ is diverging as $\nicefrac{1}{\omega}$ for $\omega \to 0$, since we are limited in the frequency resolution. 
We conclude that, as in the experiments, the question whether the accumulation of spectral weight near $\omega=0$ is due to induced \gls{SC} or an 
enhanced metallicity after a pump pulse is hard to decide~\cite{supmat}.
However, in contrast to the experimental situation we have direct access to the time dependence of the current induced by the probe pulse.
The properties of this current are further illustrated by the inset of \cref{fig:optical-conductivity:3}, where we display the response electron density $\Delta n_i(t) = \braket{\hat{n}_{i}(t)}_{\mathrm {probe}} - \braket{\hat{n}_{i}}_0$, which compares the time evolution of the local density in the \gls{SC} phase with and without probe pulse.
As can be seen, the effect of the probe pulse is to accumulate charge at the edges of the system.
After passage of the probe pulse, this causes the measured current.
In the \gls{SC} phase the probe pulse induces a long\hyp living current, while in non\hyp equilibrium we find no clear evidence for a comparable response (see \cref{fig:optical-conductivity:3}).
In turn, in our simulations the induced charge flow decays on time scales of at least $t\sim 25/t_{\mathrm{hop}}$ which sets the scale of a low-frequency response $2\pi/t\approx 0.25$ in the imaginary part $\sigma_2$.
Thus, a strengthening of the response at $\omega \to 0$ alone, as observed here, does not suffice to demonstrate \gls{SC}.
This is supported by studying a quench within the CDW phase, in which a similar behavior of the optical conductivity is observed, but without further indication for \gls{SC} in other observables \cite{supmat}.

\paragraph{Spectral Functions}
From now on we consider postquench states at $\Delta t = 15/t_{\mathrm{hop}}$.
This is justified, since for times $t > 4/t_{\mathrm{hop}}$ a transient state is reached, as seen in the time evolution of the eigenvalues of the correlation matrix, which we find to be non-thermal~\cite{supmat}. 
Motivated by \gls{tr-ARPES}, we consider the in\hyp{} and out\hyp of\hyp equilibrium time\hyp dependent lesser Greens functions for $t > \Delta t$: $C_{\hat O}(j,t,\Delta t) = \braket{\psi(\Delta t)| \hat{O}^{\dagger}_{j}(t)\hat{O}^{\nodagger}_{\nicefrac{L}{2}}(0) |\psi(\Delta t)}$ in the single\hyp{} and two\hyp particle channel, i.e., $\hat{O}^{\nodagger}_{j} = \hat{c}^{\nodagger}_{j,\uparrow}$ and $\hat{O}^{\nodagger}_{j} = \hat{d}^{\nodagger}_{j}$, respectively, and we indicate the equilibrium case by setting $\Delta t \equiv 0$.
We refer to the Fourier transform to momentum and frequency space~\cite{supmat}
\begin{align}
	S_{\hat O}(q, \omega, \Delta t) 
	&= 
	\int\limits_{-\infty}^{\infty} \frac{dt}{2\pi}\sum_{j}e^{-\mathrm i (q r_j - (\omega + \mathrm i \eta)t)} C_{\hat O}(j,t,\Delta t) \; ,
	\label{eq:def-spec-func}
\end{align}
as the differential spectral function, where we have introduced a spectral broadening $\eta = 0.1$~\cite{Paeckel2019}.
Note that here we explicitly do not restrict ourselves to single\hyp electron excitations, but also study processes that may excite double occupations, i.e., doublons. 
This can be related to ongoing experiments~\cite{Truetzschler2017}, in which photoemission of pairs of electrons is studied, and is theoretically corroborated in a BCS-type picture~\cite{Kouzakov2003}.
\begin{figure}
	\subfloat[\label{fig:eq_neq_spectral_functions:1}]
	{
		\centering
		\tikzsetnextfilename{eq_neq_spectral_functions_1}
		\begin{tikzpicture}
			\pgfplotsset
			{
				/pgfplots/colormap={temp}{
					rgb255=(36,0,217) 		
					rgb255=(25,29,247) 		
					rgb255=(41,87,255) 		
					rgb255=(61,135,255) 	
					rgb255=(87,176,255) 	
					rgb255=(117,211,255) 	
					rgb255=(153,235,255) 	
					rgb255=(189,249,255) 	
					rgb255=(235,255,255) 	
					rgb255=(255,255,235) 	
					rgb255=(255,242,189) 	
					rgb255=(255,214,153) 	
					rgb255=(255,172,117) 	
					rgb255=(255,120,87) 	
					rgb255=(255,61,61) 		
					rgb255=(247,40,54) 		
					rgb255=(217,22,48) 		
					rgb255=(166,0,33)		
				}
			}
			\begin{groupplot}
				[
					group style = 
					{
						group size 			=	1 by 2,
						vertical sep		=	0.0pt,
						x descriptions at	=	edge bottom,
						y descriptions at	=	edge left,
					},
					xtick			= {0, 16, 32, 48},
					xticklabels		= {,,,},
					xlabel			= {},
					axis on top,
					enlargelimits	= false,
					width			= 0.4\textwidth-2.71pt,
				]
				\nextgroupplot
				[
					ytick={0},
					axis x line*	= top,
					height		= 5.75em,
					ymin=-1.0, ymax=0.0,
					xmin=0, xmax=31.5,
					ytick		= {0},
					xticklabels	= {,,},
				]
				\addplot graphics
				[ 
					xmin	= 0,
					xmax	= 31.5,
					ymin 	= -7.0,
					ymax	= 0.0
				]
				{../spectral_functions/precompiled/eq_U_m4p0_V_m0p25_d.eps};
				\nextgroupplot
				[
					height		= 0.14\textheight,
					axis y discontinuity=parallel,
					ylabel		= {$\omega$},
					ylabel shift	= -7.5pt,
					ytick		= {-6,-4},
					ymin=-6.5, ymax=-2.5,
					xmin=0, xmax=31.5,
					axis x line*=bottom,
					title style={at={(1,0.85)},anchor=east,xshift=-0.75em},
					title	= {\color{white}\textbf{SC phase}},
					colorbar right,
					colormap name	= temp,
					point meta min=0.0,
					point meta max=0.4,
					every colorbar/.append style =
					{
						height			=	\pgfkeysvalueof{/pgfplots/parent axis height}+0.885em,
						ylabel			=	{$S^{\mathrm{SC}}_{\hat d}(q,\omega,0)$},
						width			=	3mm,
						yshift			=	0.885em,
						scaled y ticks	= 	false,
						ytick			= 	{0,0.4},
						yticklabels		=	{$0\phantom{.06}$, $0.4$},
						ylabel shift 	=	-4pt,
					},
				]
				\addplot graphics
				[ 
					xmin	= 0,
					xmax	= 31.5,
					ymin 	= -7.0,
					ymax	= 0.0,
				]
				{../spectral_functions/precompiled/eq_U_m4p0_V_m0p25_d.eps};
			\end{groupplot}
		\end{tikzpicture}
	}

	\vspace{-0.8em}
	\subfloat[\label{fig:eq_neq_spectral_functions:2}]
	{
		\centering
		\tikzsetnextfilename{eq_neq_spectral_functions_2}
		\begin{tikzpicture}
			\pgfplotsset
			{
				/pgfplots/colormap={temp}{
					rgb255=(36,0,217) 		
					rgb255=(25,29,247) 		
					rgb255=(41,87,255) 		
					rgb255=(61,135,255) 	
					rgb255=(87,176,255) 	
					rgb255=(117,211,255) 	
					rgb255=(153,235,255) 	
					rgb255=(189,249,255) 	
					rgb255=(235,255,255) 	
					rgb255=(255,255,235) 	
					rgb255=(255,242,189) 	
					rgb255=(255,214,153) 	
					rgb255=(255,172,117) 	
					rgb255=(255,120,87) 	
					rgb255=(255,61,61) 		
					rgb255=(247,40,54) 		
					rgb255=(217,22,48) 		
					rgb255=(166,0,33)		
				}
			}
			\begin{groupplot}
				[
					group style = 
					{
						group size 		=	1 by 2,
						vertical sep		=	0.0pt,
						x descriptions at	=	edge bottom,
						y descriptions at	=	edge left,
					},
					xtick		= {0, 16, 32, 48},
					xticklabels	= {,,,},
					xlabel	= {},
					axis on top,
					enlargelimits	= false,
					width		= 0.4\textwidth-2.71pt,
				]
				\nextgroupplot
				[
					ytick={0},
					axis x line*	= top,
					height		= 5.75em,
					ymin=-1.0, ymax=0.0,
					xmin=0, xmax=31.5,
					ytick		= {0},
				]
				\addplot graphics
				[ 
					xmin	= 0,
					xmax	= 31.5,
					ymin 	= -7.0,
					ymax	= 0.0
				]
				{../spectral_functions/precompiled/eq_U_m4p0_V_0p25_d.eps};
				\nextgroupplot
				[
					height		= 0.14\textheight,
					axis y discontinuity=parallel,
					ylabel		= {$\omega$},
					ylabel shift	= -7.5pt,
					ytick		= {-6,-4},
					point meta min=0.0,
					point meta max=0.03,
					colorbar right,
					colormap name	= temp,
					every colorbar/.append style =
					{
						height			=	\pgfkeysvalueof{/pgfplots/parent axis height}+0.885em,
						ylabel			=	{$S^{\mathrm{CDW}}_{\hat d}(q,\omega,0)$},
						width			=	3mm,
						yshift			=	0.885em,
						scaled y ticks	= 	false,
						ytick			= 	{0,0.03},
						yticklabels		=	{$0\phantom{.06}$, $0.03$},
						ylabel shift 	=	-4pt,
					},
					ymin=-6.5, ymax=-2.5,
					xmin=0, xmax=31.5,
					axis x line* = bottom,
					title style={at={(1,0.85)},anchor=east,xshift=-0.75em},
					title	= {\color{white}\textbf{CDW phase}},
				]
				\addplot graphics
				[ 
					xmin	= 0,
					xmax	= 31.5,
					ymin 	= -7.0,
					ymax	= 0.0
				]
				{../spectral_functions/precompiled/eq_U_m4p0_V_0p25_d.eps};
			\end{groupplot}
		\end{tikzpicture}
	}

	\vspace{-0.8em}
	\subfloat[\label{fig:eq_neq_spectral_functions:3}]
	{
		\centering
		\tikzsetnextfilename{eq_neq_spectral_functions_3}
		\begin{tikzpicture}
			\pgfplotsset
			{
				/pgfplots/colormap={temp}{
					rgb255=(36,0,217) 		
					rgb255=(25,29,247) 		
					rgb255=(41,87,255) 		
					rgb255=(61,135,255) 	
					rgb255=(87,176,255) 	
					rgb255=(117,211,255) 	
					rgb255=(153,235,255) 	
					rgb255=(189,249,255) 	
					rgb255=(235,255,255) 	
					rgb255=(255,255,235) 	
					rgb255=(255,242,189) 	
					rgb255=(255,214,153) 	
					rgb255=(255,172,117) 	
					rgb255=(255,120,87) 	
					rgb255=(255,61,61) 		
					rgb255=(247,40,54) 		
					rgb255=(217,22,48) 		
					rgb255=(166,0,33)		
				}
			}
			\begin{groupplot}
				[
					group style = 
					{
						group size 		=	1 by 2,
						vertical sep		=	0.0pt,
						x descriptions at	=	edge bottom,
						y descriptions at	=	edge left,
					},
					xtick		= {0, 16, 32, 48},
					xticklabels	= {$0$,$\nicefrac{\pi}{2}$,$\pi$,$\nicefrac{3\pi}{2}$},
					xlabel	= {$q$},
					axis on top,
					enlargelimits	= false,
					width		= 0.4\textwidth-2.71pt,
				]
				\nextgroupplot
				[
					ytick={0},
					axis x line* 	= top,
					height		= 5.75em,
					ymin=-1.0, ymax=0.0,
					xmin=0, xmax=31.5,
					ytick		= {0},
				]
				\addplot graphics
				[ 
					xmin	= 0,
					xmax	= 31.5,
					ymin 	= -7.0,
					ymax	= 0.0
				]
				{../spectral_functions/precompiled/neq_U_m4p0_V_0p25_dV_m0p5_d.eps};
				\nextgroupplot
				[
					axis y discontinuity=parallel,
					height		= 0.14\textheight,
					ylabel		= {$\omega$},
					ytick		= {-6,-4},
					ylabel shift	= -7.5pt,
					point meta min=0.0,
					point meta max=0.06,
					colorbar right,
					colormap name	= temp,
					every colorbar/.append style =
					{
						height			=	\pgfkeysvalueof{/pgfplots/parent axis height}+0.885em,
						ylabel			=	{$S^{\phantom{\mathrm{C}}}_{\hat d}(q,\omega,15)$},
						width			=	3mm,
						yshift			=	0.885em,
						scaled y ticks		= 	false,
						ytick			= 	{0,0.06},
						yticklabels		=	{$0\phantom{.06}$, $0.06$},
						ylabel shift 		=	-4pt,
					},
					ymin=-6.5, ymax=-2.5,
					xmin=0, xmax=31.5,
					axis x line* = bottom,
					title style={at={(1,0.85)},anchor=east,xshift=-0.75em},
					title	= {\color{white}\textbf{quenched state}},
				]
				\addplot graphics
				[ 
					xmin	= 0,
					xmax	= 31.5,
					ymin 	= -7.0,
					ymax	= 0.0
				]
				{../spectral_functions/precompiled/neq_U_m4p0_V_0p25_dV_m0p5_d.eps};
			\end{groupplot}
		\end{tikzpicture}
	}
	\caption
	{%
		\label{fig:eq-neq:spectral-functions:doublons}%
		Spectral functions of two\hyp particle excitations in equilibrium \gls{SC} phase \protect\subref{fig:eq_neq_spectral_functions:1}, CDW phase \protect\subref{fig:eq_neq_spectral_functions:2}, and in non\hyp equilibrium after quenching from the CDW ground state into the \gls{SC} phase evaluated at time $t =15$ \protect\subref{fig:eq_neq_spectral_functions:3}. Note that only regions with significant spectral weight are displayed.
	}
\end{figure}
\cref{fig:eq-neq:spectral-functions:doublons} displays the  spectral functions for double occupations $S_{\hat d}(q,\omega,\Delta t)$ for both equilibrium phases (\gls{SC} and \gls{CDW}) and after the quench.
We obtain a clear accumulation of weight at $q=0$ in the postquench state, which renders the result similar to the one of the \gls{SC} ground state displayed.  
This is the central statement of this Letter: A (quasi\hyp{})condensate of s\hyp wave (Cooper\hyp{})pairs seems to form after the quench, which is clearly detectable in $S_{\hat d}(q,\omega,\Delta t)$, and which has a striking similarity to the spectral function in the \gls{SC} ground state. 
We emphasize that this coherence between the charges is dynamically created after the quench, as seen in the comparison with \cref{fig:eq_neq_spectral_functions:2}, which shows dispersive, incoherent doublons. 
We observe similar behavior in the corresponding pairing spectral function of the $t$-$J_\perp$ model \cite{supmat}, so that we expect this to be a generic feature, at least for quenches to \gls{SC} phases.
\begin{figure}
	\subfloat[\label{fig:eq_neq_spectral_functions:4}]
	{
		\centering
		\tikzsetnextfilename{eq_neq_spectral_functions_4}
		\begin{tikzpicture}
			\pgfplotsset
			{
				/pgfplots/colormap={temp}{
					rgb255=(36,0,217) 		
					rgb255=(25,29,247) 		
					rgb255=(41,87,255) 		
					rgb255=(61,135,255) 	
					rgb255=(87,176,255) 	
					rgb255=(117,211,255) 	
					rgb255=(153,235,255) 	
					rgb255=(189,249,255) 	
					rgb255=(235,255,255) 	
					rgb255=(255,255,235) 	
					rgb255=(255,242,189) 	
					rgb255=(255,214,153) 	
					rgb255=(255,172,117) 	
					rgb255=(255,120,87) 	
					rgb255=(255,61,61) 		
					rgb255=(247,40,54) 		
					rgb255=(217,22,48) 		
					rgb255=(166,0,33)		
				}
			}
			\begin{groupplot}
				[
					group style = 
					{
						group size 		=	1 by 2,
						vertical sep		=	0.0pt,
						x descriptions at	=	edge bottom,
						y descriptions at	=	edge left,
					},
					xtick		= {0, 16, 32, 48},
					xticklabels	= {,,,},
					xlabel	= {},
					axis on top,
					enlargelimits	= false,
					width		= 0.4\textwidth-2.71pt,
				]
				\nextgroupplot
				[
					axis x line*	= top,
					height		= 5.75em,
					ymin=-1.0, ymax=0.0,
					xmin=0, xmax=31.5,
					ytick		= {0},
					xticklabels	= {,,},
				]
				\addplot graphics
				[ 
					xmin	= 0,
					xmax	= 31.5,
					ymin 	= -10.0,
					ymax	= 0.0
				]
				{../spectral_functions/precompiled/eq_U_m4p0_V_m0p25_f.eps};
				\nextgroupplot
				[
					axis y discontinuity=parallel,
					height		= 0.14\textheight,
					colorbar right,
					colormap name	= temp,
					xticklabels	= {,,},
					ylabel		= {$\omega$},
					ylabel shift	= -7.5pt,
					ytick		= {-8,-4},
					point meta min=0.0,
					point meta max=0.06,
					every colorbar/.append style =
					{
						height			=	\pgfkeysvalueof{/pgfplots/parent axis height}+0.885em,
						ylabel			= 	{$S^{\mathrm{SC}}_{\hat c}(q,\omega,0)$},
						width			=	3mm,
						yshift			=	0.885em,
						scaled y ticks		= 	false,
						ytick			= 	{0,0.06},
						yticklabels		=	{$0\phantom{.00}$, $0.06$},
						ylabel shift 		=	-4pt,
					},
					ymin=-8.0,  ymax=-2.0,
					xmin=0, xmax=31.5,
					axis x line*	= bottom,
					title style	= {at={(1,0.85)},anchor=east,xshift=-0.75em},
					title		= {\color{white}\textbf{SC phase}},
				]
				\addplot graphics
				[ 
					xmin	= 0,
					xmax	= 31.5,
					ymin 	= -10.0,
					ymax	= 0.0
				]
				{../spectral_functions/precompiled/eq_U_m4p0_V_m0p25_f.eps};
			\end{groupplot}
		\end{tikzpicture}
	}

	\vspace{-1.5em}
	\subfloat[\label{fig:eq_neq_spectral_functions:5}]
	{
		\centering
		\tikzsetnextfilename{eq_neq_spectral_functions_5}
		\begin{tikzpicture}
			\pgfplotsset
			{
				/pgfplots/colormap={temp}{
					rgb255=(36,0,217) 		
					rgb255=(25,29,247) 		
					rgb255=(41,87,255) 		
					rgb255=(61,135,255) 	
					rgb255=(87,176,255) 	
					rgb255=(117,211,255) 	
					rgb255=(153,235,255) 	
					rgb255=(189,249,255) 	
					rgb255=(235,255,255) 	
					rgb255=(255,255,235) 	
					rgb255=(255,242,189) 	
					rgb255=(255,214,153) 	
					rgb255=(255,172,117) 	
					rgb255=(255,120,87) 	
					rgb255=(255,61,61) 		
					rgb255=(247,40,54) 		
					rgb255=(217,22,48) 		
					rgb255=(166,0,33)		
				}
			}
			\begin{groupplot}
				[
					group style = 
					{
						group size 		=	1 by 2,
						vertical sep		=	0.0pt,
						x descriptions at	=	edge bottom,
						y descriptions at	=	edge left,
					},
					xtick		= {0, 16, 32, 48},
					xticklabels	= {,,,},
					xlabel	= {},
					axis on top,
					enlargelimits	= false,
					width		= 0.4\textwidth-2.71pt,
				]
				\nextgroupplot
				[				
					ytick		= {0},
					axis x line*	= top,
					height		= 5.75em,
					ymin=-1.0, ymax=0.0,
					xmin=0, xmax=31.5,
					ytick		= {0},
				]
				\addplot graphics
				[ 
					xmin	= 0,
					xmax	= 31.5,
					ymin 	= -10.0,
					ymax	= 0.0
				]
				{../spectral_functions/precompiled/eq_U_m4p0_V_0p25_f.eps};
				\nextgroupplot
				[
					axis y discontinuity = parallel,
					title style	= {at={(1,0.85)},anchor=east,xshift=-0.75em},
					title		= {\color{white}\textbf{CDW phase}},
					height		= 0.14\textheight,
					ylabel		= {$\omega$},
					ylabel shift	= -7.5pt,
					ytick		= {-8,-4},
					colorbar right,
					colormap name	= temp,
					point meta min=0.0,
					point meta max=0.06,
					every colorbar/.append style =
					{
						height			=	\pgfkeysvalueof{/pgfplots/parent axis height}+0.885em,
						ylabel			= 	{$S^{\mathrm{CDW}}_{\hat c}(q,\omega,0)$},
						width			=	3mm,
						yshift			=	0.885em,
						scaled y ticks		= 	false,
						ytick			= 	{0,0.06},
						yticklabels		=	{$0\phantom{.00}$, $0.06$},
						ylabel shift 		=	-4pt,
					},
					ymin=-8.0, ymax=-2.0,
					xmin=0, xmax=31.5,
					axis x line*	= bottom,
				]
				\addplot graphics
				[ 
					xmin	= 0,
					xmax	= 31.5,
					ymin 	= -10.0,
					ymax	= 0.0
				]
				{../spectral_functions/precompiled/eq_U_m4p0_V_0p25_f.eps};
			\end{groupplot}
		\end{tikzpicture}
	}

	\vspace{-1.5em}
	\subfloat[\label{fig:eq_neq_spectral_functions:6}]
	{
		\centering
		\tikzsetnextfilename{eq_neq_spectral_functions_6}
		\begin{tikzpicture}
			\pgfplotsset
			{
				/pgfplots/colormap={temp}{
					rgb255=(36,0,217) 		
					rgb255=(25,29,247) 		
					rgb255=(41,87,255) 		
					rgb255=(61,135,255) 	
					rgb255=(87,176,255) 	
					rgb255=(117,211,255) 	
					rgb255=(153,235,255) 	
					rgb255=(189,249,255) 	
					rgb255=(235,255,255) 	
					rgb255=(255,255,235) 	
					rgb255=(255,242,189) 	
					rgb255=(255,214,153) 	
					rgb255=(255,172,117) 	
					rgb255=(255,120,87) 	
					rgb255=(255,61,61) 		
					rgb255=(247,40,54) 		
					rgb255=(217,22,48) 		
					rgb255=(166,0,33)		
				}
			}
			\begin{groupplot}
				[
					group style = 
					{
						group size 		=	1 by 2,
						vertical sep		=	0.0pt,
						x descriptions at	=	edge bottom,
						y descriptions at	=	edge left,
					},
					xtick		= {0, 16, 32, 48},
					xticklabels	= {$0$,$\nicefrac{\pi}{2}$,$\pi$,$\nicefrac{3\pi}{2}$},
					xlabel	= {$q$},
					axis on top,
					enlargelimits	= false,
					width		= 0.4\textwidth-2.71pt,
				]
				\nextgroupplot
				[
					ytick		= {0},
					axis x line*	= top,
					height		= 5.75em,
					ymin=-1.0, ymax=0.0,
					xmin=0, xmax=31.5,
				]
				\addplot graphics
				[ 
					xmin	= 0,
					xmax	= 31.5,
					ymin 	= -10.0,
					ymax	= 0.0
				]
				{../spectral_functions/precompiled/neq_U_m4p0_V_0p25_dV_m0p5_f.eps};
				\nextgroupplot
				[
					axis y discontinuity=parallel,
					title style	= {at={(1,0.85)},anchor=east,xshift=-0.75em},
					title		= {\color{white}\textbf{quenched state}},
					height		= 0.14\textheight,
					ylabel		= {$\omega$},
					ylabel shift	= -7.5pt,
					ytick		= {-8,-4},
					colorbar right,
					colormap name	= temp,
					point meta min=0.0,
					point meta max=0.02,
					every colorbar/.append style =
					{
						height			=	\pgfkeysvalueof{/pgfplots/parent axis height}+0.885em,
						ylabel			= 	{$S^{\phantom{\mathrm{C}}}_{\hat c}(q,\omega,15)$},
						width			=	3mm,
						yshift			=	0.885em,
						scaled y ticks		= 	false,
						ytick			= 	{0,0.02},
						yticklabels		=	{$0\phantom{.06}$, $0.02$},
						ylabel shift 		=	-4pt,
					},
					ymin=-8.0, ymax=-2.0,
					xmin=0, xmax=31.5,
					axis x line*	= bottom,
				]
				\addplot graphics
				[ 
					xmin	= 0.0,
					xmax	= 31.5,
					ymin 	= -10.0,
					ymax	= 0.0
				]
				{../spectral_functions/precompiled/neq_U_m4p0_V_0p25_dV_m0p5_f.eps};
			\end{groupplot}
		\end{tikzpicture}
	}
	\caption
	{%
		\label{fig:eq-neq:spectral-functions:electrons}%
		Spectral functions of single-particle excitations in equilibrium \gls{SC} phase \protect\subref{fig:eq_neq_spectral_functions:4}, \gls{CDW} phase \protect\subref{fig:eq_neq_spectral_functions:5}, and in non\hyp equilibrium after quenching from the \gls{CDW} ground state into the \gls{SC} phase evaluated at time $t=15$ \protect\subref{fig:eq_neq_spectral_functions:6}. 
		Note that only regions with significant spectral weight are displayed.
	}
\end{figure}
Current \gls{tr-ARPES} experiments usually investigate the time evolution of the spectral functions for single\hyp electron excitations $S_{\hat c}(q,\omega,\Delta t)$, which we show in
\cref{fig:eq-neq:spectral-functions:electrons}.
The signatures to discriminate the \gls{SC} phase from the \gls{CDW} phase are not as prominent as for the double occupations.
Nevertheless, we find that in the \gls{SC} ground state (\cref{fig:eq_neq_spectral_functions:4}) there is a shift of spectral weight towards $q=\nicefrac{\pi}{2}$.
This is to be contrasted with the distribution of weights in the \gls{CDW} ground state, where the maximum value is at $q=0$.
Comparing to the spectral function after the quench (\cref{fig:eq_neq_spectral_functions:6}), a weak transfer of spectral weight from $q = 0$ towards $q = \nicefrac{\pi}{2}$ can be identified.
However, from the numerical data it is not possible to clearly identify this as a signal for a (transient) \gls{SC} phase.
Note that the observed behavior is not reproduced for a quench within the CDW phase, as the two-electron excitations do not show a sharp signal at small momenta, indicating the absence of a transient \gls{SC} phase there \cite{supmat}.
\paragraph{Correlation Matrices}
Now we turn to the question on how to further characterize the observed transient state.
To do so, we attempt a standard analysis for the investigation of \gls{LRO} in equilibrium systems and apply it to the present non\hyp equilibrium situation.
%
In Ref.~\onlinecite{supersolid_penrose}, Onsager and Penrose suggest to detect off\hyp diagonal \gls{LRO} by determining the eigenvalues $\lambda_{\nu}$ of single\hyp particle correlation matrices,
\begin{align}
	\chi_{\hat O} &= \sum_{i,j} \mathbf{e}_i\braket{\psi | \hat{O}^{\dagger}_i \hat{O}^{\nodagger}_j | \psi} \mathbf{e}^{t}_j \equiv \sum_{i,j} \mathbf{e}_i \chi_{\hat O}(i,j) \mathbf{e}^{t}_j \; ,
\end{align}
i.e., $\chi_{\hat O} \mathbf{v}_{\nu} = \lambda_{\nu}\mathbf{v}_{\nu}$\footnote{We use the bold symbol $\mathbf e_j$ to denote canonical unit vectors}.
This concept was generalized by Yang \cite{RevModPhys.34.694} to include also two-particle correlation matrices to describe superconductivity.
The correlation matrix determines the order parameter if the dominating eigenvalue scales extensively in the system size $\lim_{L\rightarrow \infty}\frac{\lambda_{L}}{L} \rightarrow \mathcal{O}(1)$, which also implies an extensive separation of the dominating eigenvalue $\lambda_{L}$ from the bulk~\cite{supmat}.

Due to the Mermin-Wagner-Hohenberg theorem~\cite{merminwagner_orig,merminwagner_erratum,hohenberg_orig}, in \gls{1D} and in equilibrium, for pair formation only \gls{qLRO} can be realized, which translates to a \gls{SC} order parameter vanishing in the thermodynamic limit. 
Here, we studied the time evolution of the correlation matrix $\chi_{\hat d}(i,j) = \braket{\psi(t)| \hat{d}^{\dagger}_{i}\hat{d}^{\nodagger}_{j} | \psi(t)}$, which provides the \gls{SC} order parameter. 
\begin{figure}
	\centering
	\tikzsetnextfilename{neq_natural_orbital_occupations_separation}
	\pgfdeclarelayer{foreground}
	\pgfdeclarelayer{background}
	\pgfsetlayers{background,main,foreground}
	\begin{tikzpicture}
		[
			declare function = 
			{
				g1(\x) = \NeqScale*\x^\NeqExp + \NeqShift;
				g2(\x) = \EqScale*\x^\EqExp + \EqShift;
			},
		]
		\begin{pgfonlayer}{background}
			\begin{semilogyaxis}
			[
				width		= 0.5\textwidth-7.08pt,
				height		= 0.3\textheight,
				xmin 		= 0,
				xmax 		= 32,
				ymin		= 6e-06,
				ymax		= 1.05e-02,
				xlabel		= {$\text{time }t$},
				ylabel		= {$\Delta \lambda_/L$},
				smooth,
				legend style	= {font=\scriptsize,legend pos=south east},
			]
				\addplot
				[
					color=colorA,
					mark=*,
					unbounded coords=jump, 
					mark size = 2pt,
					mark repeat=8,
					thick
				]
				table
				[
					x expr = \thisrowno{0},
					y expr = (\thisrowno{1}-\thisrowno{3})/32
				]
				{../spectral_functions/neq/global_quench/L_32/1TDVP/U_m4p0_V_0p25_dV_m0p5/two_particle_dm/results/max.noo};
				\addplot
				[
					color=colorF,
					mark=*,
					unbounded coords=jump, 
					mark size = 2pt,
					mark repeat=8,
					thick
				]
				table
				[
					x expr = \thisrowno{0},
					y expr = (\thisrowno{1}-\thisrowno{3})/40
				]
				{../spectral_functions/neq/global_quench/L_40/1TDVP/U_m4p0_V_0p25_dV_m0p5/two_particle_dm/results/max.noo};
				\addplot
				[
					color=colorB,
					mark=*,
					unbounded coords=jump, 
					mark size = 2pt,
					mark repeat=8,
					thick
				]
				table
				[
					x expr = \thisrowno{0},
					y expr = (\thisrowno{1}-\thisrowno{3})/48
				]
				{../spectral_functions/neq/global_quench/L_48/1TDVP/U_m4p0_V_0p25_dV_m0p5/two_particle_dm/results/max.noo};
				\addplot
				[
					color=colorG,
					mark=*,
					unbounded coords=jump, 
					mark size = 2pt,
					mark repeat=8,
					thick
				]
				table
				[
					x expr = \thisrowno{0},
					y expr = (\thisrowno{1}-\thisrowno{3})/56
				]
				{../spectral_functions/neq/global_quench/L_56/1TDVP/U_m4p0_V_0p25_dV_m0p5/two_particle_dm/results/max.noo};
				\addplot
				[
					color=colorC,
					mark=*,
					unbounded coords=jump, 
					mark size = 2pt,
					mark repeat=8,
					thick
				]
				table
				[
					x expr = \thisrowno{0},
					y expr = (\thisrowno{1}-\thisrowno{3})/64
				]
				{../spectral_functions/neq/global_quench/L_64/1TDVP/U_m4p0_V_0p25_dV_m0p5/two_particle_dm/results/max.noo};
				\addplot
				[
					color=colorD,
					mark=*,
					unbounded coords=jump, 
					mark size = 2pt,
					mark repeat=8,
					thick
				]
				table
				[
					x expr = \thisrowno{0},
					y expr = (\thisrowno{1}-\thisrowno{3})/72
				]
				{../spectral_functions/neq/global_quench/L_72/1TDVP/U_m4p0_V_0p25_dV_m0p5/two_particle_dm/results/max.noo};
				\addplot
				[
					color=colorE,
					mark=*,
					unbounded coords=jump, 
					mark size = 2pt,
					mark repeat=8,
					thick
				]
				table
				[
					x expr = \thisrowno{0},
					y expr = (\thisrowno{1}-\thisrowno{3})/80
				]
				{../spectral_functions/neq/global_quench/L_80/1TDVP/U_m4p0_V_0p25_dV_m0p5/two_particle_dm/results/max.noo};
				\addlegendentry{$L=32$}
				\addlegendentry{$L=40$}
				\addlegendentry{$L=48$}
				\addlegendentry{$L=56$}
				\addlegendentry{$L=64$}
				\addlegendentry{$L=72$}
				\addlegendentry{$L=80$}
				\coordinate (insetPosition) at (axis cs:4.7,2e-5);
			\end{semilogyaxis}
		\end{pgfonlayer}
		\begin{pgfonlayer}{foreground}
			\begin{axis}
			[
				width		= 0.3\textwidth,
				height		= 0.175\textheight,
				at		= {(insetPosition)},
				xmin 		= 0.0,
				xmax 		= 0.05,
				ymin 		= 0.0,
				ymax		= 0.125,
				scaled x ticks	= false,
				scaled y ticks	= false,
				xtick		= {0,0.05,0.15},
				xticklabels	= {$0$, $0.05$, $0.15$},
				ytick		= {0,0.1},
				xlabel		= {$\nicefrac{1}{L}$},
				ylabel		= {$\nicefrac{\overbar{\lambda}_{L}}{L}$},
				smooth,
				ticklabel style	= {font=\scriptsize},
				xlabel style	= {font=\scriptsize,yshift=5pt},
				ylabel style	= {font=\tiny,yshift=-10pt},
				legend style	= {font=\tiny,legend pos=north west},
			]
				\addplot
				[
					color = colorA,
					mark = o,
					mark size = 2pt,
					thick,
					only marks,
				]
				plot
				[
					error bars/.cd,
					y dir=both,
					y explicit,
				]
				table
				[
					x expr = 1.0/(\thisrowno{0}),
					y expr = \thisrowno{1},
					y error index=2,
				]
				{../spectral_functions/neq/global_quench/results/two_particle_noo_extrapolation};
				\addplot
				[
					color = colorD,
					mark = square,
					mark size = 2pt,
					thick,
					only marks,
				]
				table
				[
					x expr = 1.0/(\thisrowno{0}),
					y expr = \thisrowno{1},
				]
				{../spectral_functions/eq/U_m4p0_V_m0p25/results/two_particle_extrapolation};
				\addplot
				[
					color = colorA,
					thick, 
					dotted,
					mark=none,
					domain=0.0:0.05,
					samples=1000,
				]
				{g1(x)};
				\addplot
				[
					color = colorD,
					thick,
					dotted,
					mark=none,
					domain=0.0:0.05,
					samples=1000,
				]
				{g2(x)}; 
				\addlegendentry{Post-Quench}
				\addlegendentry{SC-GS}
				\node at (axis cs: 0.035,0.055) {\tiny $\frac{\overbar \lambda^{\noprime}_L}{L} \sim L^{\EqExpShort} $};
				\node at (axis cs: 0.035,0.02) {\tiny $\frac{\overbar \lambda^{\noprime}_L}{L} \sim L^{\NeqExpShort}$};
			\end{axis}
		\end{pgfonlayer}
	\end{tikzpicture}
	\caption
	{
		\label{fig:neq:natural-orbital-occupations:separation}
		Separation of largest natural orbital occupation $\Delta \lambda/L = \frac{\lambda_{L}-\lambda_{L-1}}{L}$ during time evolution after quench for various system sizes.
		%
		The inset shows the scaling behavior of the normalized dominating eigenvalue $\overbar \lambda_L/L$ with system size $L$.
	}
\end{figure}
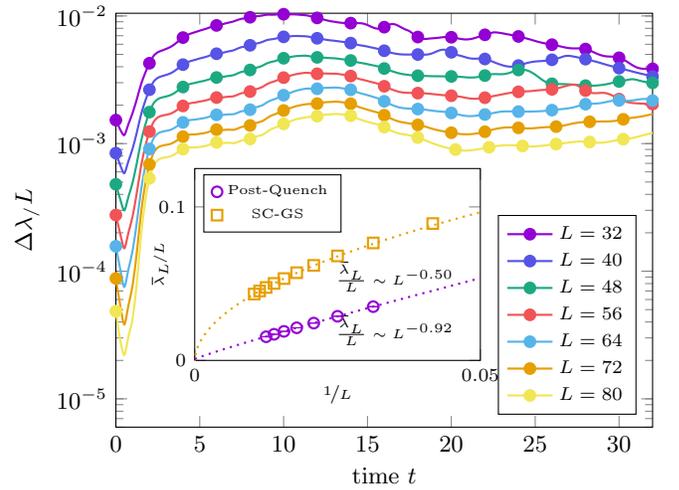
\Cref{fig:neq:natural-orbital-occupations:separation} shows the difference between the two largest eigenvalues $\Delta \lambda/L = \frac{\lambda_{L}-\lambda_{L-1}}{L}$ as a function of time. 
At later times we find this separation to be about an order of magnitude larger than in the initial state. 
We have estimated the saturation values of the dominating eigenvalue by averaging over the accessible time scales, in which we assumed quasi-stationarity,  $\overbar{\lambda}_{L} = \frac{1}{t_{1} - t_{0}}\sum_{t=t_0}^{t_1} \lambda_L(t)$ with $t_0 = 10, t_1=32$.
In the inset of \cref{fig:neq:natural-orbital-occupations:separation}, we compare the scaling behavior to the one of the \gls{SC} ground state, where $\overbar \lambda_L/L \sim L^{-0.50} = \nicefrac{1}{\sqrt{L}}$ is realized, with saturation value $\beta_{\mathrm{SC}} = -1\times 10^{-3} \approx 0$, as expected. 
After the quench, the scaling is best described by fitting the asymptotic behavior with $\overbar \lambda_L/L \sim L^{-0.92} \sim \nicefrac{1}{L}$, and in this case we extracted a value $\beta = 9\times 10^{-4}$, whose magnitude is even smaller than the one obtained in equilibrium, and hence zero within the error bars of our scaling analysis.
This scaling to zero indicates that no \gls{LRO} is obtained on the time scales investigated but cannot be excluded for later times~\cite{Lemonik2017,Lemonik2018}.
\paragraph{Conclusion and Outlook}
Our results for the differential optical conductivity in the extended Hubbard model after a quench indicate that the enhancement of $\sigma_2(\omega,\Delta t)$ in the low\hyp frequency regime does not suffice to uniquely identify \gls{SC}: The currents induced by a probe pulse cannot unambiguously be identified as supercurrents. 
In contrast, the pairing spectral function shows a clear accumulation of weights at $q \to 0$, which is absent in our initial state, but present in the \gls{SC} ground state.
This provides stronger evidence for the formation of a non\hyp equilibrium \gls{SC} state than the reflectivity measurements.
%
%
%
%

%
Our study indicates that reflectivity measurements need to be complemented by \gls{tr-ARPES}\hyp type experiments in future investigations.
Readily available \gls{tr-ARPES} setups measure the single\hyp particle spectral function, which shows a shift of weights towards the \gls{SC} state. 
We propose to also investigate the time evolution of pairing spectral functions, which provide the clearest evidence for the formation of a transient \gls{SC} state.
Our work therefore supplies an important theoretical result for the growing field of double photoemission spectroscopy.

\paragraph{Acknowledgements} 
We thank B. Lenz, L. Cevolani, K. Harms, N. Abeling, F. Heidrich-Meisner, S. Kaiser, G. Uhrig, and K. Sch\"onhammer for insightful discussions. 
B.F. and D.M. thank the Max Planck-UBC-UTokyo Center for Quantum Materials for financial support.
T.K.'s contribution was partially funded through an ERC Starting Grant from the European Union's Horizon 2020 research and innovation program under grant agreement No. 758935.
We acknowledge financial support by the Deutsche Forschungsgemeinschaft (DFG) through SFB/CRC1073  (projects  B03 and B07)  and  Research  Unit FOR  1807  (project  P7), and TU Clausthal for providing access to the Nuku computational cluster
as well as the GWDG for providing HPC resources. 
This work was supported in part by the Los Alamos National Laboratory LDRD Program.
S.P., T.K., and S.R.M. thank MPI-FKF for their hospitality. 

\bibliographystyle{prsty}
\bibliography{Literatur}

\begin{thebibliography}{10}

\bibitem{highTc_original}
J.~G. Bednorz and K.~A. M\"uller, Zeitschrift f\"ur Physik B Condensed Matter
  {\bf 64},  189  (1986).

\bibitem{Wu1987}
M.~K. Wu, J.~R. Ashburn, C.~J. Torng, P.~H. Hor, R.~L. Meng, L. Gao, Z.~J.
  Huang, Y.~Q. Wang, and C.~W. Chu, Phys. Rev. Lett. {\bf 58},  908  (1987).

\bibitem{dagotto}
E. Dagotto, Rev. Mod. Phys. {\bf 66},  763  (1994).

\bibitem{Lee2006}
P.~A. Lee, N. Nagaosa, and X.-G. Wen, Rev. Mod. Phys. {\bf 78},  17  (2006).

\bibitem{Foerst2014}
M. F\"orst, A. Frano, S. Kaiser, R. Mankowsky, C.~R. Hunt, J.~J. Turner, G.~L.
  Dakovski, M.~P. Minitti, J. Robinson, T. Loew, M. Le~Tacon, B. Keimer, J.~P.
  Hill, A. Cavalleri, and S.~S. Dhesi, Phys. Rev. B {\bf 90},  184514  (2014).

\bibitem{Mankowsky2014}
R. Mankowsky, A. Subedi, M. Först, S.~O. Mariager, M. Chollet, H.~T. Lemke,
  J.~S. Robinson, J.~M. Glownia, M.~P. Minitti, A. Frano, M. Fechner, N.~A.
  Spaldin, T. Loew, B. Keimer, A. Georges, and A. Cavalleri, Nature {\bf 516},
  71  (2014).

\bibitem{Hu2014}
W. Hu, S. Kaiser, D. Nicoletti, C.~R. Hunt, I. Gierz, M.~C. Hoffmann, M.
  Le~Tacon, T. Loew, B. Keimer, and A. Cavalleri, Nature Materials {\bf 13},
  705  (2014).

\bibitem{Hunt2016}
C.~R. Hunt, D. Nicoletti, S. Kaiser, D. Pr\"opper, T. Loew, J. Porras, B.
  Keimer, and A. Cavalleri, Phys. Rev. B {\bf 94},  224303  (2016).

\bibitem{Mitrano2016a}
M. Mitrano, A. Cantaluppi, D. Nicoletti, S. Kaiser, A. Perucchi, S. Lupi, P.
  Di~Pietro, D. Pontiroli, M. Riccò, S.~R. Clark, D. Jaksch, and A. Cavalleri,
  Nature {\bf 530},  461  (2016).

\bibitem{Foerst2011}
M. Först, C. Manzoni, S. Kaiser, Y. Tomioka, Y. Tokura, R. Merlin, and A.
  Cavalleri, Nature Physics {\bf 7},  854  (2011).

\bibitem{Fausti189}
D. Fausti, R.~I. Tobey, N. Dean, S. Kaiser, A. Dienst, M.~C. Hoffmann, S. Pyon,
  T. Takayama, H. Takagi, and A. Cavalleri, Science {\bf 331},  189  (2011).

\bibitem{Kaiser2014}
S. Kaiser, C.~R. Hunt, D. Nicoletti, W. Hu, I. Gierz, H.~Y. Liu, M. Le~Tacon,
  T. Loew, D. Haug, B. Keimer, and A. Cavalleri, Phys. Rev. B {\bf 89},  184516
   (2014).

\bibitem{buzzi2020photomolecular}
M. Buzzi, D. Nicoletti, M. Fechner, N. Tancogne-Dejean, M.~A. Sentef, A.
  Georges, M. Dressel, A. Henderson, T. Siegrist, J.~A. Schlueter, K. Miyagawa,
  K. Kanoda, M.~S. Nam, A. Ardavan, J. Coulthard, J. Tindall, F. Schlawin, D.
  Jaksch, and A. Cavalleri, arXiv e-prints  arXiv:2001.05389  (2020).

\bibitem{Singla2015}
R. Singla, G. Cotugno, S. Kaiser, M. F\"orst, M. Mitrano, H.~Y. Liu, A.
  Cartella, C. Manzoni, H. Okamoto, T. Hasegawa, S.~R. Clark, D. Jaksch, and A.
  Cavalleri, Phys. Rev. Lett. {\bf 115},  187401  (2015).

\bibitem{Eckstein2010}
M. Eckstein, M. Kollar, and P. Werner, Phys. Rev. B {\bf 81},  115131  (2010).

\bibitem{PhysRevB.96.054506}
D.~M. Kennes, E.~Y. Wilner, D.~R. Reichman, and A.~J. Millis, Phys. Rev. B {\bf
  96},  054506  (2017).

\bibitem{PhysRevB.96.235142}
Y. Wang, M. Claassen, B. Moritz, and T.~P. Devereaux, Phys. Rev. B {\bf 96},
  235142  (2017).

\bibitem{Bittner2019}
N. Bittner, T. Tohyama, S. Kaiser, and D. Manske, Journal of the Physical
  Society of Japan {\bf 88},  044704  (2019).

\bibitem{Fauseweh2017}
B. {Fauseweh}, L. {Schwarz}, N. {Tsuji}, N. {Cheng}, N. {Bittner}, H. {Krull},
  M. {Berciu}, G.~S. {Uhrig}, A.~P. {Schnyder}, S. {Kaiser}, and D. {Manske},
  Nat. Commun. {\bf 11},  287  (2020).

\bibitem{schwarz2020momentumresolved}
L. Schwarz, B. Fauseweh, and D. Manske, arXiv e-prints  arXiv:2002.05904
  (2020).

\bibitem{White1992}
S.~R. White, Phys. Rev. Lett. {\bf 69},  2863  (1992).

\bibitem{Schollwoeck2010}
U. Schollw{\"o}ck, Annals of Physics {\bf 326},  96   (2011).

\bibitem{Paeckel2019}
S. {Paeckel}, T. {K{\"o}hler}, A. {Swoboda}, S.~R. {Manmana}, U.
  {Schollw{\"o}ck}, and C. {Hubig}, Annals of Physics {\bf 411},  167998
  (2019).

\bibitem{Damascelli2003}
A. Damascelli, Z. Hussain, and Z.-X. Shen, Rev. Mod. Phys. {\bf 75},  473
  (2003).

\bibitem{Damascelli2004}
A. Damascelli, Physica Scripta {\bf T109},  61  (2004).

\bibitem{Lynch2005}
D. Lynch and C. Olson, {\em Photoemission Studies of High-Temperature
  Superconductors}, {\em Cambridge Studies in Low Temperature Physics}
  (Cambridge University Press, Cambridge, 2005).

\bibitem{Eckstein2008}
M. Eckstein and M. Kollar, Phys. Rev. B {\bf 78},  245113  (2008).

\bibitem{Freericks2009}
J.~K. Freericks, H.~R. Krishnamurthy, and T. Pruschke, Phys. Rev. Lett. {\bf
  102},  136401  (2009).

\bibitem{Ligges2018}
M. Ligges, I. Avigo, D. Gole\ifmmode~\check{z}\else \v{z}\fi{}, H.~U.~R.
  Strand, Y. Beyazit, K. Hanff, F. Diekmann, L. Stojchevska, M. Kall\"ane, P.
  Zhou, K. Rossnagel, M. Eckstein, P. Werner, and U. Bovensiepen, Phys. Rev.
  Lett. {\bf 120},  166401  (2018).

\bibitem{wang2018theoretical}
Y. Wang, M. Claassen, C.~D. Pemmaraju, C. Jia, B. Moritz, and T.~P. Devereaux,
  Nature Reviews Materials  1  (2018).

\bibitem{merminwagner_orig}
N.~D. Mermin and H. Wagner, Phys. Rev. Lett. {\bf 17},  1133  (1966).

\bibitem{merminwagner_erratum}
N.~D. Mermin and H. Wagner, Phys. Rev. Lett. {\bf 17},  1307  (1966).

\bibitem{hohenberg_orig}
P.~C. Hohenberg, Phys. Rev. {\bf 158},  383  (1967).

\bibitem{voit}
J. Voit, Rep. Prog. Phys. {\bf 58},  977  (1995).

\bibitem{PRBVoit2}
J. Voit, Phys. Rev. B {\bf 45},  4027  (1992).

\bibitem{Jeckelmann2002}
E. Jeckelmann, Phys. Rev. Lett. {\bf 89},  236401  (2002).

\bibitem{Tsuchiizu2002}
M. Tsuchiizu and A. Furusaki, Phys. Rev. Lett. {\bf 88},  056402  (2002).

\bibitem{Sandvik2004}
A.~W. Sandvik, L. Balents, and D.~K. Campbell, Phys. Rev. Lett. {\bf 92},
  236401  (2004).

\bibitem{Ejima2007}
S. Ejima and S. Nishimoto, Phys. Rev. Lett. {\bf 99},  216403  (2007).

\bibitem{PhysRevB.85.205127}
F. Hofmann and M. Potthoff, Phys. Rev. B {\bf 85},  205127  (2012).

\bibitem{tJoriginal0}
K.~A. Chao, J. Spalek, and A.~M. Oles, Journal of Physics C: Solid State
  Physics {\bf 10},  L271  (1977).

\bibitem{tJoriginal1}
P.~W. Anderson, Science {\bf 235},  1196  (1987).

\bibitem{tJoriginal2}
F.~C. Zhang and T.~M. Rice, Phys. Rev. B {\bf 37},  3759  (1988).

\bibitem{PhysRevB.83.205113NO}
A. Moreno, A. Muramatsu, and S.~R. Manmana, Phys. Rev. B {\bf 83},  205113
  (2011).

\bibitem{PhysRevLett.107.115301}
A.~V. Gorshkov, S.~R. Manmana, G. Chen, J. Ye, E. Demler, M.~D. Lukin, and
  A.~M. Rey, Phys. Rev. Lett. {\bf 107},  115301  (2011).

\bibitem{PhysRevA.96.043618}
S.~R. Manmana, M. M\"oller, R. Gezzi, and K.~R.~A. Hazzard, Phys. Rev. A {\bf
  96},  043618  (2017).

\bibitem{Hubbard1963}
J. Hubbard, Proceedings of the Royal Society of London A: Mathematical,
  Physical and Engineering Sciences {\bf 276},  238  (1963).

\bibitem{Gutzwiller1963}
M.~C. Gutzwiller, Phys. Rev. Lett. {\bf 10},  159  (1963).

\bibitem{Kanamori1963}
J. Kanamori, Progress of Theoretical Physics {\bf 30},  275  (1963).

\bibitem{HubbBook}
F.~H.~L. Essler, H. Frahm, F. G{\"o}hmann, A. Kl{\"u}mper, and V.~E. Korepin,
  {\em The One-Dimensional {H}ubbard Model} (Cambridge University Press,
  Cambridge, 2005).

\bibitem{Eisert2015a}
J. Eisert, M. Friesdorf, and C. Gogolin, Nature Physics {\bf 11},  124  (2015).

\bibitem{BittnerPhD}
N. Bittner, Novel Non–Equilibrium Dynamics in Superconductors: Induced
  Superconductivity and Higgs Modes, ~PhD thesis, Freie Universit{\"a}t Berlin,
  2017.

\bibitem{supmat}
See Supplemental Material at [URL will be inserted by publisher] for a detailed
  description of the definition and numerical details of the spectral
  functions, (non-)thermal properties of the postquench state, scaling analysis
  of the eigenvalues of the correlation matrices (natural orbitals), and
  results for the $t-J_\perp$ model.

\bibitem{Nomurae1500568}
Y. Nomura, S. Sakai, M. Capone, and R. Arita, Science Advances {\bf 1},
  (2015).

\bibitem{PhysRevB.94.165116}
J. Haegeman, C. Lubich, I. Oseledets, B. Vandereycken, and F. Verstraete, Phys.
  Rev. B {\bf 94},  165116  (2016).

\bibitem{Lenar2014}
Z. Lenar\ifmmode \check{c}\else \v{c}\fi{}i\ifmmode~\check{c}\else \v{c}\fi{},
  D. Gole\ifmmode~\check{z}\else \v{z}\fi{}, J. Bon\ifmmode~\check{c}\else
  \v{c}\fi{}a, and P. Prelov\ifmmode~\check{s}\else \v{s}\fi{}ek, Phys. Rev. B
  {\bf 89},  125123  (2014).

\bibitem{Shao2016}
C. Shao, T. Tohyama, H.-G. Luo, and H. Lu, Phys. Rev. B {\bf 93},  195144
  (2016).

\bibitem{supersolid_penrose}
O. Penrose and L. Onsager, Phys. Rev. {\bf 104},  576  (1956).

\bibitem{rigol:031603}
M. Rigol and A. Muramatsu, Phys. Rev. A {\bf 70},  031603  (2004).

\bibitem{Kouzakov2003}
K.~A. Kouzakov and J. Berakdar, Phys. Rev. Lett. {\bf 91},  257007  (2003).

\bibitem{Truetzschler2017}
A. Tr\"utzschler, M. Huth, C.-T. Chiang, R. Kamrla, F.~O. Schumann, J.
  Kirschner, and W. Widdra, Phys. Rev. Lett. {\bf 118},  136401  (2017).

\bibitem{Stahl2019}
C. Stahl and M. Eckstein, Phys. Rev. B {\bf 99},  241111  (2019).

\bibitem{giamarchi}
T. Giamarchi, {\em Quantum Physics in One Dimension}, Vol.~121 of {\em
  International Series of Monographs on Physics} (Oxford University Press,
  Oxford, 2004).

\bibitem{Note1}
The operators $S_j^\pm $ are the usual $S-\protect \nicefrac {1}{2}$ ladder
  operators on site $j$. Note that double occupancies are forbidden in the
  $t$-$J$ model.

\bibitem{Note2}
For the width of the probe pulse we chose $\tau =0.05$ throughout this letter.

\bibitem{PhysRevB.14.2239}
D.~R. Hofstadter, Phys. Rev. B {\bf 14},  2239  (1976).

\bibitem{Note3}
In~\cite {Shao2016} the non\protect \hyp equilibrium response function is
  obtained by considering the Fourier transform of the difference $j(t,\Delta
  t) - j_{0}(t)$ with $j_0(t)$ the current without probe pulse. As in our
  quench protocol without probe pulse there is always $j_0(t)\equiv 0$ we drop
  this term for brevity.

\bibitem{Note4}
We use the bold symbol $\protect \mathbf e_j$ to denote canonical unit vectors.

\bibitem{RevModPhys.34.694}
C.~N. Yang, Rev. Mod. Phys. {\bf 34},  694  (1962).

\bibitem{Lemonik2017}
Y. Lemonik and A. Mitra, Phys. Rev. B {\bf 96},  104506  (2017).

\bibitem{Lemonik2018}
Y. Lemonik and A. Mitra, Phys. Rev. Lett. {\bf 121},  067001  (2018).

\end{thebibliography}


\begin{thebibliography}{10}

\bibitem{White1992}
S.~R. White, Phys. Rev. Lett. {\bf 69},  2863  (1992).

\bibitem{Schollwoeck2010}
U. Schollw{\"o}ck, Annals of Physics {\bf 326},  96   (2011).

\bibitem{Paeckel2019}
S. {Paeckel}, T. {K{\"o}hler}, A. {Swoboda}, S.~R. {Manmana}, U.
  {Schollw{\"o}ck}, and C. {Hubig}, Annals of Physics {\bf 411},  167998
  (2019).

\bibitem{hastings_arealaw}
M.~B. Hastings, Journal of Statistical Mechanics: Theory and Experiment {\bf
  2007},  P08024  (2007).

\bibitem{Verstraete2006}
F. Verstraete and J.~I. Cirac, Phys. Rev. B {\bf 73},  094423  (2006).

\bibitem{PhysRevB.94.165116}
J. Haegeman, C. Lubich, I. Oseledets, B. Vandereycken, and F. Verstraete, Phys.
  Rev. B {\bf 94},  165116  (2016).

\bibitem{Paeckel}
S. Paeckel and T. K\"ohler, The \textsc{SymMPS} Toolkit,
  \url{https://www.symmps.eu}, accessed: 2019-12-29.

\bibitem{supersolid_penrose}
O. Penrose and L. Onsager, Phys. Rev. {\bf 104},  576  (1956).

\bibitem{kagome_ioannisNO}
I. Rousochatzakis, S.~R. Manmana, A.~M. L\"auchli, B. Normand, and F. Mila,
  Phys. Rev. B {\bf 79},  214415  (2009).

\bibitem{merminwagner_orig}
N.~D. Mermin and H. Wagner, Phys. Rev. Lett. {\bf 17},  1133  (1966).

\bibitem{merminwagner_erratum}
N.~D. Mermin and H. Wagner, Phys. Rev. Lett. {\bf 17},  1307  (1966).

\bibitem{hohenberg_orig}
P.~C. Hohenberg, Phys. Rev. {\bf 158},  383  (1967).

\bibitem{rigol:031603}
M. Rigol and A. Muramatsu, Phys. Rev. A {\bf 70},  031603  (2004).

\bibitem{rigol:013604}
M. Rigol and A. Muramatsu, Phys. Rev. A {\bf 72},  013604  (2005).

\bibitem{PhysRevLett.107.115301}
A.~V. Gorshkov, S.~R. Manmana, G. Chen, J. Ye, E. Demler, M.~D. Lukin, and
  A.~M. Rey, Phys. Rev. Lett. {\bf 107},  115301  (2011).

\bibitem{PhysRevA.96.043618}
S.~R. Manmana, M. M\"oller, R. Gezzi, and K.~R.~A. Hazzard, Phys. Rev. A {\bf
  96},  043618  (2017).

\bibitem{giamarchi}
T. Giamarchi, {\em Quantum Physics in One Dimension}, Vol.~121 of {\em
  International Series of Monographs on Physics} (Oxford University Press,
  Oxford, 2004).

\end{thebibliography}

\end{document}


\title{Supplementary Material for ``Detecting superconductivity out-of-equilibrium''}

\author{S.~Paeckel}
\affiliation{Fakult\"at f\"ur Physik, Ludwig-Maximilians-Universit\"at M\"unchen, D-80333 M\"nchen, Germany}
\affiliation{Institut f\"ur Theoretische Physik, Georg-August-Universit\"at G\"ottingen, D-37077 G\"ottingen, Germany}

\author{B.~Fauseweh}
\affiliation{Max-Planck-Institut f\"ur Festk\"orperforschung, Heisenbergstra\ss{}e 1, D-70569 Stuttgart, Germany}
\affiliation{Theoretical Division, Los Alamos National Laboratory, Los Alamos, New Mexico 87545, USA}

\author{A.~Osterkorn}
\affiliation{Institut f\"ur Theoretische Physik, Georg-August-Universit\"at G\"ottingen, D-37077 G\"ottingen, Germany}

\author{T.~K\"ohler}
\affiliation{Department of Physics and Astronomy, Uppsala University, Box 516, S-751 20 Uppsala, Sweden}
\affiliation{Institut f\"ur Theoretische Physik, Georg-August-Universit\"at G\"ottingen, D-37077 G\"ottingen, Germany}

\author{D.~Manske}
\affiliation{Max-Planck-Institut f\"ur Festk\"orperforschung, Heisenbergstra\ss{}e 1, D-70569 Stuttgart, Germany}

\author{S.R.~Manmana}
\affiliation{Institut f\"ur Theoretische Physik, Georg-August-Universit\"at G\"ottingen, D-37077 G\"ottingen, Germany}

\maketitle
\section{Spectral Function}\label{app:spectral-function}
\paragraph{General theory}
We defined the differential spectral function $S_{\hat O}(q,\omega,\Delta t)$ by the power spectrum in momentum space of the propagator of the operators $\hat{O}_{j}$
\begin{align}
	S_{\hat O}(q, \omega ,\Delta t) 
	&= 
	\int_{-\infty}^{\infty} \frac{d t}{2\pi} e^{\mathrm i \omega t} C_{\hat O}(q,t,\Delta t) \; , \label{eq:def-spec-func}\\
	C_{\hat O}(q,t,\Delta t)
	&=
	\frac{1}{L}\sum_{r_{ij}} e^{-\mathrm iqr_{ij}}\braket{\psi(\Delta t) | \hat{O}^{\dagger}_{i} (t) \hat{O}^{\nodagger}_{j} (0) | \psi(\Delta t)} \notag \\
	&= \braket{\hat{O}^{\dagger}_q (t) \hat{O}^{\nodagger}_{q} (0)}_{\Delta t}\; ,
\end{align}
where we defined $r_{ij}= r_i -r_j = (i-j)\cdot a$ in terms of the lattice spacing $a$ and abbreviated the expectation values at time delay $\Delta t$ by $\braket{\psi(\Delta t) | \cdots | \psi(\Delta t)} \equiv \braket{\cdots}_{\Delta t}$.
The $\hat{O}_i$'s are operators obeying canonical (anti-)commutation relations $\left[ \hat{O}^{\nodagger}_i, \hat{O}^{\dagger}_j \right]_{\pm} \propto \delta_{ij}$ and so do their Fourier transformed counterparts 
\begin{align}
	\hat{O}^{\nodagger}_q = \frac{1}{\sqrt{L}}\sum_{r_i}e^{-\mathrm i q r_i}\hat{O}^{\nodagger}_i\;, 
	\quad 
	\left[ \hat{O}^{\nodagger}_q, \hat{O}^{\dagger}_k \right]_{\pm} \propto \delta_{qk} \; .
\end{align}
The differential spectral function is reformulated to simplify the numerical evaluation:
\begin{widetext}
	\begin{align}
		S_{\hat O}(q, \omega ,\Delta t) 
		&=
		\int_{0}^{\infty} \frac{d t}{2\pi} \left[ e^{\mathrm i \omega t} C_{\hat O}(q,t,\Delta t) + e^{-\mathrm i \omega t} C_{\hat O}(q,-t,\Delta t) \right]
		=
		\int_{0}^{\infty} \frac{d t}{2\pi} \left[ e^{\mathrm i \omega t} C_{\hat O}(q,t,\Delta t) + e^{-\mathrm i \omega t} C^{*}_{\hat O}(q,t,\Delta t) \right] \label{eq:appendix:timetranslational-invariant-propagator}\; ,
	\end{align}
\end{widetext}
where we assumed a (quasi\hyp )steady state to exploit time translational invariance to shift the argument: $\braket{\hat{O}^{\dagger}_q(-t)\hat{O}^{\nodagger}_{q}(0)} = \braket{\hat{O}^{\dagger}_q(0) \hat{O}^{\nodagger}_{q}(t)} = \braket{\hat{O}^{\dagger}_q(t) \hat{O}^{\nodagger}_q(0)}^{*}$.
Thus, we can evaluate the time integral by taking the real part of the Fourier transformation and restrict the integration domain to $t \geq 0$.
Note that in order to discretize the Fourier transformation, we defined the limit as
\begin{align}
	S_{\hat O}(q, \omega, \Delta t) 
	&= 
	\lim_{T\rightarrow \infty} \frac{1}{T} \int_{-\nicefrac{T}{2}}^{\nicefrac{T}{2}} dt \; e^{\mathrm i \omega t} C_{\hat O}(q,t,\Delta t) \; .
\end{align}
Using \cref{eq:appendix:timetranslational-invariant-propagator}, we thus numerically evaluated the Fourier transformation via
\begin{align}
	S_{\hat O}(q, \omega_n, \Delta t) &= \frac{\delta}{T} \Re \sum_{m=0}^{N_T - 1} e^{\mathrm i \omega_n t_m} C_{\hat O}(q,t_m,\Delta t) \; ,
	\label{eq:appendix:spectral_function_numerics}
\end{align}
with discretized frequencies $\omega_n = n\frac{2\pi}{T}$ ($n=0,\ldots, N_T-1$), $t_m = m \delta$ and the summation range fixed by the time step $\delta = \nicefrac{T}{N_T}$.
Note that this way we only obtain the positive frequency part which we accounted for by shifting the time arguments, relabeling $t_m \rightarrow t_m - \nicefrac{T}{2} \Rightarrow \omega_n \rightarrow \omega_n - \nicefrac{\pi}{T}$.
\paragraph{Numerical technique}
We calculated the time-evolution of the lesser Greens function
\begin{align}
	C_{\hat O}(j,t,\Delta t)
	&=
	\braket{\psi(\Delta t)| \hat O^\dagger_j(t)\hat O^{\nodagger}_{\nicefrac{L}{2}}(0) |\psi(\Delta t)}\; ,
\end{align}
by representing the states and operators in the \gls{MPS} and \gls{MPO} framework~\cite{White1992,Schollwoeck2010,Paeckel2019}.
Within this framework the wavefunctions $\ket{\psi} = \sum_{\sigma_1,\ldots \sigma_L} \psi_{\sigma_1,\ldots,\sigma_L} \ket{\sigma_1,\ldots,\sigma_L}$ are approximated in terms of a tensor-network decomposition of the coefficients of the states
\begin{align}
	\psi_{\sigma_1,\ldots,\sigma_L}
	&=
	\sum_{\left\{\alpha_{j}\right\}} M^{\sigma_1}_{\alpha_1} M^{\sigma_2}_{\alpha_1,\alpha_2} \cdots M^{\sigma_{L-1}}_{\alpha_{L-2},\alpha_{L-1}} M^{\sigma_L}_{\alpha_{L-1}} \; ,
\end{align}
and a similar expansion exists for operators.
The range of the so-called bond indices $\alpha_j$ is usually known as bond dimension of the \gls{MPS} and is closely related to the van-Neumann entanglement entropy $S_N$ of the state for a bipartition of the Hilbert space at bond $(j-1,j)$.
It can be shown that by fixing the maximal bond dimesion to a certain value $m$ states can be faithfully approximated as long as their bipartite entanglement entropy is not significantly exceeding values of $\log_{\sigma}(m)$ where $\sigma$ denotes the dimension of the local degrees of freedom, i.e., the Hubbard ($\sigma=4$) or the $t$-$J$-basis ($\sigma=3$) of spin-$\nicefrac{1}{2}$ fermions in the systems treated by us.
%
This condition is usually fulfilled for ground states of gapped systems in one dimension~\cite{hastings_arealaw,Verstraete2006} but is no longer valid out\hyp of\hyp equilibrium.
In order to evaluate the time-dependent Greens function we used the Schr\"odinger picture to decompose the scalar product
\begin{align}
	C_{\hat O}(j,t,\Delta t)
	&\equiv
	\braket{\psi(t,\Delta t) | \hat O^\dagger_j | \phi(t,\Delta t)} \notag \\
	\ket{\psi(t,\Delta t)} 
	&= \hat U (t)\ket{\psi(\Delta t)} \\
	\ket{\phi(t,\Delta t)}
	&=
	\hat U(t) \hat O^{\nodagger}_{\nicefrac{L}{2}}(0) \ket{\psi(\Delta t)}
\end{align}
and calculated the time evolutions independently.
The maximally achieved time scales when evaluating the time evolution of the lesser Greens function were $t = 32$.
For determining the time evolution itself we used a hybrid version of the \gls{TDVP} algorithm~\cite{PhysRevB.94.165116,Paeckel2019} where we combined a two-site solver in the initial stage of the time evolution to grow the bond dimensions until the maximum value was reached on all bonds.
Then we switched to a faster single-site solver of the local \gls{TDVP} equations to continue the time evolution.
The nature of the problem demands a high numerical effort to obtain the results presented in this letter.
Therefore, we used a state-of the art implementation exploiting $U(1)$-symmetries~\cite{Paeckel}.
\paragraph{Useful identities}
Integrating over the frequency domain, the propagator fulfills
\begin{widetext}
	\begin{align}
		\int_{-\infty}^{\infty} d\omega \int_{-\infty}^{\infty} \frac{dt}{2\pi} e^{i\omega t}C_{\hat O}(q,t,\Delta t)
		&=
		\int_{-\infty}^{\infty} d\omega \int_{-\infty}^{\infty} \frac{dt}{2\pi} e^{i\omega t} \sum_{n,m} \braket{\psi(\Delta t) | n} \braket{n | \hat{O}^{\dagger}_{q}(t) | m } \braket{m | \hat{O}^{\nodagger}_{q} | \psi(\Delta t)} 
		=
		\braket{\hat{O}^{\dagger}_{q}\hat{O}^{\nodagger}_{q}}_{\Delta t}\; ,
	\end{align}
\end{widetext}
which implies the sum rule
\begin{align}
	2\sum_{q} \int_{0}^{\infty} d\omega \Re S_{\hat O}(q,\omega,\Delta t) &= \sum_{q} \braket{\hat{O}^{\dagger}_{q}\hat{O}^{\nodagger}_q}_{\Delta t} \; .
\end{align}
Thus, for $\hat{O}^{\nodagger}_{i} = \hat{c}^{\nodagger}_{i,\sigma}$ this yields the total number of particles with local spin projection $\sigma$, while for $\hat{O}^{\nodagger}_i = \hat{d}^{\nodagger}_{i}$ we expect the overall number of doublons in the system.
Note that the overall doublon occupation $\hat{D} = \sum_{i}\hat{d}^{\dagger}_{i}\hat{d}^{\nodagger}_{i}$ in general is not conserved by $\hat{H}$.
However, the sum rule holds at each time delay $\Delta t$. In particular, if the correlation function is invariant under time-translations on some time interval $I$, $\hat D$ will be conserved on $I$.
\section{Natural Orbitals}\label{app:natural-orbitals}
\paragraph{General Theory} 
We consider a correlation matrix for some local observable $\hat{O}_j$
\begin{equation}
	\chi_{\hat O}(i,j) = \braket{\psi| \hat{O}^{\dagger}_i\hat{O}^{\nodagger}_j |\psi}\; .
\end{equation}
Then, off\hyp diagonal \gls{LRO} in the thermodynamic limit is defined according to
\begin{equation}
	\lim_{\lvert i-j \rvert \rightarrow \infty} \chi_{\hat O}(i,j) > 0\; ,
\end{equation}
i.e., the correlation function $\chi_{\hat O}(i,j)$ saturates at a finite value for arbitrary large separations $\vert i-j \vert$.
In the following we will briefly summarize the discussion on natural orbitals (see, for instance, \cite{supersolid_penrose,kagome_ioannisNO}) to review the tools exploited in the main text to identify off\hyp diagonal \gls{LRO}.
For a finite system with $L$ lattice sites, $\chi_{\hat O}(i,j)$ is a Hermitian $L\times L$ matrix, which we can formally diagonalize to obtain real eigenvalues $\lambda_{\nu}$ and corresponding eigenvectors $\mathbf{v}_{\nu}$.
Expanding $\chi_{\hat O}(i,j)$ as matrix $\chi_{\hat O}$ in its eigenbasis, we can write
\begin{align}
	\chi_{\hat O} 
	&= 
	\sum_{\nu} \mathbf{v}^{\nodagger}_{\nu}\underbrace{\braket{\psi|\left( \sum_i v^{*}_{\nu,i}\hat{O}^{\dagger}_i\right) \left(\sum_j v_{\nu,j} \hat{O}^{\nodagger}_{j} \right)| \psi}}_{\lambda_{\nu}} \mathbf{v}^{\dagger}_{\nu} \;.\label{eq:corr-matrix:nat-orb}
\end{align}
Introducing field operators $\hat{\eta}_{\nu} = \sum_{i}v_{\nu,i}\hat{O}_{i}$, their squared expectation values are related to the eigenvalues $\lambda_{\nu}$ of $\chi_{\hat O}$
\begin{equation}
	\mathbf{v}^{\dagger}_{\nu}\chi_{\hat O}\mathbf{v}_{\nu}
	=
	\braket{\psi|\hat{\eta}^{\dagger}_{\nu} \hat{\eta}^{\nodagger}_{\nu} |\psi}
	=
	\lambda_{\nu} \geq 0\;,
\end{equation}
i.e., the eigenvalues are strictly positive. 
If we impose $\left[ \hat{O}^{\nodagger}_i, \hat{O}^{\dagger}_j \right]_{\epsilon} = \delta_{ij}$ with $\epsilon = \pm 1$ choosing between the commutator ($+1$) and anticommutator ($-1$), it follows that
\begin{align}
	\left[\hat{\eta}^{\nodagger}_{\nu}, \hat{\eta}^{\dagger}_{\mu}\right]_{\epsilon}
	&=
	\sum_{i,j} \left[\hat{O}^{\nodagger}_i, \hat{O}^{\dagger}_j \right]_{\epsilon} v^{\phantom{*}}_{\nu,i} v^{*}_{\mu,j} = \delta_{\nu\mu}\; .
\end{align}
This suggests to call the eigenvectors $\mathbf v_{\nu}$ natural orbitals and the eigenvalues natural orbital occupations $\lambda_{\nu}$ of the correlation matrix $\chi_{\hat O}$.
The existence of off-diagonal \gls{LRO} can now be related to properties of the natural orbitals.
To see this, note that due to the canonical commutator relations of $\hat O_j$ the trace over $\chi_{\hat O}$ corresponds to the total occupation $N_{\hat O} = \sum_j \braket{\hat n_j} \equiv \sum_j \braket{\hat O^{\dagger}_j \hat O^{\nodagger}_j}$ of the single particle states generated by the $\hat O_j$'s.
If we assume that there is a natural orbital $\mathbf v_L$ that is macroscopically occupied, i.e., $\nicefrac{\lambda_L}{N_{\hat O}} \sim \mathcal{O}(1)$, it is instructive to pull out this contribution from the sum in \cref{eq:corr-matrix:nat-orb}.
Written out componentwise, we find
\begin{align}
	\chi_{\hat O}(i,j)
	&=
	\lambda_L v^{\phantom{*}}_{L,i} v^{*}_{L,j} + \sum_{\nu<L} \lambda_{\nu} v^{\phantom{*}}_{\nu,i} v^{*}_{\nu,j}\; .
\end{align}
The natural orbitals are normalized so that in general the components fulfill $v_{\nu,j} \sim \nicefrac{1}{\sqrt{L}}$.
Therefore, an extensively scaling natural orbital occupation $\lambda_L$ yields a dominating leading summand in the above expansion of the correlation matrix, while the remaining summands are suppressed as $\nicefrac{1}{L}$.
In this situation, for increasing system sizes $L$ the correlation matrix is well approximated by neglecting the latter,
\begin{align}
	\chi_{\hat O}(i,j) 
	&= 
	\lambda_L v^{\phantom{*}}_{L,i} v^{*}_{L,j} + \mathcal{O}(\nicefrac{1}{L})
	\approx
	\lambda_L v^{\phantom{*}}_{L,i} v^{*}_{L,j} \; .
\end{align}
The correlation matrix obtained this way factorizes in the spatial components $(i,j)$.
Since, in a state with explicitely broken symmetry with respect to $\hat O_j$, one expects $\braket{\hat O_j} \neq 0$, this permits to identify
\begin{align}
	\chi_{\hat O}(i,j)
	&=
	\braket{\hat O^{\dagger}_i}\braket{\hat O^{\nodagger}_j}
	\equiv
	\sqrt{\lambda_L}v^{\phantom{*}}_{L,i} \sqrt{\lambda_L}v^{*}_{L,i} \; ,
\end{align}
so that $\sqrt{\lambda_L}v_{L,j}$ constitues the microscopic local order parameter.
\paragraph{\Acrlong{qLRO} and error estimates}
In the following paragraphs, we focus on our results for the correlation matrix of double occupancies $\chi_{\hat d}(i,j) = \braket{\psi | \hat{d}^{\dagger}_{i} \hat{d}^{\nodagger}_{j} | \psi}$.
The problem under consideration is one-dimensional so that due to the Mermin-Wagner-Hohenberg theorem~\cite{merminwagner_orig,merminwagner_erratum,hohenberg_orig} at least in the ground state there can be \gls{qLRO} only.
This translates to a still extensively scaling, dominating natural orbital but with vanishing value in the thermodynamic limit.
In the case of a non-interacting gas of hard-core bosons on a lattice, it is well-known~\cite{rigol:031603,rigol:013604} that
\begin{align}
	\frac{\lambda_{\tilde{\nu}}}{L}
	&\stackrel{L \rightarrow \infty}{\sim}
	\frac{A}{\sqrt{L}} + \beta_{\mathrm{SC}}
\end{align}
with $\beta_{\mathrm{SC}} \equiv 0$, which is in excellent aggreement with our ground\hyp state calculations in the superconducting phase at $U=-4, V=-\nicefrac{1}{4}$ and $\hat{O}^{\nodagger}_j = \hat{d}^{\nodagger}_j$.
We performed a finite size scaling of the eigenvalue of the dominant natural orbital by fitting the function $\frac{\lambda_{\tilde{\nu}}}{L}$ over $L^{-\gamma}$ with $\gamma=0.50$ in the superconducting  ground state.
The obtained value for the order parameter is $\beta_{\mathrm{SC}} = -1\times 10^{-3}$, which we used to gauge the precision of our method to be  $\delta \sim \mathcal{O}(10^{-3})$.
In the non\hyp equilibrium situation, we time-averaged the dominating natural orbital for times $t_0 \geq t_0$ via
\begin{align}
	\overbar{\lambda}_{\tilde{\nu}} &= \frac{1}{t_1 - t_0} \sum_{n} \lambda_{\tilde{\nu}}(t_0+n\delta) ,
\end{align}
with $\delta$ being the time step of the time\hyp evolution scheme and with $t_0 = 10, t_1=32$.
Performing the finite\hyp size analysis, we found the data to be best described by a scaling $\sim \alpha \cdot L^{-\gamma} + \beta_{\mathrm{SC}}$ with $\gamma = 0.92$ and $\beta_{\mathrm{SC}}= 9 \cdot 10^{-4}$.
Within the error bounds deduced from the ground\hyp state analysis, this implies $\beta_{\mathrm{SC}} \equiv 0$, and hence no \gls{LRO}.
\paragraph{Momentum space}
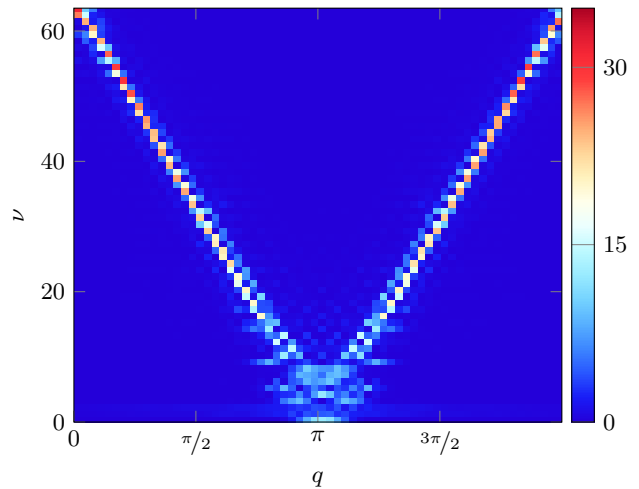
\begin{figure}
	\tikzsetnextfilename{neq_natural_orbital_fft}
	\begin{tikzpicture}
		\pgfplotsset
		{
			/pgfplots/colormap={temp}{
				rgb255=(36,0,217) 		
				rgb255=(25,29,247) 		
				rgb255=(41,87,255) 		
				rgb255=(61,135,255) 	
				rgb255=(87,176,255) 	
				rgb255=(117,211,255) 	
				rgb255=(153,235,255) 	
				rgb255=(189,249,255) 	
				rgb255=(235,255,255) 	
				rgb255=(255,255,235) 	
				rgb255=(255,242,189) 	
				rgb255=(255,214,153) 	
				rgb255=(255,172,117) 	
				rgb255=(255,120,87) 	
				rgb255=(255,61,61) 		
				rgb255=(247,40,54) 		
				rgb255=(217,22,48) 		
				rgb255=(166,0,33)		
			}
		}
		\begin{axis}
		[
			axis on top,
			enlargelimits	= false,
			width		= 0.45\textwidth,
			height		= 0.3\textheight,
			xlabel		= {$q$},
			ylabel		= {$\nu$},
			colorbar right,
			colormap name	= temp,
			xmin		= 0,
			xmax		= 64,
			ymin		= 0,
			ymax		= 63.5,
			xtick		= {0, 16, 32, 48},
			xticklabels	= {$0$,$\nicefrac{\pi}{2}$,$\pi$,$\nicefrac{3\pi}{2}$},
			point meta min	= 0,
			point meta max	= 35,
			every colorbar/.append style =
				{
					width			=	3mm,
					xshift			=	-5pt,
					scaled y ticks		= 	false,
					ytick			= 	{0,15,30},
					ylabel shift 		=	-4pt,
				},
		]
		\addplot graphics
		[
			xmin		= 0,
			xmax		= 64,
			ymin		= 0,
			ymax		= 64,
		]
		{../spectral_functions/precompiled/neq_U_m4p0_V_0p25_dV_m0p5_ft_no.eps};
		\end{axis}
	\end{tikzpicture}
	
	\caption{\label{fig:neq:natural-orbital:fft} Absolute value of the Fourier transform of the natural orbitals $\tilde{\mathbf{v}}_{\nu}(t=32)$ after quenching from the CDW to the \gls{SC} phase.}
\end{figure}
To complete the discussion, we calculated the Fourier transformation of the natural orbitals $\tilde{v}_{\nu,q} = \sum_{j}v_{\nu,j}e^{\mathrm i qj}$.
As shown in \cref{fig:neq:natural-orbital:fft}, the Fourier modes $\tilde{\mathbf{v}}_{\nu}$ are nearly diagonal in $k$-space, i.e., the correlation matrix is approximately diagonalized by a Fourier transformation and the dominating \gls{NO} is characterized by a mode with $q=0$.
The off-diagonal contributions around $q = \pi$ arise from highly excited states (cf. the band spectral function in the main text). 
Inspecting their real-space distributions $|v_{\nu,j}|^2$ (not shown here) these are located near the boundaries so that we conjecture them to vanish in the thermodynamic limit.
We observe these properties consistently for times $t>4$ in all considered system sizes $L\in\left[32,80\right]$.
\section{$J$-quench in the $t$-$J_\perp$ Chain}\label{app:tj_quench}
Next to the extended Hubbard model, we considered the quench dynamics of the one\hyp dimensional $t$-$J_\perp$ Hamiltonian~\cite{PhysRevLett.107.115301,PhysRevA.96.043618}
\begin{align}
	\hat{H}_\text{$t$-$J_\perp$} 
	&= 
	-t_{\mathrm{hop}} \sum_{j, \sigma} \left( \hat{c}_{j, \sigma}^\dagger \hat{c}^{\nodagger}_{j+1,\sigma} + \text{H. c.} \right) \notag \\
	&\phantom{=-t \sum_{i, \sigma} }
	+ \frac{J_\perp}{2} \sum_j \left( \hat S_j^+ \hat S_{j+1}^- + \hat S_j^- \hat S_{j+1}^+ \right) \;,
\end{align}
at filling $n = 0.2$.
Its phase diagram~\cite{PhysRevA.96.043618} at $n = 0.2$ features a metallic (Luttinger liquid~\cite{giamarchi}) spin-density wave phase (SDW) at $J_\perp = 2t_{\mathrm{hop}}$ and a spin-gapped singlet \gls{SC} phase at $J_\perp = 6t_{\mathrm{hop}}$.
In the following, we compute the equilibrium one-particle as well as singlet\hyp pair spectral functions at these two points in parameter space and also the non\hyp equilibrium spectral functions at late times for a quench $J_\perp = 2t_{\mathrm{hop}} \longrightarrow J_\perp = 6t_{\mathrm{hop}}$.
Furthermore, we discuss the time-evolution of the natural orbital occupations.
\paragraph{Spectral functions}
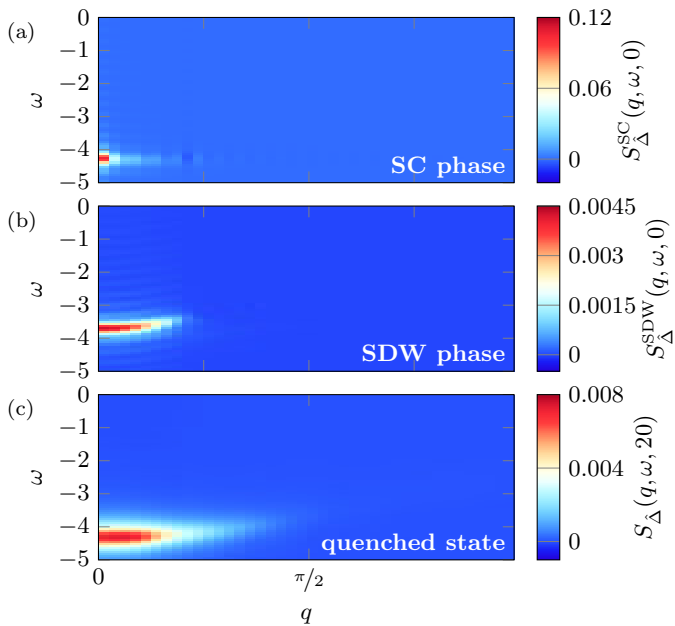
\begin{figure}[!t]
	\subfloat[\label{fig:tJ:spectral-functions:two-particle:eq:sc}]
	{
		\tikzsetnextfilename{tJ_spec_func_two_particle_1}
		\begin{tikzpicture}
			\pgfplotsset
			{
				/pgfplots/colormap={temp}{
					rgb255=(36,0,217) 		
					rgb255=(25,29,247) 		
					rgb255=(41,87,255) 		
					rgb255=(61,135,255) 	
					rgb255=(87,176,255) 	
					rgb255=(117,211,255) 	
					rgb255=(153,235,255) 	
					rgb255=(189,249,255) 	
					rgb255=(235,255,255) 	
					rgb255=(255,255,235) 	
					rgb255=(255,242,189) 	
					rgb255=(255,214,153) 	
					rgb255=(255,172,117) 	
					rgb255=(255,120,87) 	
					rgb255=(255,61,61) 		
					rgb255=(247,40,54) 		
					rgb255=(217,22,48) 		
					rgb255=(166,0,33)		
				}
			}
			\begin{axis}
			[
				ylabel		= {$\omega$},
				xlabel		= {},
				xtick		= {},
				xticklabels	= {,,},
				ytick		= {-5, -4, -3, -2, -1, 0},
				ymin=-5.0, ymax=0.0,
				xmin=0.0, xmax=39.5,
				height		= 0.16\textheight,
				width		= 0.4\textwidth-1.86pt,
				colorbar right,
				colormap name	= temp,	
				axis on top,	
				enlargelimits	= false,
				point meta min	= -0.02,
				point meta max	= 0.12,
				every colorbar/.append style =
				{
					ylabel			= 	{$S^{\mathrm{SC}}_{\hat \Delta}(q,\omega,0)$},
					width			=	3mm,
					scaled y ticks		= 	false,
					ytick			= 	{0, 0.06, 0.12},
					yticklabels		=	{$0\phantom{.00}$, $0.06$, $0.12$},
					ylabel shift 		=	-4pt,
				},
				title style={at={(1,0)},anchor=east},
				title	= {\color{white}\textbf{SC phase}},
			]
				\addplot graphics
				[ 
					xmin	= 0,
					xmax	= 39.5,
					ymin 	= -5.0,
					ymax	= 0.0
				]
				{../tj_data/sf/raw_plots/kw-L80scstate_delta.eps};
			\end{axis}
		\end{tikzpicture}
	}
	
	\vspace{-0.8em}
	\subfloat[\label{fig:tJ:spectral-functions:two-particle:eq:m}]
	{
		\tikzsetnextfilename{tJ_spec_func_two_particle_2}
		\begin{tikzpicture}
			\pgfplotsset
			{
				/pgfplots/colormap={temp}{
					rgb255=(36,0,217) 		
					rgb255=(25,29,247) 		
					rgb255=(41,87,255) 		
					rgb255=(61,135,255) 	
					rgb255=(87,176,255) 	
					rgb255=(117,211,255) 	
					rgb255=(153,235,255) 	
					rgb255=(189,249,255) 	
					rgb255=(235,255,255) 	
					rgb255=(255,255,235) 	
					rgb255=(255,242,189) 	
					rgb255=(255,214,153) 	
					rgb255=(255,172,117) 	
					rgb255=(255,120,87) 	
					rgb255=(255,61,61) 		
					rgb255=(247,40,54) 		
					rgb255=(217,22,48) 		
					rgb255=(166,0,33)		
				}
			}
			\begin{axis}
			[
				ylabel		= {$\omega$},
				xlabel		= {},
				xtick		= {},
				xticklabels	= {,,},
				ytick		= {-5, -4, -3, -2, -1, 0},
				ymin=-5.0, ymax=0.0,
				xmin=0, xmax=39.5,
				height		= 0.16\textheight,
				width		= 0.4\textwidth-1.86pt,
				colorbar right,
				colormap name	= temp,	
				axis on top,	
				enlargelimits	= false,
				point meta min	= -0.0005,
				point meta max	= 0.0045,
				every colorbar/.append style =
				{
					ylabel			= 	{$S^{\mathrm{SDW}}_{\hat \Delta}(q,\omega,0)$},
					width			=	3mm,
					scaled y ticks		= 	false,
					ytick			= 	{0.0, 0.0015, 0.003, 0.0045},
					yticklabels		=	{$0\phantom{.0000}$, $0.0015$, $0.003\phantom{0}$, $0.0045$},
					ylabel shift 		=	-4pt,
				},
				title style={at={(1,0)},anchor=east},
				title	= {\color{white}\textbf{SDW phase}},
			]
				\addplot graphics
				[ 
					xmin	= 0,
					xmax	= 39.5,
					ymin 	= -5.0,
					ymax	= 0.0
				]
				{../tj_data/sf/raw_plots/kw-L80groundstate_delta.eps};
			\end{axis}
		\end{tikzpicture}
	}
	
	\vspace{-0.8em}
	\subfloat[\label{fig:tJ:spectral-functions:two-particle:neq}]
	{
		\tikzsetnextfilename{tJ_spec_func_two_particle_3}
		\begin{tikzpicture}
			\pgfplotsset
			{
				/pgfplots/colormap={temp}{
					rgb255=(36,0,217) 		
					rgb255=(25,29,247) 		
					rgb255=(41,87,255) 		
					rgb255=(61,135,255) 	
					rgb255=(87,176,255) 	
					rgb255=(117,211,255) 	
					rgb255=(153,235,255) 	
					rgb255=(189,249,255) 	
					rgb255=(235,255,255) 	
					rgb255=(255,255,235) 	
					rgb255=(255,242,189) 	
					rgb255=(255,214,153) 	
					rgb255=(255,172,117) 	
					rgb255=(255,120,87) 	
					rgb255=(255,61,61) 		
					rgb255=(247,40,54) 		
					rgb255=(217,22,48) 		
					rgb255=(166,0,33)		
				}
			}
			\begin{axis}
			[
				ylabel		= {$\omega$},
				xlabel		= $q$,
				xtick		= {0, 20, 40},
				xticklabels	= {$0$,$\nicefrac{\pi}{2}$,$\pi$,$\nicefrac{3\pi}{2}$},
				ytick		= {-5, -4, -3, -2, -1, 0},
				ymin=-5.0, ymax=0.0,
				xmin=0, xmax=39.5,
				height		= 0.16\textheight,
				width		= 0.4\textwidth-1.86pt,
				colorbar right,
				colormap name	= temp,	
				axis on top,	
				enlargelimits	= false,
				point meta min	= -0.001,
				point meta max	= 0.008,
				every colorbar/.append style =
				{
					ylabel			= 	{$S^{\phantom{\mathrm{I}}}_{\hat \Delta}(q,\omega,20)$},
					width			=	3mm,
					scaled y ticks		= 	false,
					ytick			= 	{0.0, 0.004, 0.008},
					yticklabels		=	{$0\phantom{.000}$, $0.004$, $0.008$},
					ylabel shift 		=	-4pt,
				},
				title style={at={(1,0)},anchor=east},
				title	= {\color{white}\textbf{quenched state}},
			]
				\addplot graphics
				[ 
					xmin	= 0,
					xmax	= 39.5,
					ymin 	= -5.0,
					ymax	= 0.0
				]
				{../tj_data/sf/raw_plots/kw-L80latestate_delta.eps};
			\end{axis}
		\end{tikzpicture}
	}
	\caption
	{%
		\label{fig:tJ_spec_func_two_particle}%
		Two-particle singlet spectral functions for the $t$-$J_\perp$ model	. 
		Top: \gls{SC} phase, middle: SDW phase,	bottom: non\hyp equilibrium state 20 time units after the quench.
	}
\end{figure}

\begin{figure}[!t]
	\subfloat[\label{fig:tJ:spectral-functions:single-particle:eq:sc}]
	{
	\tikzsetnextfilename{tJ_spec_func_one_particle_1}
		\begin{tikzpicture}
			\pgfplotsset
			{
				/pgfplots/colormap={temp}{
					rgb255=(36,0,217) 		
					rgb255=(25,29,247) 		
					rgb255=(41,87,255) 		
					rgb255=(61,135,255) 	
					rgb255=(87,176,255) 	
					rgb255=(117,211,255) 	
					rgb255=(153,235,255) 	
					rgb255=(189,249,255) 	
					rgb255=(235,255,255) 	
					rgb255=(255,255,235) 	
					rgb255=(255,242,189) 	
					rgb255=(255,214,153) 	
					rgb255=(255,172,117) 	
					rgb255=(255,120,87) 	
					rgb255=(255,61,61) 		
					rgb255=(247,40,54) 		
					rgb255=(217,22,48) 		
					rgb255=(166,0,33)		
				}
			}
			\begin{axis}
			[
				ylabel		= {$\omega$},
				xlabel		= {},
				xtick		= {},
				xticklabels	= {,,},
				ytick		= {-4,-3,-2,-1,0},
				ymin=-4.0, ymax=0.0,
				xmin=0, xmax=39.5,
				height		= 0.16\textheight,
				width		= 0.4\textwidth-1.86pt,
				colorbar right,
				colormap name	= temp,	
				axis on top,	
				enlargelimits	= false,
				point meta min	= -0.01,
				point meta max	= 0.06,
				every colorbar/.append style =
				{
					ylabel			= 	{$S^{\mathrm{SC}}_{\hat c}(q,\omega,0)$},
					width			=	3mm,
					scaled y ticks		= 	false,
					ytick			= 	{0, 0.03, 0.06},
					yticklabels		=	{$0\phantom{.00}$, $0.03$, $0.06$},
					ylabel shift 		=	-4pt,
				},
				title style={at={(0,0)},anchor=west},
				title	= {\color{white}\textbf{SC phase}},
			]
				\addplot graphics
				[ 
					xmin	= 0,
					xmax	= 39.5,
					ymin 	= -4.0,
					ymax	= 0.0
				]
				{../tj_data/sf/raw_plots/kw-L80scstate_c.eps};
			\end{axis}
		\end{tikzpicture}
	}

	\vspace{-0.8em}
	\subfloat[\label{fig:tJ:spectral-functions:single-particle:eq:m}]
	{
		\tikzsetnextfilename{tJ_spec_func_one_particle_2}
		\begin{tikzpicture}
			\pgfplotsset
			{
				/pgfplots/colormap={temp}{
					rgb255=(36,0,217) 		
					rgb255=(25,29,247) 		
					rgb255=(41,87,255) 		
					rgb255=(61,135,255) 	
					rgb255=(87,176,255) 	
					rgb255=(117,211,255) 	
					rgb255=(153,235,255) 	
					rgb255=(189,249,255) 	
					rgb255=(235,255,255) 	
					rgb255=(255,255,235) 	
					rgb255=(255,242,189) 	
					rgb255=(255,214,153) 	
					rgb255=(255,172,117) 	
					rgb255=(255,120,87) 	
					rgb255=(255,61,61) 		
					rgb255=(247,40,54) 		
					rgb255=(217,22,48) 		
					rgb255=(166,0,33)		
				}
			}
			\begin{axis}
			[
				ylabel		= {$\omega$},
				xlabel		= {},
				xtick		= {},
				xticklabels	= {,,},
				ytick		= {-4,-3,-2,-1,0},
				ymin=-4.0, ymax=0.0,
				xmin=0, xmax=39.5,
				height		= 0.16\textheight,
				width		= 0.4\textwidth-1.86pt,
				colorbar right,
				colormap name	= temp,	
				axis on top,	
				enlargelimits	= false,
				point meta min	= -0.01,
				point meta max	= 0.08,
				every colorbar/.append style =
				{
					ylabel			= 	{$S^{\mathrm{SDW}}_{\hat c}(q,\omega,0)$},
					width			=	3mm,
					scaled y ticks		= 	false,
					ytick			= 	{0.0, 0.04,0.08},
					yticklabels		=	{$0\phantom{.00}$, $0.04$, $0.08$},
					ylabel shift 		=	-4pt,
				},
				title style={at={(0,0)},anchor=west},
				title	= {\color{white}\textbf{SDW phase}},
			]
				\addplot graphics
				[ 
					xmin	= 0,
					xmax	= 39.5,
					ymin 	= -4.0,
					ymax	= 0.0
				]
				{../tj_data/sf/raw_plots/kw-L80groundstate_c.eps};
			\end{axis}
		\end{tikzpicture}
	}

	\vspace{-0.8em}
	\subfloat[\label{fig:tJ:spectral-functions:single-particle:neq}]
	{
		\tikzsetnextfilename{tJ_spec_func_one_particle_3}
		\begin{tikzpicture}
			\pgfplotsset
			{
				/pgfplots/colormap={temp}{
					rgb255=(36,0,217) 		
					rgb255=(25,29,247) 		
					rgb255=(41,87,255) 		
					rgb255=(61,135,255) 	
					rgb255=(87,176,255) 	
					rgb255=(117,211,255) 	
					rgb255=(153,235,255) 	
					rgb255=(189,249,255) 	
					rgb255=(235,255,255) 	
					rgb255=(255,255,235) 	
					rgb255=(255,242,189) 	
					rgb255=(255,214,153) 	
					rgb255=(255,172,117) 	
					rgb255=(255,120,87) 	
					rgb255=(255,61,61) 		
					rgb255=(247,40,54) 		
					rgb255=(217,22,48) 		
					rgb255=(166,0,33)		
				}
			}
			\begin{axis}
			[
				ylabel		= {$\omega$},
				xlabel		= $q$,
				xtick		= {0, 20, 40},
				xticklabels	= {$0$,$\nicefrac{\pi}{2}$,$\pi$,$\nicefrac{3\pi}{2}$},
				ytick		= {-4,-3,-2,-1,0},
				ymin=-4.0, ymax=0.0,
				xmin=0, xmax=39.5,
				height		= 0.16\textheight,
				width		= 0.4\textwidth-1.86pt,
				colorbar right,
				colormap name	= temp,	
				axis on top,	
				enlargelimits	= false,
				point meta min	= 0.0,
				point meta max	= 0.016,
				every colorbar/.append style =
				{
					ylabel			= 	{$S^{\phantom{\mathrm{I}}}_{\hat c}(q,\omega,20)$},
					width			=	3mm,
					scaled y ticks		= 	false,
					ytick			= 	{0.0, 0.008, 0.016},
					yticklabels		=	{$0\phantom{.000}$, $0.008$, $0.016$},
					ylabel shift 		=	-4pt,
				},
				title style={at={(0,0)},anchor=west},
				title	= {\color{white}\textbf{quenched state}},
			]
				\addplot graphics
				[ 
					xmin	= 0,
					xmax	= 39.5,
					ymin 	= -4.0,
					ymax	= 0.0
				]
				{../tj_data/sf/raw_plots/kw-L80latestate_c.eps};
			\end{axis}
		\end{tikzpicture}
	}
	
	\caption
	{
		\label{fig:tJ_spec_func_one_particle}
		Single-particle spectral functions for the $t$-$J_\perp$ model.
		Top: SC phase, middle: SDW phase, bottom: non\hyp equilibrium state 20 time units after the quench.
	}
\end{figure}
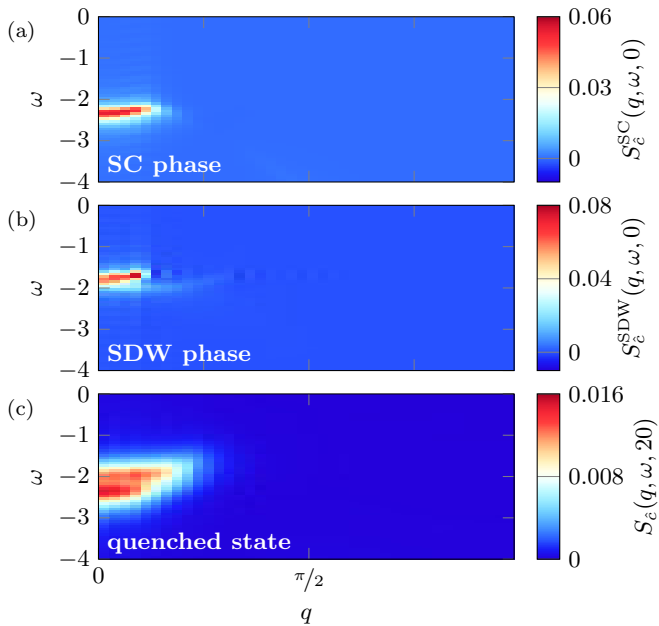
We consider the singlet\hyp pair creation and annihilation operators
\begin{equation}
	\hat{\Delta}_j^{(\dagger)} = \frac{1}{\sqrt{2}} \left( \hat c_{j,\uparrow}^{(\dagger)} \hat c_{j+1,\downarrow}^{(\dagger)} - \hat c_{j,\downarrow}^{(\dagger)} \hat c_{j+1,\uparrow}^{(\dagger)} \right)
\end{equation}
and compute the two-particle spectral function according to \cref{eq:def-spec-func,eq:appendix:spectral_function_numerics}.
The results are shown in \cref{fig:tJ_spec_func_two_particle} for the singlet-pair spectral function and in \cref{fig:tJ_spec_func_one_particle} for the one-particle spectral function (corresponding to the operators $c_{j,\uparrow}^{(\dagger)}$).
A quasi-stationary state is estimated to be reached after about 20 time units.
This is estimated from the time-evolution of the natural orbitals (cf. the corresponding section below).
The equilibrium two-particle SC spectral function shows a large weight at $q = 0$, whereas in the spin-density wave state it is much lower and a dispersion is visible.
The peak amplitude of the non\hyp equilibrium state is about doubled in comparison to the SDW and shifted towards the SC peak position.
This is consistent with our findings in the extended Hubbard model in the main text where we find an increase of spectral weight around $q=0$ in the postquench state (see \cref{fig:tJ:spectral-functions:two-particle:neq})
and a shift of weight towards the SC state.
In the single-particle case the equilibrium spectral functions show the occupied states of a cosine dispersion (the unoccupied part of the spectrum is seen in the spectral function obtained from the greater Green's function).
The spectral weights are of comparable amplitude but clearly differ in their $\omega$-axis position.
The non\hyp equilibrium state is a kind of hybrid that contains weight around both the equilibrium branches.
\paragraph{Shift of the spectral peak}

\begin{figure}[!t]
	\subfloat[\label{fig:tJ:sf_peaks_delta}]
	{
		\tikzset{external/export next=false}
		\tikzsetnextfilename{tJ_sf_peaks_delta}
		\begin{tikzpicture}
			\begin{axis}
			[
				width			=	0.495\textwidth,
				height			=	0.25\textheight,
				grid			=	major,
				xlabel			=	$k$,
				ylabel			=	$\omega$,
				zmin			=	0.0,
				zmax			=	0.054,
				xtick		= {0, 0.785398163, 1.5708},
				xticklabels	= {$0$,$\nicefrac{\pi}{4}$,$\nicefrac{\pi}{2}$},
			]
				\addplot3 graphics
				[
					points	=
					{
						(1.57,-6,0) => (44+2,180-106-18) 
						(1.57,-2,0) => (227,180-117-18)
						(1.57,-6,0.025) => (44+2,180-52-18-1)
						(0,-2,0) => (305-2,180-90-18-1)
					}
				] {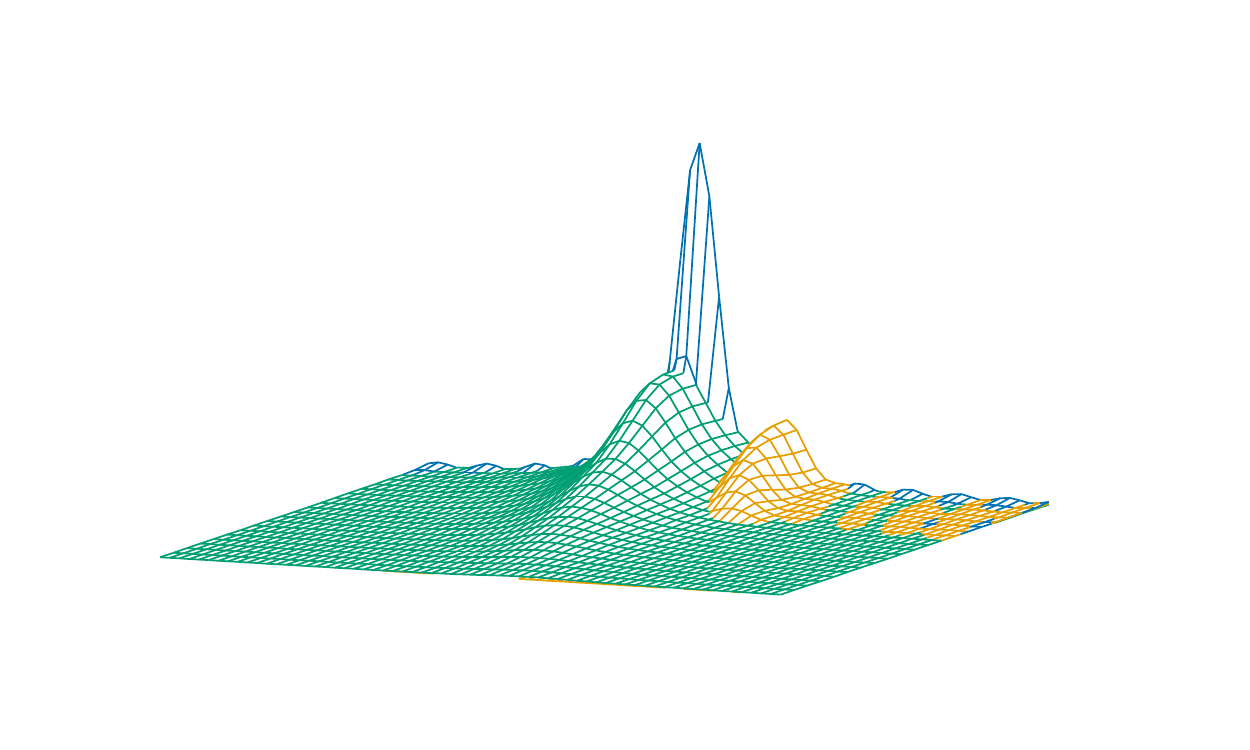};
				
				\node[align=left] at (axis cs:0,-4.7,0.045) {SC ($J=6$)\\$\times 0.2$};
				\node[align=left] at (axis cs:0,-3,0.02) {SDW ($J=2$)};
				\node[align=left] at (axis cs:0.75,-6,0.005) {noneq.};
			\end{axis}
		\end{tikzpicture}
	}
	
	\vspace{1em}
	\subfloat[\label{fig:tJ:sf_peaks_c}]
	{
		\tikzset{external/export next=false}
		\tikzsetnextfilename{tJ_sf_peaks_c}
		\begin{tikzpicture}
			\begin{axis}
			[
				width			=	0.495\textwidth,
				height			=	0.25\textheight,
				grid			=	major,
				xlabel			=	$k$,
				ylabel			=	$\omega$,
				zmin			=	0.0,
				zmax			=	0.054,
				xtick		= {0, 0.785398163, 1.5708},
				xticklabels	= {$0$,$\nicefrac{\pi}{4}$,$\nicefrac{\pi}{2}$},
			]
				\addplot3 graphics
				[
					points	=
					{
						(1.57,-4,0) => (44,180-106-18) 
						(1.57,0,0) => (227,180-117-18)
						(1.57,-4,0.025) => (44,180-52-18)
						(0,0,0) => (305,180-90-18)
					}
				] {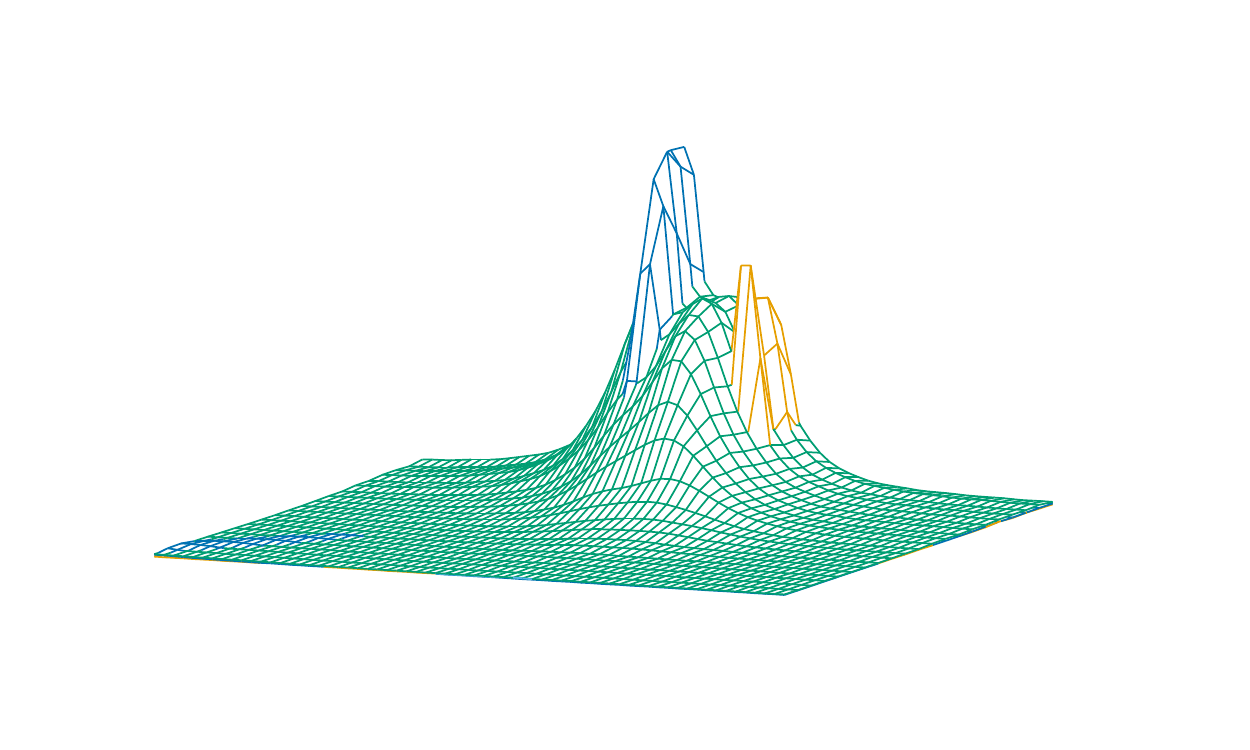};
				
				\node[align=left] at (axis cs:0,-2.7,0.045) {SC ($J=6$)\\$\times 0.4$};
				\node[align=left] at (axis cs:0,-1,0.02) {SDW ($J=2$)\\$\times 0.2$};
				\node[align=left] at (axis cs:0.75,-4,0.005) {noneq.};
			\end{axis}
		\end{tikzpicture}
	}
	
	\caption
	{
		\label{fig:tJ_sf_peaks}
		Comparison of the peak positions of \protect\subref{fig:tJ:sf_peaks_delta} the singlet pair spectral function (cf. \cref{fig:tJ_spec_func_two_particle}) and \protect\subref{fig:tJ:sf_peaks_c} the single-particle spectral function (cf. \cref{fig:tJ_spec_func_one_particle}) for the spin-density wave and superconducting equilibrium states as well as the nonequilibrium state, respectively.
		In \protect\subref{fig:tJ:sf_peaks_delta} the spectral peak position is a hybrid of the SDW and SC peaks, whereas in \protect\subref{fig:tJ:sf_peaks_c} the spectral weight has clearly shifted towards the SC peak.
	}
\end{figure}
As a guide to the eye, the peaks of the one- and two\hyp particle spectral functions are replotted in \cref{fig:tJ_sf_peaks}.
In addition to a slight increase in weight one can clearly see that the two-particle spectral peak in the post-quench state has shifted towards the superconducting peak position whereas
in the one-particle case the picture is more complicated.
Hence, like in the main text the two-particle spectral function seems to be a more reliable indicator for the formation of transient superconducting spectral weights.
\paragraph{Natural orbitals}
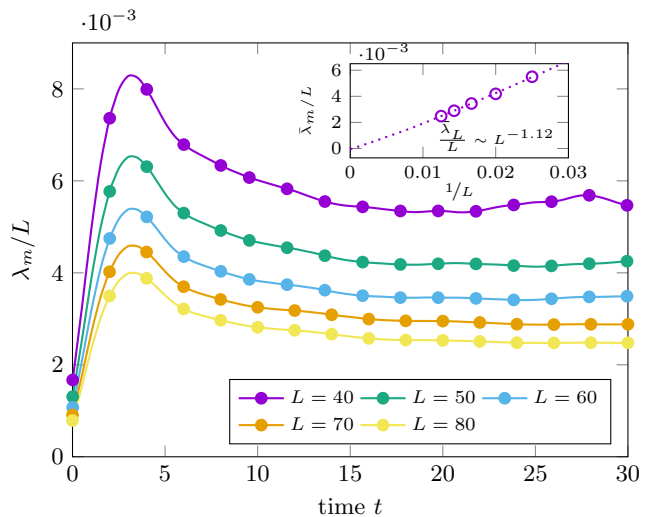
\begin{figure}[!t]
	\centering
	\tikzsetnextfilename{tJ_neq_max_natorb}
	\pgfdeclarelayer{foreground}
	\pgfdeclarelayer{background}
	\pgfsetlayers{background,main,foreground}
	\begin{tikzpicture}
		[
			declare function = 
			{
				g(\x) = \NeqMtJ*\x^\NeqBtJ + \NeqCtJ;
			},
		]
		\begin{pgfonlayer}{background}
			\begin{axis}
			[
				width		= 0.5\textwidth,
				height		= 0.3\textheight,
				xmin 		= 0,
				xmax 		= 30,
				ymin		= 0,
				ymax		= 0.009,
				xlabel		= {$\text{time }t$},
				ylabel		= {$\lambda_m/L$},
				smooth,
				legend style	= {font=\scriptsize,legend pos=south east},
				legend columns	=	3, 
			]
				\addplot
				[
					color=colorA,
					mark=*,
					unbounded coords=jump, 
					mark size = 2pt,
					mark repeat=4,
					thick
				]
				table
				[
					x expr = \thisrowno{0},
					y expr = (\thisrowno{1})/40
				]
				{../tj_data/natorb/max_natorb_L40};
				\addplot
				[
					color=colorB,
					mark=*,
					unbounded coords=jump, 
					mark size = 2pt,
					mark repeat=4,
					thick
				]
				table
				[
					x expr = \thisrowno{0},
					y expr = (\thisrowno{1})/50
				]
				{../tj_data/natorb/max_natorb_L50};
				\addplot
				[
					color=colorC,
					mark=*,
					unbounded coords=jump, 
					mark size = 2pt,
					mark repeat=4,
					thick
				]
				table
				[
					x expr = \thisrowno{0},
					y expr = (\thisrowno{1})/60
				]
				{../tj_data/natorb/max_natorb_L60};
				\addplot
				[
					color=colorD,
					mark=*,
					unbounded coords=jump, 
					mark size = 2pt,
					mark repeat=4,
					thick
				]
				table
				[
					x expr = \thisrowno{0},
					y expr = (\thisrowno{1})/70
				]
				{../tj_data/natorb/max_natorb_L70};
				\addplot
				[
					color=colorE,
					mark=*,
					unbounded coords=jump, 
					mark size = 2pt,
					mark repeat=4,
					thick
				]
				table
				[
					x expr = \thisrowno{0},
					y expr = (\thisrowno{1})/80
				]
				{../tj_data/natorb/max_natorb_L80};
				\addlegendentry{$L=40$}
				\addlegendentry{$L=50$}
				\addlegendentry{$L=60$}
				\addlegendentry{$L=70$}
				\addlegendentry{$L=80$}
				\coordinate (insetPosition) at (axis cs:15.0,0.0065);
			\end{axis}
		\end{pgfonlayer}
		\begin{pgfonlayer}{foreground}
			\begin{axis}
			[
				width		= 0.25\textwidth,
				height		= 0.12\textheight,
				at		= {(insetPosition)},
				xmin 		= 0.0,
				xmax 		= 0.03,
				ymax		= 0.0065,
				scaled x ticks	= false,
				xtick		= {0.0, 0.01, 0.02, 0.03},
				xticklabels	= {$0$, $0.01$, $0.02$, $0.03$},
				ytick		= {0, 0.002, 0.004, 0.006},
				xlabel		= {$\nicefrac{1}{L}$},
				ylabel		= {$\bar \lambda_{m}/L$},
				smooth,
				ticklabel style	= {font=\scriptsize},
				xlabel style	= {font=\scriptsize,yshift=5pt},
				ylabel style	= {font=\tiny,yshift=0pt},
			]
				\addplot
				[
					color = colorA,
					mark = o,
					mark size = 2pt,
					thick,
					only marks,
				]
				plot
				[
					error bars/.cd,
					y dir=both,
					y explicit,
				]
				table
				[
					x expr = \thisrowno{0},
					y expr = \thisrowno{1},
				]
				{../tj_data/natorb/max_natorb_avg_scaling};
				\addplot
				[
					color = colorA,
					thick, 
					dotted,
					mark=none,
					domain=0.0:0.18,
				]
				{g(x)};
				\node at (axis cs: 0.02,0.001) {\tiny $\frac{\overbar \lambda^{\noprime}_L}{L} \sim L^{\NeqBtJShort} $};
			\end{axis}
		\end{pgfonlayer}
	\end{tikzpicture}
	\caption
	{
		\label{fig:tJ_neq_max_natorb}
		Time evolution of the largest natural orbital for the quench in the $t$-$J_\perp$ model; the transient state is reached after about 20 time units.
		Inset: extrapolated scaling of the normalized dominating eigenvalue $\overbar \lambda_L/L$ of the correlation matrix with the system size $L$.
	}
\end{figure}
At the end, analogous to the main text, we consider the question of whether true long-ranged order out\hyp of\hyp equilibrium is detectable from the correlation matrices.
In \cref{fig:tJ_neq_max_natorb} the time-dependent largest eigenvalue of the singlet-pair correlation matrix $\langle \hat \Delta_i^\dagger \hat \Delta_j \rangle_t$ is shown.
As for the extended Hubbard model in the main text a saturation value can be made out after 20 time units.
We use this as the criterion for the choice of the non\hyp equilibrium reference state.

In order to perform a scaling analysis $\bar \lambda_m / L = \alpha \cdot L^{-\gamma} + \beta_{\mathrm{SC}}$, we average over all largest natural orbitals for times $20 \leq t \leq 30$.
The results shown in the inset of \cref{fig:tJ_neq_max_natorb} yield a value of $\beta \approx -6 \cdot 10^{-5}$, such that $\beta_{\mathrm{SC}} \approx 0$ within the estimated error bounds.
This supports the conclusion that no true long-ranged order is realized in the transient state also for this model.
The scaling exponent is found to be $\gamma \approx 1.12$ which is somewhat close to the scaling exponent of $0.92$ found in the extended Hubbard model.
Note that by definition $\gamma \leq 1$ so that the obtained scaling exponent has an error of $\approx 10 \%$.
\section{Comparison to Finite-temperature States}\label{app:finite-temperature}
\begin{figure}[!t]
	\tikzsetnextfilename{P1_Energy_Thermal_vs_Nonthermal}
	\pgfdeclarelayer{foreground}
	\pgfdeclarelayer{background}
	\pgfsetlayers{background,main,foreground}
	\begin{tikzpicture}
		\begin{pgfonlayer}{background}
			\begin{axis}
			[
					name		= main,
					width		= 0.5\textwidth-9.5pt,
					height		= 0.3\textheight,
					xmin 		= 0,
					xmax 		= 9,
					ylabel style	= {yshift=-15pt},
					xlabel		= $\beta$,
					ylabel		= {$\left \langle \hat{H}\right\rangle_{\beta}$},
					smooth,
					extra y ticks={-84},
					extra y tick style=
					{
						yticklabel style=
						{
							font=\small,
						},
						yticklabel pos = left,
						yticklabels = {$E_{\text{quench}}$}
					}
			]
				\coordinate (insetPosition) at (axis cs:3.5,-75);
				\addplot
				[
					color	= colorC,
					mark	= none,
					thick
				]
				table
				[
					x expr	= 2*(\thisrowno{0}),
					y expr	= \thisrowno{1},
				]
				{data/L_32/sc_finite_temp_U_m4p0_V_m0p25/energy_coarse};\label{pgfplots:L_32_sc_imag_time_evolution}
				\draw[dashed] ({axis cs:0,-84}-|{rel axis cs:0,0}) -- ({axis cs:9,-84}-|{rel axis cs:1,0});
			\end{axis}
		\end{pgfonlayer}
		\begin{pgfonlayer}{foreground}
			\begin{loglogaxis}
				[
					name		= inset,
					width		= 0.325\textwidth-10pt,
					height		= 0.225\textheight,
					at		= {(insetPosition)},
					ticklabel style	= {font=\scriptsize},
					xlabel style	= {font=\scriptsize,yshift=2.5pt},
					ylabel style	= {font=\scriptsize,yshift=-2.5pt},
					xmin 		= 32,
					xmax 		= 64,
					ymin		= 1e-08,
					xlabel		= $L/2 + i$,
					ylabel		= {$\log_{10} |\Re (P_{1})|$},
					xtick		= {32, 40, 48, 54, 64},
					xticklabels	= {$32$, $40$, $48$, $54$, $64$},
					ytick		= {1e-2, 1e-4, 1e-6},
					legend style	= {font=\scriptsize,legend pos=south west},
					log basis x	= 10,
				]
				\addplot
				[
					color=colorA,
					mark=*,
					unbounded coords=jump, 
					mark size = 2pt,
					mark repeat=2,
					each nth point	= 2,
					restrict expr to domain = {x}{32:63}
				]
				table
				[
					x expr = \coordindex,
					y expr = abs(\thisrowno{0})
				]
				{data/L_32/sc_finite_temp_U_m4p0_V_m0p25/P1_fine/P1_fine_t_0p01};\label{pgfplots:L_32_sc_finite_temp_P1_t_0p01}
				\addlegendentry{$\beta=2.9$}
				\addplot
				[
					color=colorB,
					mark=*,
					unbounded coords=jump, 
					mark size = 2pt,
					mark repeat=2,
				]
				table
				[
					x expr = 2*(\coordindex+1),
					y expr = abs(\thisrowno{0})
				]
				{data/L_32/cdw_sc_quench_U_m4p0_V_0p25_dV_m0p5/P1/P1_t_10};\label{pgfplots:L_32_cdw_sc_quench_P1_t_10}
				\addlegendentry{quench: $t=10$}
			\end{loglogaxis}
		\end{pgfonlayer}
		
		\begin{pgfonlayer}{main}
			\node[rectangle, fill=black!0,fit=(inset)] {};
		\end{pgfonlayer}
	\end{tikzpicture}
	\caption
	{
Energy $\left\langle \hat{H} \right\rangle_{\beta}$ during imaginary time evolution of a system with $L=32$ sites over imaginary time $\beta$. The inset shows the superconducting correlation function $P_{1}(L/2)$ during the time evolution after a global quench from the CDW to the \gls{SC} phase at time step $t=10$ (green), and in a thermal state with $\left\langle \hat{H} \right\rangle_{\beta}$ corresponding to the energy of the quenched system $E_{\text{quench}}$ (purple).
	}
	\label{fig:compare-thermal}
\end{figure}
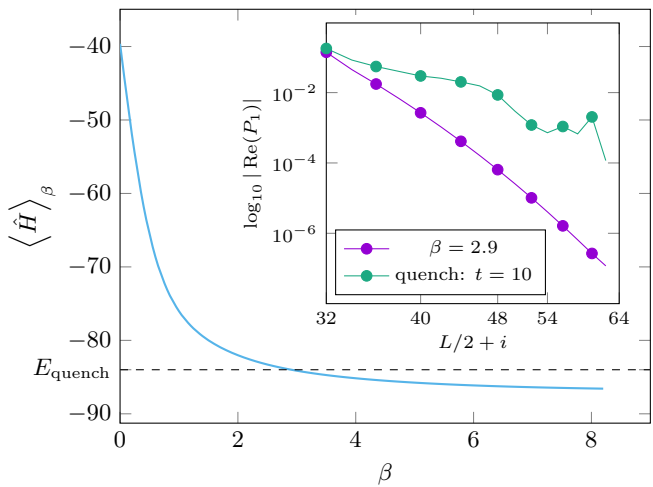
All our simulations were performed on pure states that are either ground\hyp states of the models under consideration or quenches from the latter. 
The question arises whether, following the quench, the system reaches a state in which local observables are thermalized towards their values in the canonical ensemble.
To address this issue, we calculated the density matrix $\hat{\rho}(\beta)$ in the \gls{SC} phase where we choose the inverse temperature $\beta \equiv \nicefrac{1}{T}$ in a way that $\braket{\hat{H}}_{\beta} \equiv \braket{H}_{0}$.
Here, $\braket{\cdots}_{0}$ denotes the expectation value of $\hat{H}$ after the quench and $\braket{\cdots}_{\beta}$ is the expectation value of $\hat{H}$ with respect to a thermal state $\hat{\rho}(\beta)$ in the superconducting phase.
In order to obtain the correct inverse temperature, we performed an imaginary time evolution on an infinite\hyp temperature state $\hat{\rho}(0)$ and cooled down the system this way until the energy matched the one of the quenched state~\cite{Paeckel2019}. 
In \cref{fig:compare-thermal}, the obtained relation between energy and inverse temperature is displayed with the energy of the quenched state marked as dashed line.
Having found the respective density matrix $\hat{\rho}(\beta_Q)$, we plot the expectation value of the superconducting correlation function 
\begin{align}
	\hat{P}_{1}(i,j) &= \hat{c}^{\dagger}_{i,\uparrow}\hat{c}^{\dagger}_{i,\downarrow}\hat{c}^{\nodagger}_{j,\uparrow}\hat{c}^{\nodagger}_{j,\downarrow}
\end{align}
in both the thermal state and the quenched state at $t=10$ in the inset of \cref{fig:compare-thermal}.
As can be seen, we do not obtain a state that is characterized by a thermal density matrix with inverse temperature $\beta_Q$ chosen in a way such that its energy matches the one after the quench.

\section{Quench within the CDW Phase}
In this section we aim to demonstrate that enhanced superconductivity and an enhanced mobility of charge carriers are hard to distinguish if the only information at hand is the optical conductivity.
For this purpose we discuss a complementary quench from the \gls{CDW} phase ($U=-4$, $V=\nicefrac{1}{4}$) -- as in the main text -- to a point in the \gls{CDW} phase ($U=-2$, $V=\nicefrac{1}{4}$) which exhibits a higher metalliticity.
Similar analysis of the optical conductivity and the spectral functions of the one- and two-particle excitations as in the main text are done.
Furthermore, we show the energy scales after this quench compared to the quench in the main text.

\subsection{Optical Conductivity}
\vspace{-2em}
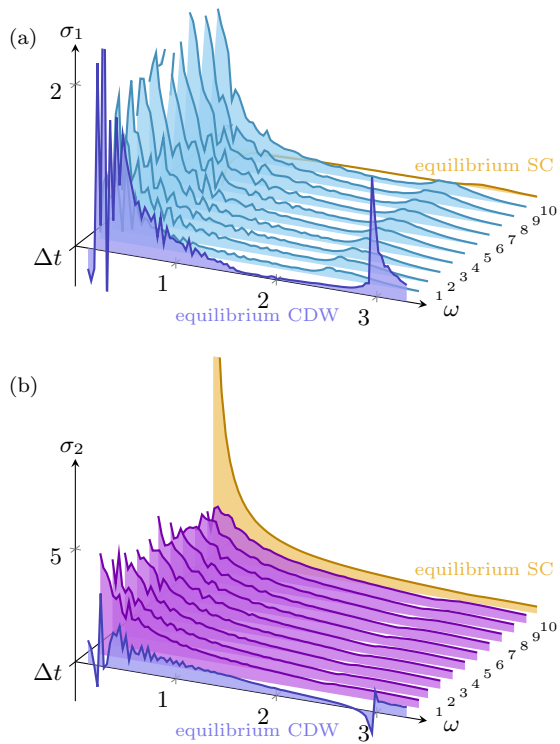
\begin{figure}[!t]
	\subfloat[\label{fig:optical-conductivity:dU:1}]
	{
		\centering
		\tikzsetnextfilename{optical_conductivity_real_CDW_CDW_quench_U_m4p0_V_0p25_dU_2p0}
		\begin{tikzpicture}
			\begin{axis}
			[
				axis lines=center,
				xlabel={$\omega$},
				xlabel style = {at=(xticklabel cs:1.15),yshift=10},
				ylabel={$\Delta t$},
				ylabel style = {at=(yticklabel cs:-0.0),yshift=5pt},
				zlabel={$\sigma_{1}$}, 
				zlabel style = {at=(zticklabel cs:1.125), xshift=5pt},
				yticklabels = {,,},
				xmin=1e-6,
				xmax=3.5,
				ymin=-1,
				ymax=26,
				zmax=2.5,
				zmin=-0.5,
				grid,
			]
			\addplot3
				[
					fill opacity = 0.75,
					draw opacity = 0.0,
					fill=colorD!60!white,
					restrict expr to domain = {x}{0.1:3.3},
					restrict expr to domain = {z}{-2:2},
				]
				table
				[
					x expr 	= \thisrowno{0},
					y expr	= 22,
					z expr	= -1.0*\thisrowno{2},
				]
				{../spectral_functions/eq/optical_conductivity/U_m4p0_V_m0p25/results/optical_conductivity_merged_A0_0p5_t0_0p0_tau_0p05_omega_2p38} \closedcycle;
			\addplot3
				[
					draw=colorD!80!black,
					thick,
					restrict expr to domain = {x}{0.1:3.3},
					restrict expr to domain = {z}{-2:2},
				]
				table
				[
					x expr 	= \thisrowno{0},
					y expr	= 22,
					z expr	= -1.0*\thisrowno{2},
				]
				{../spectral_functions/eq/optical_conductivity/U_m4p0_V_m0p25/results/optical_conductivity_merged_A0_0p5_t0_0p0_tau_0p05_omega_2p38};
				\node[text=colorD!80] at (axis cs:2.5,2*12+2,0) {\scriptsize equilibrium SC};
			\foreach \dt [count=\x from 1] in {{10p0},{9p0},{8p0},{7p0},{6p0},{5p0},{4p0},{3p0},{2p0},{1p0}}
			{
				\addplot3
					[
						fill opacity = 0.75,
						draw opacity = 0.0,
						fill=colorC!60!white,
						restrict expr to domain = {x}{0.1:3.3},
						restrict expr to domain = {z}{-2:2},
					]
					table
					[
						x expr 	= \thisrowno{0},
						y expr	= {2*(11-\x)},
						z expr	= -1.0*\thisrowno{2},
					]
					{../spectral_functions/global_quench/1TDVP/probe_pulse/U_m4p0_V_0p25_dU_2p0/results/optical_conductivity_merged_A0_0p5_t0_\dt_tau_0p05_omega_2p38} \closedcycle;
				\addplot3
					[
						draw=colorC!80!black,
						thick,
						restrict expr to domain = {x}{0.1:3.3},
						restrict expr to domain = {z}{-2:2},
					]
					table
					[
						x expr 	= \thisrowno{0},
						y expr	= {2*(11-\x)},
						z expr	= -1.0*\thisrowno{2},
					]
					{../spectral_functions/global_quench/1TDVP/probe_pulse/U_m4p0_V_0p25_dU_2p0/results/optical_conductivity_merged_A0_0p5_t0_\dt_tau_0p05_omega_2p38};
				\pgfmathparse{int(round(11 - \x))}
				\edef\PrintLabel{
				    \noexpand\node at (axis cs:3.5,2*\pgfmathresult,0.0) {\noexpand\tiny$\pgfmathresult$};
				}
				\
				\PrintLabel
			}
			\begin{scope}
				\clip (axis cs:1e-6,0.01,-0.5) -- (axis cs:3.5,0.01,-0.5) -- (axis cs:3.5,0.01,2.5) -- (axis cs:1e-6,0.01,2.5) -- (axis cs:1e-6,0.01,-0.5);
				\addplot3
					[
						fill opacity = 0.75,
						draw opacity = 0.0,
						fill=colorF!60!white,
						restrict expr to domain = {x}{0.1:3.3},
						restrict expr to domain = {z}{-6:6},
					]
					table
					[
						x expr 	= \thisrowno{0},
						y expr	= 0.01,
						z expr	= -1.0*\thisrowno{2},
					]
					{../spectral_functions/eq/optical_conductivity/U_m4p0_V_0p25/results/optical_conductivity_merged_A0_0p5_t0_0p0_tau_0p05_omega_2p38} \closedcycle;
				\addplot3
					[
						draw=colorF!80!black,
						thick,
						restrict expr to domain = {x}{0.1:3.3},
						restrict expr to domain = {z}{-6:6},
					]
					table
					[
						x expr 	= \thisrowno{0},
						y expr	= 0.01,
						z expr	= -1.0*\thisrowno{2},
					]
					{../spectral_functions/eq/optical_conductivity/U_m4p0_V_0p25/results/optical_conductivity_merged_A0_0p5_t0_0p0_tau_0p05_omega_2p38};
			\end{scope}
			\end{axis}
			\node[text=colorF!80] at (2.5,0.45) {\scriptsize equilibrium CDW};
		\end{tikzpicture}
	}
	
	\subfloat[\label{fig:optical-conductivity:dU:2}]
	{
		\centering
		\tikzsetnextfilename{optical_conductivity_imag_CDW_CDW_quench_U_m4p0_V_0p25_dU_2p0}
		\begin{tikzpicture}
			\pgfplotsset
			{
				/pgfplots/colormap={moreland}{
					rgb255=(59,76,192)
					rgb255=(77,104,215)
					rgb255=(98,130,234)
					rgb255=(119,154,247)
					rgb255=(141,176,254)
					rgb255=(163,194,255)
					rgb255=(184,208,249)
					rgb255=(204,217,238)
					rgb255=(221,221,221)
					rgb255=(236,211,197)
					rgb255=(245,196,173)
					rgb255=(247,177,148)
					rgb255=(244,154,123)
					rgb255=(236,127,99)
					rgb255=(222,96,77)
					rgb255=(203,62,56)
					rgb255=(180,4,38)
				}
			}
			\begin{axis}
			[
				axis lines=center,
				xlabel={$\omega$},
				xlabel style = {at=(xticklabel cs:1.15),yshift=10},
				ylabel={$\Delta t$},
				ylabel style = {at=(yticklabel cs:-0.0),yshift=5pt},
				zlabel={$\sigma_{2}$}, 
				zlabel style = {at=(zticklabel cs:1.125), xshift=5pt},
				yticklabels = {,,},
				xmin=1e-6,
				xmax=3.5,
				ymin=-1,
				ymax=26,
				zmax=9,
				zmin=-1.75,
				grid,
			]
				\coordinate (insetPosition) at (axis cs: 0.0,38.0,0.0);
				\begin{scope}
					\clip (axis cs:1e-6,22,-0.5) -- (axis cs:3.5,22,-0.5) -- (axis cs:3.5,22,9) -- (axis cs:1e-6,22,9) -- (axis cs:1e-6,22,-0.5);
					\addplot3
						[
							fill opacity = 0.75,
							draw opacity = 0.0,
							fill=colorD!60!white,
							restrict expr to domain = {x}{0.01:3.3},
						]
						table
						[
							x expr 	= \thisrowno{0},
							y expr	= 22,
							z expr	= \thisrowno{3},
						]
						{../spectral_functions/eq/optical_conductivity/U_m4p0_V_m0p25/results/optical_conductivity_merged_A0_0p5_t0_0p0_tau_0p05_omega_2p38} \closedcycle;
					\addplot3
						[
							draw=colorD!80!black,
							thick,
							restrict expr to domain = {x}{0.01:3.3},
						]
						table
						[
							x expr 	= \thisrowno{0},
							y expr	= 22,
							z expr	= \thisrowno{3},
						]
						{../spectral_functions/eq/optical_conductivity/U_m4p0_V_m0p25/results/optical_conductivity_merged_A0_0p5_t0_0p0_tau_0p05_omega_2p38};
				\end{scope}
				\node[text=colorD!80] at (axis cs:2.5,2*12+2,0.5) {\scriptsize equilibrium SC};
			\foreach \dt [count=\x from 1] in {{10p0},{9p0},{8p0},{7p0},{6p0},{5p0},{4p0},{3p0},{2p0},{1p0}}
			{
				\addplot3
					[
						fill opacity = 0.75,
						draw opacity = 0.0,
						fill=colorA!60!white,
						restrict expr to domain = {x}{0.1:3.3},
					]
					table
					[
						x expr 	= \thisrowno{0},
						y expr	= {2*(11-\x)},
						z expr	= \thisrowno{3},
					]
					{../spectral_functions/global_quench/1TDVP/probe_pulse/U_m4p0_V_0p25_dU_2p0/results/optical_conductivity_merged_A0_0p5_t0_\dt_tau_0p05_omega_2p38} \closedcycle;
				\addplot3
					[
						draw=colorA!80!black,
						thick,
						restrict expr to domain = {x}{0.1:3.3},
					]
					table
					[
						x expr 	= \thisrowno{0},
						y expr	= {2*(11-\x)},
						z expr	= \thisrowno{3},
					]
					{../spectral_functions/global_quench/1TDVP/probe_pulse/U_m4p0_V_0p25_dU_2p0/results/optical_conductivity_merged_A0_0p5_t0_\dt_tau_0p05_omega_2p38};
				\pgfmathparse{int(round(11 - \x))}
				\edef\PrintLabel{
				    \noexpand\node at (axis cs:3.5,2*\pgfmathresult,0,0.0) {\noexpand\tiny$\pgfmathresult$};
				}
				\
				\PrintLabel
			}
			\addplot3
				[
					fill opacity = 0.75,
					draw opacity = 0.0,
					fill=colorF!60!white,
					restrict expr to domain = {x}{0.1:3.3},
				]
				table
				[
					x expr 	= \thisrowno{0},
					y expr	= 0.01,
					z expr	= \thisrowno{3},
				]
				{../spectral_functions/eq/optical_conductivity/U_m4p0_V_0p25/results/optical_conductivity_merged_A0_0p5_t0_0p0_tau_0p05_omega_2p38} \closedcycle;
			\addplot3
				[
					draw=colorF!80!black,
					thick,
					restrict expr to domain = {x}{0.1:3.3},
				]
				table
				[
					x expr 	= \thisrowno{0},
					y expr	= 0.01,
					z expr	= \thisrowno{3},
				]
				{../spectral_functions/eq/optical_conductivity/U_m4p0_V_0p25/results/optical_conductivity_merged_A0_0p5_t0_0p0_tau_0p05_omega_2p38};
			\end{axis}
			\node[text=colorF!80] at (2.5,0.45) {\scriptsize equilibrium CDW};
		\end{tikzpicture}
	}
	\caption
	{
		Real \protect\subref{fig:optical-conductivity:dU:1} and imaginary \protect\subref{fig:optical-conductivity:dU:2} part of the optical conductivity $\sigma_{1,2}(\omega,\Delta t)$ with probe pulses applied at different time delays $\Delta t$ for the quench within the CDW phase ($U=-4t \rightarrow U=-2t$) and the corresponding quantities in the \gls{CDW} ground state (violet) and the \gls{SC} ground state (orange).
	}
	\label{fig:optical-conductivity}
\end{figure}
\vspace{2em}
In \cref{fig:optical-conductivity} the optical conductivity after the quench within the \gls{CDW} phase is shown.
The data is more noisy compared to the case of a quench into the \gls{SC} phase which is related to the observation that here the response current exhibits oscillations with larger amplitude and slower decay compared to a quench into the \gls{SC} phase.
We address this to the fact that in the insulating \gls{CDW} phase there are various scattering processes between the particles.
In order to properly resolve the frequency dependencies of these processes the simulation time-range after applying the probe pulse needed to be increased way beyond what was computationally feasible.
\Cref{fig:optical-conductivity:dU:1} displays the real part of the optical conductivity where we find a weak transfer of spectral weight towards smaller frequencies.
The imaginary part \cref{fig:optical-conductivity:dU:2} also shows an increased spectral weight towards smaller frequencies compared to the equilibrium case.
Therefore, without having a reference curve of the system being in the superconducting phase the only possible observation would be a redistribution of spectral weight towards smaller frequencies which may also be interpreted as a sign for enhanced superconductivity.
In particular $\sigma_2$ suggests that a $1/\omega$ behavior can be realized after the quench if we also take into consideration that due to the finite simulation time $t=25/t_{\mathrm{hop}}$ after applying the probe pulse the minimally achieved frequency resolution is bounded by $2\pi/25t_{\mathrm{hop}}$.
%
%

%
\subsection{Spectral Functions}
\begin{figure}[!t]
	\subfloat[\label{fig:eq-neq-neq-eq-spectral-functions-2particle:sc}]
	{
		\centering
		\tikzsetnextfilename{eq-neq-neq-eq-spectral-functions-2particle_sc}
		\begin{tikzpicture}
			\pgfplotsset
			{
				/pgfplots/colormap={temp}{
					rgb255=(36,0,217) 		
					rgb255=(25,29,247) 		
					rgb255=(41,87,255) 		
					rgb255=(61,135,255) 	
					rgb255=(87,176,255) 	
					rgb255=(117,211,255) 	
					rgb255=(153,235,255) 	
					rgb255=(189,249,255) 	
					rgb255=(235,255,255) 	
					rgb255=(255,255,235) 	
					rgb255=(255,242,189) 	
					rgb255=(255,214,153) 	
					rgb255=(255,172,117) 	
					rgb255=(255,120,87) 	
					rgb255=(255,61,61) 		
					rgb255=(247,40,54) 		
					rgb255=(217,22,48) 		
					rgb255=(166,0,33)		
				}
			}
			\begin{groupplot}
				[
					group style = 
					{
						group size 			=	1 by 2,
						vertical sep		=	0.0pt,
						x descriptions at	=	edge bottom,
						y descriptions at	=	edge left,
					},
					xtick			= {0, 16, 32, 48},
					xticklabels		= {,,,},
					xlabel			= {},
					axis on top,
					enlargelimits	= false,
					width			= 0.4\textwidth-2.71pt,
				]
				\nextgroupplot
				[
					ytick={0},
					axis x line*	= top,
					height		= 5.75em,
					ymin=-1.0, ymax=0.0,
					xmin=0, xmax=31.5,
					ytick		= {0},
					xticklabels	= {,,},
				]
				\addplot graphics
				[ 
					xmin	= 0,
					xmax	= 31.5,
					ymin 	= -7.0,
					ymax	= 0.0
				]
				{../spectral_functions/precompiled/eq_U_m4p0_V_m0p25_d.eps};
				\nextgroupplot
				[
					height		= 0.14\textheight,
					axis y discontinuity=parallel,
					ylabel		= {$\omega$},
					ylabel shift	= -7.5pt,
					ytick		= {-6,-4},
					ymin=-6.5, ymax=-2.5,
					xmin=0, xmax=31.5,
					axis x line*=bottom,
					title style={at={(1,0.85)},anchor=east,xshift=-0.75em},
					title	= {\color{white}\textbf{SC phase}},
					colorbar right,
					colormap name	= temp,
					point meta min=0.0,
					point meta max=0.4,
					every colorbar/.append style =
					{
						height			=	\pgfkeysvalueof{/pgfplots/parent axis height}+0.885em,
						ylabel			=	{$S^{\mathrm{SC}}_{\hat d}(q,\omega,0)$},
						width			=	3mm,
						yshift			=	0.885em,
						scaled y ticks	= 	false,
						ytick			= 	{0,0.4},
						yticklabels		=	{$0\phantom{.06}$, $0.4$},
						ylabel shift 	=	-4pt,
					},
				]
				\addplot graphics
				[ 
					xmin	= 0,
					xmax	= 31.5,
					ymin 	= -7.0,
					ymax	= 0.0,
				]
				{../spectral_functions/precompiled/eq_U_m4p0_V_m0p25_d.eps};
			\end{groupplot}
		\end{tikzpicture}
	}

	\vspace{-0.8em}
	\subfloat[\label{fig:eq-neq-neq-eq-spectral-functions-2particle:cdw2sc}]
	{
		\centering
 		\tikzsetnextfilename{eq-neq-neq-eq-spectral-functions-2particle_cdw2sc}
		\begin{tikzpicture}
			\pgfplotsset
			{
				/pgfplots/colormap={temp}{
					rgb255=(36,0,217) 		
					rgb255=(25,29,247) 		
					rgb255=(41,87,255) 		
					rgb255=(61,135,255) 	
					rgb255=(87,176,255) 	
					rgb255=(117,211,255) 	
					rgb255=(153,235,255) 	
					rgb255=(189,249,255) 	
					rgb255=(235,255,255) 	
					rgb255=(255,255,235) 	
					rgb255=(255,242,189) 	
					rgb255=(255,214,153) 	
					rgb255=(255,172,117) 	
					rgb255=(255,120,87) 	
					rgb255=(255,61,61) 		
					rgb255=(247,40,54) 		
					rgb255=(217,22,48) 		
					rgb255=(166,0,33)		
				}
			}
			\begin{groupplot}
				[
					group style = 
					{
						group size 		=	1 by 2,
						vertical sep		=	0.0pt,
						x descriptions at	=	edge bottom,
						y descriptions at	=	edge left,
					},
					xtick		= {0, 16, 32, 48},
					xticklabels	= {,,,},
					xlabel	= {},
					axis on top,
					enlargelimits	= false,
					width		= 0.4\textwidth-2.71pt,
				]
				\nextgroupplot
				[
					ytick={0},
					axis x line* 	= top,
					height		= 5.75em,
					ymin=-1.0, ymax=0.0,
					xmin=0, xmax=31.5,
					ytick		= {0},
				]
				\addplot graphics
				[ 
					xmin	= 0,
					xmax	= 31.5,
					ymin 	= -7.0,
					ymax	= 0.0
				]
				{../spectral_functions/precompiled/neq_U_m4p0_V_0p25_dV_m0p5_d.eps};
				\nextgroupplot
				[
					axis y discontinuity=parallel,
					height		= 0.14\textheight,
					ylabel		= {$\omega$},
					ytick		= {-6,-4},
					ylabel shift	= -7.5pt,
					point meta min=0.0,
					point meta max=0.06,
					colorbar right,
					colormap name	= temp,
					every colorbar/.append style =
					{
						height			=	\pgfkeysvalueof{/pgfplots/parent axis height}+0.885em,
						ylabel			=	{$S^{\phantom{\mathrm{C}}}_{\hat d}(q,\omega,15)$},
						width			=	3mm,
						yshift			=	0.885em,
						scaled y ticks		= 	false,
						ytick			= 	{0,0.06},
						yticklabels		=	{$0\phantom{.06}$, $0.06$},
						ylabel shift 		=	-4pt,
					},
					ymin=-6.5, ymax=-2.5,
					xmin=0, xmax=31.5,
					axis x line* = bottom,
					title style={at={(1,0.85)},anchor=east,xshift=-0.75em},
					title	= {\color{white}\textbf{CDW $\rightarrow$ SC}},
				]
				\addplot graphics
				[ 
					xmin	= 0,
					xmax	= 31.5,
					ymin 	= -7.0,
					ymax	= 0.0
				]
				{../spectral_functions/precompiled/neq_U_m4p0_V_0p25_dV_m0p5_d.eps};
			\end{groupplot}
		\end{tikzpicture}
	}

	\vspace{-0.8em}
	\subfloat[\label{fig:eq-neq-neq-eq-spectral-functions-2particle:cdw2cdw}]
	{
		\centering
 		\tikzsetnextfilename{eq-neq-neq-eq-spectral-functions-2particle_cdw2cdw}
		\begin{tikzpicture}
			\pgfplotsset
			{
				/pgfplots/colormap={temp}{
					rgb255=(36,0,217) 		
					rgb255=(25,29,247) 		
					rgb255=(41,87,255) 		
					rgb255=(61,135,255) 	
					rgb255=(87,176,255) 	
					rgb255=(117,211,255) 	
					rgb255=(153,235,255) 	
					rgb255=(189,249,255) 	
					rgb255=(235,255,255) 	
					rgb255=(255,255,235) 	
					rgb255=(255,242,189) 	
					rgb255=(255,214,153) 	
					rgb255=(255,172,117) 	
					rgb255=(255,120,87) 	
					rgb255=(255,61,61) 		
					rgb255=(247,40,54) 		
					rgb255=(217,22,48) 		
					rgb255=(166,0,33)		
				}
			}
			\begin{groupplot}
				[
					group style = 
					{
						group size 		=	1 by 1,
						vertical sep		=	0.0pt,
						x descriptions at	=	edge bottom,
						y descriptions at	=	edge left,
					},
					xtick		= {0, 16, 32, 48},
					xticklabels	= {,,,},
					xlabel	= {},
					axis on top,
					enlargelimits	= false,
					width		= 0.4\textwidth-2.71pt,
				]
				\nextgroupplot
				[
					height		= 0.14\textheight+0.885em,
					ylabel		= {$\omega$},
					ytick		= {-6,-4,-2,0},
					ylabel shift	= -7.5pt,
					point meta min=0.0,
					point meta max=0.02,
					colorbar right,
					colormap name	= temp,
					every colorbar/.append style =
					{
						height			=	\pgfkeysvalueof{/pgfplots/parent axis height},
						ylabel			=	{$S^{\phantom{\mathrm{C}}}_{\hat d}(q,\omega,15)$},
						width			=	3mm,
						xshift			=	8pt,
						scaled y ticks		= 	false,
						ytick			= 	{0,0.02},
						yticklabels		=	{$0\phantom{.06}$, $0.02$},
						ylabel shift 		=	-4pt,
					},
					ymin=-6.5, ymax=-0.0,
					xmin=0, xmax=31.5,
					axis x line* = bottom,
					title style={at={(1,0.0)},anchor=south east,xshift=-0.75em},
					title	= {\color{white}\textbf{CDW $\rightarrow$ CDW}},
				]
				\addplot graphics
				[ 
					xmin	= 0,
					xmax	= 31.5,
					ymin 	= -7.0,
					ymax	= 0.0
				]
				{../spectral_functions/precompiled/neq_U_m4p0_V_0p25_dU_2p0_d.eps};
			\end{groupplot}
		\end{tikzpicture}
	}

	\vspace{-0.8em}
	\subfloat[\label{fig:eq-neq-neq-eq-spectral-functions-2particle:cdw}]
	{
		\centering
		\tikzsetnextfilename{eq-neq-neq-eq-spectral-functions-2particle_cdw}
		\begin{tikzpicture}
			\pgfplotsset
			{
				/pgfplots/colormap={temp}{
					rgb255=(36,0,217) 		
					rgb255=(25,29,247) 		
					rgb255=(41,87,255) 		
					rgb255=(61,135,255) 	
					rgb255=(87,176,255) 	
					rgb255=(117,211,255) 	
					rgb255=(153,235,255) 	
					rgb255=(189,249,255) 	
					rgb255=(235,255,255) 	
					rgb255=(255,255,235) 	
					rgb255=(255,242,189) 	
					rgb255=(255,214,153) 	
					rgb255=(255,172,117) 	
					rgb255=(255,120,87) 	
					rgb255=(255,61,61) 		
					rgb255=(247,40,54) 		
					rgb255=(217,22,48) 		
					rgb255=(166,0,33)		
				}
			}
			\begin{groupplot}
				[
					group style = 
					{
						group size 		=	1 by 2,
						vertical sep		=	0.0pt,
						x descriptions at	=	edge bottom,
						y descriptions at	=	edge left,
					},
					xtick		= {0, 16, 32, 48},
					xticklabels	= {$0$,$\nicefrac{\pi}{2}$,$\pi$,$\nicefrac{3\pi}{2}$},
					xlabel	= {$q$},
					axis on top,
					enlargelimits	= false,
					width		= 0.4\textwidth-2.71pt,
				]
				\nextgroupplot
				[
					ytick={0},
					axis x line*	= top,
					height		= 5.75em,
					ymin=-1.0, ymax=0.0,
					xmin=0, xmax=31.5,
					ytick		= {0},
				]
				\addplot graphics
				[ 
					xmin	= 0,
					xmax	= 31.5,
					ymin 	= -7.0,
					ymax	= 0.0
				]
				{../spectral_functions/precompiled/eq_U_m4p0_V_0p25_d.eps};
				\nextgroupplot
				[
					height		= 0.14\textheight,
					axis y discontinuity=parallel,
					ylabel		= {$\omega$},
					ylabel shift	= -7.5pt,
					ytick		= {-6,-4},
					point meta min=0.0,
					point meta max=0.03,
					colorbar right,
					colormap name	= temp,
					every colorbar/.append style =
					{
						height			=	\pgfkeysvalueof{/pgfplots/parent axis height}+0.885em,
						ylabel			=	{$S^{\mathrm{CDW}}_{\hat d}(q,\omega,0)$},
						width			=	3mm,
						yshift			=	0.885em,
						scaled y ticks	= 	false,
						ytick			= 	{0,0.03},
						yticklabels		=	{$0\phantom{.06}$, $0.03$},
						ylabel shift 	=	-4pt,
					},
					ymin=-6.5, ymax=-2.5,
					xmin=0, xmax=31.5,
					axis x line* = bottom,
					title style={at={(1,0.85)},anchor=east,xshift=-0.75em},
					title	= {\color{white}\textbf{CDW phase}},
				]
				\addplot graphics
				[ 
					xmin	= 0,
					xmax	= 31.5,
					ymin 	= -7.0,
					ymax	= 0.0
				]
				{../spectral_functions/precompiled/eq_U_m4p0_V_0p25_d.eps};
			\end{groupplot}
		\end{tikzpicture}
	}

	\caption
	{%
		\label{fig:eq-neq-neq-eq-spectral-functions-2particle}%
		Comparison between spectral functions of two-particle excitations%
		\protect\subref{fig:eq-neq-neq-eq-spectral-functions-2particle:sc} in the SC ground state, %
		\protect\subref{fig:eq-neq-neq-eq-spectral-functions-2particle:cdw2sc} after a quench from the CDW phase to the SC phase,%
		\protect\subref{fig:eq-neq-neq-eq-spectral-functions-2particle:cdw2cdw} after a quench from the CDW phase to a more metalic point in the CDW phase ($U=-2$, $V=\nicefrac{1}{4}$).
		Note that we do not show the full frequency scale, because there is no significant spectral weight in the region we left out.
	}
\end{figure}
\begin{figure}[!t]
	\subfloat[\label{fig:eq-neq-neq-eq-spectral-functions-1particle:sc}]
	{
		\centering
		\tikzsetnextfilename{eq-neq-neq-eq-spectral-functions-1particle_sc}
		\begin{tikzpicture}
			\pgfplotsset
			{
				/pgfplots/colormap={temp}{
					rgb255=(36,0,217) 		
					rgb255=(25,29,247) 		
					rgb255=(41,87,255) 		
					rgb255=(61,135,255) 	
					rgb255=(87,176,255) 	
					rgb255=(117,211,255) 	
					rgb255=(153,235,255) 	
					rgb255=(189,249,255) 	
					rgb255=(235,255,255) 	
					rgb255=(255,255,235) 	
					rgb255=(255,242,189) 	
					rgb255=(255,214,153) 	
					rgb255=(255,172,117) 	
					rgb255=(255,120,87) 	
					rgb255=(255,61,61) 		
					rgb255=(247,40,54) 		
					rgb255=(217,22,48) 		
					rgb255=(166,0,33)		
				}
			}
			\begin{groupplot}
				[
					group style = 
					{
						group size 		=	1 by 2,
						vertical sep		=	0.0pt,
						x descriptions at	=	edge bottom,
						y descriptions at	=	edge left,
					},
					xtick		= {0, 16, 32, 48},
					xticklabels	= {,,,},
					xlabel	= {},
					axis on top,
					enlargelimits	= false,
					width		= 0.4\textwidth-2.71pt,
				]
				\nextgroupplot
				[
					axis x line*	= top,
					height		= 5.75em,
					ymin=-1.0, ymax=0.0,
					xmin=0, xmax=31.5,
					ytick		= {0},
					xticklabels	= {,,},
				]
				\addplot graphics
				[ 
					xmin	= 0,
					xmax	= 31.5,
					ymin 	= -10.0,
					ymax	= 0.0
				]
				{../spectral_functions/precompiled/eq_U_m4p0_V_m0p25_f.eps};
				\nextgroupplot
				[
					axis y discontinuity=parallel,
					height		= 0.14\textheight,
					colorbar right,
					colormap name	= temp,
					xticklabels	= {,,},
					ylabel		= {$\omega$},
					ylabel shift	= -7.5pt,
					ytick		= {-8,-4},
					point meta min=0.0,
					point meta max=0.06,
					every colorbar/.append style =
					{
						height			=	\pgfkeysvalueof{/pgfplots/parent axis height}+0.885em,
						ylabel			= 	{$S^{\mathrm{SC}}_{\hat c}(q,\omega,0)$},
						width			=	3mm,
						yshift			=	0.885em,
						scaled y ticks		= 	false,
						ytick			= 	{0,0.06},
						yticklabels		=	{$0\phantom{.00}$, $0.06$},
						ylabel shift 		=	-4pt,
					},
					ymin=-8.0,  ymax=-2.0,
					xmin=0, xmax=31.5,
					axis x line*	= bottom,
					title style	= {at={(1,0.85)},anchor=east,xshift=-0.75em},
					title		= {\color{white}\textbf{SC phase}},
				]
				\addplot graphics
				[ 
					xmin	= 0,
					xmax	= 31.5,
					ymin 	= -10.0,
					ymax	= 0.0
				]
				{../spectral_functions/precompiled/eq_U_m4p0_V_m0p25_f.eps};
			\end{groupplot}
		\end{tikzpicture}
	}

	\vspace{-2em}
	\subfloat[\label{fig:eq-neq-neq-eq-spectral-functions-1particle:cdw2sc}]
	{
		\centering
		\tikzsetnextfilename{eq-neq-neq-eq-spectral-functions-1particle_cdw2sc}
		\begin{tikzpicture}
			\pgfplotsset
			{
				/pgfplots/colormap={temp}{
					rgb255=(36,0,217) 		
					rgb255=(25,29,247) 		
					rgb255=(41,87,255) 		
					rgb255=(61,135,255) 	
					rgb255=(87,176,255) 	
					rgb255=(117,211,255) 	
					rgb255=(153,235,255) 	
					rgb255=(189,249,255) 	
					rgb255=(235,255,255) 	
					rgb255=(255,255,235) 	
					rgb255=(255,242,189) 	
					rgb255=(255,214,153) 	
					rgb255=(255,172,117) 	
					rgb255=(255,120,87) 	
					rgb255=(255,61,61) 		
					rgb255=(247,40,54) 		
					rgb255=(217,22,48) 		
					rgb255=(166,0,33)		
				}
			}
			\begin{groupplot}
				[
					group style = 
					{
						group size 		=	1 by 2,
						vertical sep		=	0.0pt,
						x descriptions at	=	edge bottom,
						y descriptions at	=	edge left,
					},
					xtick		= {0, 16, 32, 48},
					xticklabels	= {,,},
					xlabel	= {},
					axis on top,
					enlargelimits	= false,
					width		= 0.4\textwidth-2.71pt,
				]
				\nextgroupplot
				[
					ytick		= {0},
					axis x line*	= top,
					height		= 5.75em,
					ymin=-1.0, ymax=0.0,
					xmin=0, xmax=31.5,
				]
				\addplot graphics
				[ 
					xmin	= 0,
					xmax	= 31.5,
					ymin 	= -10.0,
					ymax	= 0.0
				]
				{../spectral_functions/precompiled/neq_U_m4p0_V_0p25_dV_m0p5_f.eps};
				\nextgroupplot
				[
					axis y discontinuity=parallel,
					title style	= {at={(1,0)},anchor=south east,xshift=-0.75em},
					title		= {\color{white}\textbf{CDW $\rightarrow$ SC}},
					height		= 0.14\textheight,
					ylabel		= {$\omega$},
					ylabel shift	= -7.5pt,
					ytick		= {-8,-4},
					colorbar right,
					colormap name	= temp,
					point meta min=0.0,
					point meta max=0.02,
					every colorbar/.append style =
					{
						height			=	\pgfkeysvalueof{/pgfplots/parent axis height}+0.885em,
						ylabel			= 	{$S^{\phantom{\mathrm{C}}}_{\hat c}(q,\omega,15)$},
						width			=	3mm,
						yshift			=	0.885em,
						scaled y ticks		= 	false,
						ytick			= 	{0,0.02},
						yticklabels		=	{$0\phantom{.06}$, $0.02$},
						ylabel shift 		=	-4pt,
					},
					ymin=-8.0, ymax=-2.0,
					xmin=0, xmax=31.5,
					axis x line*	= bottom,
				]
				\addplot graphics
				[ 
					xmin	= 0.0,
					xmax	= 31.5,
					ymin 	= -10.0,
					ymax	= 0.0
				]
				{../spectral_functions/precompiled/neq_U_m4p0_V_0p25_dV_m0p5_f.eps};
			\end{groupplot}
		\end{tikzpicture}
	}
	
	\vspace{-2em}
	\subfloat[\label{fig:eq-neq-neq-eq-spectral-functions-1particle:cdw2cdw}]
	{
		\centering
		\tikzsetnextfilename{eq-neq-neq-eq-spectral-functions-1particle_cdw2cdw}
		\begin{tikzpicture}
			\pgfplotsset
			{
				/pgfplots/colormap={temp}{
					rgb255=(36,0,217) 		
					rgb255=(25,29,247) 		
					rgb255=(41,87,255) 		
					rgb255=(61,135,255) 	
					rgb255=(87,176,255) 	
					rgb255=(117,211,255) 	
					rgb255=(153,235,255) 	
					rgb255=(189,249,255) 	
					rgb255=(235,255,255) 	
					rgb255=(255,255,235) 	
					rgb255=(255,242,189) 	
					rgb255=(255,214,153) 	
					rgb255=(255,172,117) 	
					rgb255=(255,120,87) 	
					rgb255=(255,61,61) 		
					rgb255=(247,40,54) 		
					rgb255=(217,22,48) 		
					rgb255=(166,0,33)		
				}
			}
			\begin{groupplot}
				[
					group style = 
					{
						group size 		=	1 by 1,
						vertical sep		=	0.0pt,
						x descriptions at	=	edge bottom,
						y descriptions at	=	edge left,
					},
					xtick		= {0, 16, 32, 48},
					xticklabels	= {,,,},
					xlabel	= {},
					axis on top,
					enlargelimits	= false,
					width		= 0.4\textwidth-2.71pt,
				]
				\nextgroupplot
				[
					title style	= {at={(1,0)},anchor=south east,xshift=-0.75em},
					title		= {\color{white}\textbf{CDW $\rightarrow$ CDW}},
					height		= 0.14\textheight+0.885em,
					ylabel		= {$\omega$},
					ylabel shift	= -7.5pt,
					ytick		= {-8,-4, 0},
					colorbar right,
					colormap name	= temp,
					point meta min=0.0,
					point meta max=0.05,
					every colorbar/.append style =
					{
						height			=	\pgfkeysvalueof{/pgfplots/parent axis height},
						ylabel			= 	{$S^{\phantom{\mathrm{C}}}_{\hat c}(q,\omega,15)$},
						width			=	3mm,
						xshift			=	8pt,
						scaled y ticks		= 	false,
						ytick			= 	{0,0.05},
						yticklabels		=	{$0\phantom{.00}$, $0.05$},
						ylabel shift 		=	-4pt,
					},
					ymin=-8.0, ymax=0.0,
					xmin=0, xmax=31.5,
					axis x line*	= bottom,
				]
				\addplot graphics
				[ 
					xmin	= 0,
					xmax	= 31.5,
					ymin 	= -10.0,
					ymax	= 0.0
				]
				{../spectral_functions/precompiled/neq_U_m4p0_V_0p25_dU_2p0_f.eps};
			\end{groupplot}
		\end{tikzpicture}
	}

	\vspace{-0.8em}
	\subfloat[\label{fig:eq-neq-neq-eq-spectral-functions-1particle:cdw}]
	{
		\centering
		\tikzsetnextfilename{eq-neq-neq-eq-spectral-functions-1particle_cdw}
		\begin{tikzpicture}
			\pgfplotsset
			{
				/pgfplots/colormap={temp}{
					rgb255=(36,0,217) 		
					rgb255=(25,29,247) 		
					rgb255=(41,87,255) 		
					rgb255=(61,135,255) 	
					rgb255=(87,176,255) 	
					rgb255=(117,211,255) 	
					rgb255=(153,235,255) 	
					rgb255=(189,249,255) 	
					rgb255=(235,255,255) 	
					rgb255=(255,255,235) 	
					rgb255=(255,242,189) 	
					rgb255=(255,214,153) 	
					rgb255=(255,172,117) 	
					rgb255=(255,120,87) 	
					rgb255=(255,61,61) 		
					rgb255=(247,40,54) 		
					rgb255=(217,22,48) 		
					rgb255=(166,0,33)		
				}
			}
			\begin{groupplot}
				[
					group style = 
					{
						group size 		=	1 by 2,
						vertical sep		=	0.0pt,
						x descriptions at	=	edge bottom,
						y descriptions at	=	edge left,
					},
					xtick		= {0, 16, 32, 48},
					xticklabels	= {$0$,$\nicefrac{\pi}{2}$,$\pi$,$\nicefrac{3\pi}{2}$},
					xlabel	= {$q$},
					axis on top,
					enlargelimits	= false,
					width		= 0.4\textwidth-2.71pt,
				]
				\nextgroupplot
				[				
					ytick		= {0},
					axis x line*	= top,
					height		= 5.75em,
					ymin=-1.0, ymax=0.0,
					xmin=0, xmax=31.5,
					ytick		= {0},
				]
				\addplot graphics
				[ 
					xmin	= 0,
					xmax	= 31.5,
					ymin 	= -10.0,
					ymax	= 0.0
				]
				{../spectral_functions/precompiled/eq_U_m4p0_V_0p25_f.eps};
				\nextgroupplot
				[
					axis y discontinuity = parallel,
					title style	= {at={(1,0)},anchor=south east,xshift=-0.75em},
					title		= {\color{white}\textbf{CDW phase}},
					height		= 0.14\textheight,
					ylabel		= {$\omega$},
					ylabel shift	= -7.5pt,
					ytick		= {-8,-4},
					colorbar right,
					colormap name	= temp,
					point meta min=0.0,
					point meta max=0.06,
					every colorbar/.append style =
					{
						height			=	\pgfkeysvalueof{/pgfplots/parent axis height}+0.885em,
						ylabel			= 	{$S^{\mathrm{CDW}}_{\hat c}(q,\omega,0)$},
						width			=	3mm,
						yshift			=	0.885em,
						scaled y ticks		= 	false,
						ytick			= 	{0,0.06},
						yticklabels		=	{$0\phantom{.00}$, $0.06$},
						ylabel shift 		=	-4pt,
					},
					ymin=-8.0, ymax=-2.0,
					xmin=0, xmax=31.5,
					axis x line*	= bottom,
				]
				\addplot graphics
				[ 
					xmin	= 0,
					xmax	= 31.5,
					ymin 	= -10.0,
					ymax	= 0.0
				]
				{../spectral_functions/precompiled/eq_U_m4p0_V_0p25_f.eps};
			\end{groupplot}
		\end{tikzpicture}
	}
	\caption
	{%
		\label{fig:eq-neq-neq-eq-spectral-functions-1particle}%
		Spectral functions of single\hyp particle excitations analog to \protect\cref{fig:eq-neq-neq-eq-spectral-functions-2particle}.
		Note that we do not show the full frequency scale, because there is no significant spectral weight in the region we left out.
	}
\end{figure}
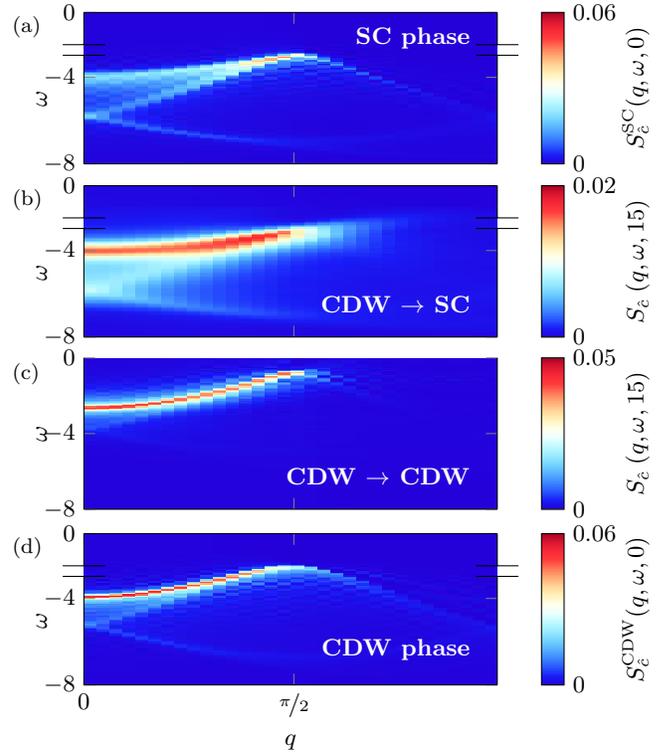
In contrast to the optical conductivity, in the spectral functions we find clearly distinguishable features belonging either to the elementary excitations of the \gls{SC} or \gls{CDW} phase.
In \cref{fig:eq-neq-neq-eq-spectral-functions-2particle,fig:eq-neq-neq-eq-spectral-functions-1particle} the spectral functions of two- and single-particle excitations are shown, respectively.
For easier comparison, the ground states of the \gls{SC} and the \gls{CDW} phase, and the quench into the \gls{SC} phase are included in both figures.
The two-particle spectral function \cref{fig:eq-neq-neq-eq-spectral-functions-2particle} shows a continuum of excitations at small frequencies which is to be contrasted with the sharp excitations in the \gls{SC} phase.
Interestingly, we also find a weak transfer of spectral weight towards $k=0$ after the quench which we relate to the enhanced metalliticity in the post-quench state.
However, the formation of a condensate is suppressed by the continuum at smaller frequencies.
On the other hand, in the single-particle spectral function \cref{fig:eq-neq-neq-eq-spectral-functions-1particle} the identification of the excitation spectrum with either the \gls{SC} or the \gls{CDW} phase is not possible, unambiguously.
Both findings are consistent with the results discussed above and in the main text supporting our picture that the two-particle spectral function is a more reliable indicator for enhanced superconductivity.
\section{Coherent and dissipative charge flow}
\begin{figure}
	\subfloat[\label{fig:eq:probe-pulse:charge-densities:sc}]
	{
		\centering
 		\tikzsetnextfilename{eq_probe-pulse_charge-densities_sc}
 		\begin{tikzpicture}
			\pgfplotsset
			{
				/pgfplots/colormap={temp}{
					rgb255=(36,0,217) 		
					rgb255=(25,29,247) 		
					rgb255=(41,87,255) 		
					rgb255=(61,135,255) 	
					rgb255=(87,176,255) 	
					rgb255=(117,211,255) 	
					rgb255=(153,235,255) 	
					rgb255=(189,249,255) 	
					rgb255=(235,255,255) 	
					rgb255=(255,255,235) 	
					rgb255=(255,242,189) 	
					rgb255=(255,214,153) 	
					rgb255=(255,172,117) 	
					rgb255=(255,120,87) 	
					rgb255=(255,61,61) 		
					rgb255=(247,40,54) 		
					rgb255=(217,22,48) 		
					rgb255=(166,0,33)		
				}
			}
			\begin{axis}
				[
					width		= 0.4\textwidth,
					axis on top,
					enlargelimits	= false,
					ticklabel style	= {font=\small},
					xlabel style	= {font=\small,yshift=2.5pt},
					ylabel style	= {font=\small,yshift=-3.5pt},
					ylabel		= {time $t\,[\nicefrac{1}{t_\mathrm{hop}}]$},
					xlabel		= {site $i$},
					colorbar right,
					colormap name	= temp,
					ytick		= {0,50,100},
					xtick		= {0,16,32,48},
					point meta min=-0.06,
					point meta max=0.06,
					every colorbar/.append style =
						{
							width			=	3mm,
							scaled y ticks		= 	false,
							ylabel style		= 	{font=\small,yshift=-2.5pt},
							ytick			= 	{-0.06,0.0,0.06},
							yticklabels		=	{$-0.06$, $\phantom{-}0.00$, $\phantom{-}0.06$},
							ylabel shift 		=	-4pt,
						},
					title		= {\small $\Delta n_{i}(t) = \langle \hat{n}_{i}(t)\rangle_{\mathrm{probe}}-\langle \hat{n}_{i} \rangle_0$},
					title style = {yshift=-0.75em, xshift=1.75em,},
				]
				\addplot graphics
				[
					xmin = 0, 
					xmax = 63, 
					ymin = 0, 
					ymax = 100
				]
				{../spectral_functions/precompiled/eq_U_m4p0_V_m0p25_pulse_response_densities_diff.eps};
			\end{axis}
 		\end{tikzpicture}
 	}%
 	
	\subfloat[\label{fig:eq:probe-pulse:charge-densities:cdw}]
	{
		\centering
 		\tikzsetnextfilename{eq_probe-pulse_charge-densities_cdw}
 		\begin{tikzpicture}
			\pgfplotsset
			{
				/pgfplots/colormap={temp}{
					rgb255=(36,0,217) 		
					rgb255=(25,29,247) 		
					rgb255=(41,87,255) 		
					rgb255=(61,135,255) 	
					rgb255=(87,176,255) 	
					rgb255=(117,211,255) 	
					rgb255=(153,235,255) 	
					rgb255=(189,249,255) 	
					rgb255=(235,255,255) 	
					rgb255=(255,255,235) 	
					rgb255=(255,242,189) 	
					rgb255=(255,214,153) 	
					rgb255=(255,172,117) 	
					rgb255=(255,120,87) 	
					rgb255=(255,61,61) 		
					rgb255=(247,40,54) 		
					rgb255=(217,22,48) 		
					rgb255=(166,0,33)		
				}
			}
			\begin{axis}
				[
					width		= 0.4\textwidth,
					axis on top,
					enlargelimits	= false,
					ticklabel style	= {font=\small},
					xlabel style	= {font=\small,yshift=2.5pt},
					ylabel style	= {font=\small,yshift=-3.5pt},
					ylabel		= {time $t\,[\nicefrac{1}{t_\mathrm{hop}}]$},
					xlabel		= {site $i$},
					colorbar right,
					colormap name	= temp,
					ytick		= {0,50,100},
					xtick		= {0,16,32,48},
					point meta min=-0.06,
					point meta max=0.06,
					every colorbar/.append style =
						{
							width			=	3mm,
							scaled y ticks		= 	false,
							ylabel style		= 	{font=\small,yshift=-2.5pt},
							ytick			= 	{-0.06,0.0,0.06},
							yticklabels		=	{$-0.06$, $\phantom{-}0.00$, $\phantom{-}0.06$},
							ylabel shift 		=	-4pt,
						},
					title		= {$\Delta n_{i}(t) = \langle \hat{n}_{i}(t)\rangle_{\mathrm{probe}}-\langle \hat{n}_{i} \rangle_0$},
					title style = {yshift=-0.75em, xshift=1.75em,},
				]
				\addplot graphics
				[
					xmin = 0, 
					xmax = 63, 
					ymin = 0, 
					ymax = 100
				]
				{../spectral_functions/precompiled/eq_U_m4p0_V_0p25_pulse_response_densities_diff.eps};
			\end{axis}
 		\end{tikzpicture}
 	}
 	\caption
 	{
		\label{fig:eq:probe-pulse:charge-densities}
		Induced charge flow $\Delta n_{i}(t) = \langle \hat{n}_{i}(t)\rangle_{\mathrm{probe}}-\langle \hat{n}_{i} \rangle_0$ after applying a probe pulse $A(t,\Delta t)=A_{0}e^{-\frac{(t-\Delta t)^2}{2\tau^2}}\cos(\omega_{0} t)$ with $\Delta = 0.0$, $\tau = 0.05$ and $\omega_0=2.38$ in \protect\subref{fig:eq:probe-pulse:charge-densities:sc} the superconducting and \protect\subref{fig:eq:probe-pulse:charge-densities:cdw} the charge density wave phase.
		Coherent transport is seen in the superconducting phase where charge is accumulated at the boundaries and flows coherently after the probe pulse has passed the system.
		In the charge density wave phase scattering processes destroy the coherence yielding dissipative transport.
	}
\end{figure}
In this section we consider the charge transport in a one-dimensional extended Hubbard model in its SC and CDW ground states, respectively, after excitation with a probe pulse.
For this purpose we calculated the time-dependent charge densities after applying a probe pulse
\begin{align}
	A(t,\Delta t)=A_{0}e^{-\frac{(t-\Delta t)^2}{2\tau^2}}\cos(\omega_{0} t) \; ,
\end{align}
with $A_0=0.5$, $\Delta t = 0.0$, $\tau = 0.05$ and $\omega_0=2.38$ in both ground states.
%
\Cref{fig:eq:probe-pulse:charge-densities} displays the induced charge flow
\begin{align}
	\Delta n_{i}(t) = \langle \hat{n}_{i}(t)\rangle_{\mathrm{probe}}-\langle \hat{n}_{i} \rangle_0
\end{align}
where we subtracted the equilibrium value $\langle \hat{n}_{i} \rangle_0$.
Being in the linear response regime this quantity constitutes a measure for the charge transport caused by an optical excitation.
Since the system we are considering is calculated using open boundary conditions, the question arises how coherent charge transport in the superconducting phase manifests itself.
In \cref{fig:eq:probe-pulse:charge-densities:sc} (see also the inset of Fig. 1c in the main text) this is demonstrated in the superconducting phase ($U/t_{\rm hop}=-4$, $V/t_{\rm hop}=-\nicefrac{1}{4}$).
The probe pulse creates charge accumulation at the boundaries of the system, which act as topological defects.
Importantly, the resulting charge flow is nearly dissipationless and we find perfect reflection at the boundaries indicating a long-lived induced current.
We interpret our findings as a one-dimensional analogy of the Meissner effect.
In contrast in \cref{fig:eq:probe-pulse:charge-densities:cdw} the induced charge flow in the charge density wave phase ($U/t_{\rm hop}=-4$, $V/t_{\rm hop}=\nicefrac{1}{4}$) is shown after exciting the system.
Again, the effect of the probe pulse is to accumulate charge at the boundaries.
However, here we find strong scattering processes rendering the charge flow dissipative.
\section{Check for Convergence}
\begin{figure}
	\subfloat[\label{fig:neq-neq-spectral-functions-2particle:1000}]
	{
		\centering
 		\tikzsetnextfilename{neq-neq-spectral-functions-2particle-1000}
		\begin{tikzpicture}
			\pgfplotsset
			{
				/pgfplots/colormap={temp}{
					rgb255=(36,0,217) 		
					rgb255=(25,29,247) 		
					rgb255=(41,87,255) 		
					rgb255=(61,135,255) 	
					rgb255=(87,176,255) 	
					rgb255=(117,211,255) 	
					rgb255=(153,235,255) 	
					rgb255=(189,249,255) 	
					rgb255=(235,255,255) 	
					rgb255=(255,255,235) 	
					rgb255=(255,242,189) 	
					rgb255=(255,214,153) 	
					rgb255=(255,172,117) 	
					rgb255=(255,120,87) 	
					rgb255=(255,61,61) 		
					rgb255=(247,40,54) 		
					rgb255=(217,22,48) 		
					rgb255=(166,0,33)		
				}
			}
			\begin{groupplot}
				[
					group style = 
					{
						group size 		=	1 by 2,
						vertical sep		=	0.0pt,
						x descriptions at	=	edge bottom,
						y descriptions at	=	edge left,
					},
					xtick		= {0, 16, 32, 48},
					xticklabels	= {,,,},
					xlabel	= {},
					axis on top,
					enlargelimits	= false,
					width		= 0.4\textwidth-2.71pt,
				]
				\nextgroupplot
				[
					ytick={0},
					axis x line* 	= top,
					height		= 5.75em,
					ymin=-1.0, ymax=0.0,
					xmin=0, xmax=31.5,
					ytick		= {0},
				]
				\addplot graphics
				[ 
					xmin	= 0,
					xmax	= 31.5,
					ymin 	= -7.0,
					ymax	= 0.0
				]
				{../spectral_functions/precompiled/neq_U_m4p0_V_0p25_dV_m0p5_d.eps};
				\nextgroupplot
				[
					axis y discontinuity=parallel,
					height		= 0.14\textheight,
					ylabel		= {$\omega$},
					ytick		= {-6,-4},
					ylabel shift	= -7.5pt,
					point meta min=0.0,
					point meta max=0.06,
					colorbar right,
					colormap name	= temp,
					every colorbar/.append style =
					{
						height			=	\pgfkeysvalueof{/pgfplots/parent axis height}+0.885em,
						ylabel			=	{$S^{\phantom{\mathrm{C}}}_{\hat d}(q,\omega,15)$},
						width			=	3mm,
						yshift			=	0.885em,
						scaled y ticks		= 	false,
						ytick			= 	{0,0.06},
						yticklabels		=	{$0\phantom{.06}$, $0.06$},
						ylabel shift 		=	-4pt,
					},
					ymin=-6.5, ymax=-2.5,
					xmin=0, xmax=31.5,
					axis x line* = bottom,
					title style={at={(1,0.85)},anchor=east,xshift=-0.75em},
					title	= {\color{white}\textbf{CDW $\rightarrow$ SC, $m_{\mathrm{max}}=1000$}},
				]
				\addplot graphics
				[ 
					xmin	= 0,
					xmax	= 31.5,
					ymin 	= -7.0,
					ymax	= 0.0
				]
				{../spectral_functions/precompiled/neq_U_m4p0_V_0p25_dV_m0p5_d.eps};
			\end{groupplot}
		\end{tikzpicture}
	}

	\vspace{-0.8em}
	\subfloat[\label{fig:neq-neq-spectral-functions-2particle:500}]
	{
		\centering
		\tikzsetnextfilename{neq-neq-spectral-functions-2particle-500}
		\begin{tikzpicture}
			\pgfplotsset
			{
				/pgfplots/colormap={temp}{
					rgb255=(36,0,217) 		
					rgb255=(25,29,247) 		
					rgb255=(41,87,255) 		
					rgb255=(61,135,255) 	
					rgb255=(87,176,255) 	
					rgb255=(117,211,255) 	
					rgb255=(153,235,255) 	
					rgb255=(189,249,255) 	
					rgb255=(235,255,255) 	
					rgb255=(255,255,235) 	
					rgb255=(255,242,189) 	
					rgb255=(255,214,153) 	
					rgb255=(255,172,117) 	
					rgb255=(255,120,87) 	
					rgb255=(255,61,61) 		
					rgb255=(247,40,54) 		
					rgb255=(217,22,48) 		
					rgb255=(166,0,33)		
				}
			}
			\begin{groupplot}
				[
					group style = 
					{
						group size 		=	1 by 2,
						vertical sep		=	0.0pt,
						x descriptions at	=	edge bottom,
						y descriptions at	=	edge left,
					},
					xtick		= {0, 16, 32, 48},
					xticklabels	= {$0$,$\nicefrac{\pi}{2}$,$\pi$,$\nicefrac{3\pi}{2}$},
					xlabel	= {$q$},
					axis on top,
					enlargelimits	= false,
					width		= 0.4\textwidth-2.71pt,
				]
				\nextgroupplot
				[
					ytick={0},
					axis x line* 	= top,
					height		= 5.75em,
					ymin=-1.0, ymax=0.0,
					xmin=0, xmax=31.5,
					ytick		= {0},
				]
				\addplot graphics
				[ 
					xmin	= 0,
					xmax	= 31.5,
					ymin 	= -7.0,
					ymax	= 0.0
				]
				{../spectral_functions/precompiled/neq_U_m4p0_V_0p25_dV_m0p5_chi_max_500_d.eps};
				\nextgroupplot
				[
					axis y discontinuity=parallel,
					height		= 0.14\textheight,
					ylabel		= {$\omega$},
					ytick		= {-6,-4},
					ylabel shift	= -7.5pt,
					point meta min=0.0,
					point meta max=0.06,
					colorbar right,
					colormap name	= temp,
					every colorbar/.append style =
					{
						height			=	\pgfkeysvalueof{/pgfplots/parent axis height}+0.885em,
						ylabel			=	{$S^{\phantom{\mathrm{C}}}_{\hat d}(q,\omega,15)$},
						width			=	3mm,
						yshift			=	0.885em,
						scaled y ticks		= 	false,
						ytick			= 	{0,0.06},
						yticklabels		=	{$0\phantom{.06}$, $0.06$},
						ylabel shift 		=	-4pt,
					},
					ymin=-6.5, ymax=-2.5,
					xmin=0, xmax=31.5,
					axis x line* = bottom,
					title style={at={(1,0.85)},anchor=east,xshift=-0.75em},
					title	= {\color{white}\textbf{CDW $\rightarrow$ SC, $m_{\mathrm{max}}=500$}},
				]
				\addplot graphics
				[ 
					xmin	= 0,
					xmax	= 31.5,
					ymin 	= -7.0,
					ymax	= 0.0
				]
				{../spectral_functions/precompiled/neq_U_m4p0_V_0p25_dV_m0p5_chi_max_500_d.eps};
			\end{groupplot}
		\end{tikzpicture}
	}

	\caption
	{%
		\label{fig:neq-neq-spectral-functions-2particle}%
		Spectral functions of two\hyp particle excitations after quenching $V=0.25 \rightarrow V=-0.25$ with \protect\subref{fig:neq-neq-spectral-functions-2particle:1000} $m_{\mathrm{max}}=1000$ and  \protect\subref{fig:neq-neq-spectral-functions-2particle:500} $m_{\mathrm{max}}=500$.
		Note that we do not show the full frequency scale, because there is no significant spectral weight in the region we left out.
	}
\end{figure}
\begin{figure}
	\subfloat[\label{fig:neq-neq-spectral-functions-1particle:1000}]
	{
		\centering
		\tikzsetnextfilename{neq-neq-spectral-functions-1particle-1000}
		\begin{tikzpicture}
			\pgfplotsset
			{
				/pgfplots/colormap={temp}{
					rgb255=(36,0,217) 		
					rgb255=(25,29,247) 		
					rgb255=(41,87,255) 		
					rgb255=(61,135,255) 	
					rgb255=(87,176,255) 	
					rgb255=(117,211,255) 	
					rgb255=(153,235,255) 	
					rgb255=(189,249,255) 	
					rgb255=(235,255,255) 	
					rgb255=(255,255,235) 	
					rgb255=(255,242,189) 	
					rgb255=(255,214,153) 	
					rgb255=(255,172,117) 	
					rgb255=(255,120,87) 	
					rgb255=(255,61,61) 		
					rgb255=(247,40,54) 		
					rgb255=(217,22,48) 		
					rgb255=(166,0,33)		
				}
			}
			\begin{groupplot}
				[
					group style = 
					{
						group size 		=	1 by 2,
						vertical sep		=	0.0pt,
						x descriptions at	=	edge bottom,
						y descriptions at	=	edge left,
					},
					xtick		= {0, 16, 32, 48},
					xticklabels	= {,,},
					xlabel	= {},
					axis on top,
					enlargelimits	= false,
					width		= 0.4\textwidth-2.71pt,
				]
				\nextgroupplot
				[
					ytick		= {0},
					axis x line*	= top,
					height		= 5.75em,
					ymin=-1.0, ymax=0.0,
					xmin=0, xmax=31.5,
				]
				\addplot graphics
				[ 
					xmin	= 0,
					xmax	= 31.5,
					ymin 	= -10.0,
					ymax	= 0.0
				]
				{../spectral_functions/precompiled/neq_U_m4p0_V_0p25_dV_m0p5_f.eps};
				\nextgroupplot
				[
					axis y discontinuity=parallel,
					title style	= {at={(1,0.85)},anchor=east,xshift=-0.75em},
					title		= {\color{white}\textbf{CDW $\rightarrow$ SC, $m_{\mathrm{max}}=1000$}},
					height		= 0.14\textheight,
					ylabel		= {$\omega$},
					ylabel shift	= -7.5pt,
					ytick		= {-8,-4},
					colorbar right,
					colormap name	= temp,
					point meta min=0.0,
					point meta max=0.02,
					every colorbar/.append style =
					{
						height			=	\pgfkeysvalueof{/pgfplots/parent axis height}+0.885em,
						ylabel			= 	{$S^{\phantom{\mathrm{C}}}_{\hat c}(q,\omega,15)$},
						width			=	3mm,
						yshift			=	0.885em,
						scaled y ticks		= 	false,
						ytick			= 	{0,0.02},
						yticklabels		=	{$0\phantom{.06}$, $0.02$},
						ylabel shift 		=	-4pt,
					},
					ymin=-8.0, ymax=-2.0,
					xmin=0, xmax=31.5,
					axis x line*	= bottom,
				]
				\addplot graphics
				[ 
					xmin	= 0.0,
					xmax	= 31.5,
					ymin 	= -10.0,
					ymax	= 0.0
				]
				{../spectral_functions/precompiled/neq_U_m4p0_V_0p25_dV_m0p5_f.eps};
			\end{groupplot}
		\end{tikzpicture}
	}

	\vspace{-2.0em}
	\subfloat[\label{fig:neq-neq-spectral-functions-1particle:500}]
	{
		\centering
		\tikzsetnextfilename{neq-neq-spectral-functions-1particle-500}
		\begin{tikzpicture}
			\pgfplotsset
			{
				/pgfplots/colormap={temp}{
					rgb255=(36,0,217) 		
					rgb255=(25,29,247) 		
					rgb255=(41,87,255) 		
					rgb255=(61,135,255) 	
					rgb255=(87,176,255) 	
					rgb255=(117,211,255) 	
					rgb255=(153,235,255) 	
					rgb255=(189,249,255) 	
					rgb255=(235,255,255) 	
					rgb255=(255,255,235) 	
					rgb255=(255,242,189) 	
					rgb255=(255,214,153) 	
					rgb255=(255,172,117) 	
					rgb255=(255,120,87) 	
					rgb255=(255,61,61) 		
					rgb255=(247,40,54) 		
					rgb255=(217,22,48) 		
					rgb255=(166,0,33)		
				}
			}
			\begin{groupplot}
				[
					group style = 
					{
						group size 		=	1 by 2,
						vertical sep		=	0.0pt,
						x descriptions at	=	edge bottom,
						y descriptions at	=	edge left,
					},
					xtick		= {0, 16, 32, 48},
					xticklabels	= {$0$,$\nicefrac{\pi}{2}$,$\pi$,$\nicefrac{3\pi}{2}$},
					xlabel	= {$q$},
					axis on top,
					enlargelimits	= false,
					width		= 0.4\textwidth-2.71pt,
				]
				\nextgroupplot
				[
					ytick		= {0},
					axis x line*	= top,
					height		= 5.75em,
					ymin=-1.0, ymax=0.0,
					xmin=0, xmax=31.5,
				]
				\addplot graphics
				[ 
					xmin	= 0,
					xmax	= 31.5,
					ymin 	= -10.0,
					ymax	= 0.0
				]
				{../spectral_functions/precompiled/neq_U_m4p0_V_0p25_dV_m0p5_chi_max_500_f.eps};
				\nextgroupplot
				[
					axis y discontinuity=parallel,
					title style	= {at={(1,0.85)},anchor=east,xshift=-0.75em},
					title		= {\color{white}\textbf{CDW $\rightarrow$ SC, $m_{\mathrm{max}}=500$}},
					height		= 0.14\textheight,
					ylabel		= {$\omega$},
					ylabel shift	= -7.5pt,
					ytick		= {-8,-4},
					colorbar right,
					colormap name	= temp,
					point meta min=0.0,
					point meta max=0.02,
					every colorbar/.append style =
					{
						height			=	\pgfkeysvalueof{/pgfplots/parent axis height}+0.885em,
						ylabel			= 	{$S^{\phantom{\mathrm{C}}}_{\hat c}(q,\omega,15)$},
						width			=	3mm,
						yshift			=	0.885em,
						scaled y ticks		= 	false,
						ytick			= 	{0,0.02},
						yticklabels		=	{$0\phantom{.06}$, $0.02$},
						ylabel shift 		=	-4pt,
					},
					ymin=-8.0, ymax=-2.0,
					xmin=0, xmax=31.5,
					axis x line*	= bottom,
				]
				\addplot graphics
				[ 
					xmin	= 0.0,
					xmax	= 31.5,
					ymin 	= -10.0,
					ymax	= 0.0
				]
				{../spectral_functions/precompiled/neq_U_m4p0_V_0p25_dV_m0p5_chi_max_500_f.eps};
			\end{groupplot}
		\end{tikzpicture}
	}

	\caption
	{%
		\label{fig:neq-neq-spectral-functions-1particle}%
		Spectral functions of single\hyp particle excitations after quenching $V=0.25 \rightarrow V=-0.25$ with \protect\subref{fig:neq-neq-spectral-functions-1particle:1000} $m_{\mathrm{max}}=1000$ and  \protect\subref{fig:neq-neq-spectral-functions-1particle:500} $m_{\mathrm{max}}=500$.
		Note that we do not show the full frequency scale, because there is no significant spectral weight in the region we left out.
	}
\end{figure}
\begin{figure}
	\centering
	\tikzsetnextfilename{compare_spectral_functions_eq_neq_quench_f}
	\begin{tikzpicture}
		\begin{axis}
		[
			grid	= both,
			xlabel	= {$\omega$},
			ylabel	= {$S_{\hat c}(q,\omega,15)$},
			restrict x to domain = -10:0,
			legend style = {font=\scriptsize},
			legend pos = north west,
		]																			
			\pgfplotstabletranspose{\data}{../spectral_functions/neq/global_quench/L_64/1TDVP/spectral_function/U_m4p0_V_0p25_dV_m0p5/single_particle/results/kw-0.1.rows-0-4.real};
			\addplot
			[
				color = colorA,
				thick,
				mark = none,
			]
			table
			[
				x expr = -2*pi*\coordindex/(0.05*2048),
				y expr = \thisrowno{1},
			]
			{\data};
			\addlegendentry{$n=0, m_{\mathrm{max}}=1000$};
			\addplot
			[
				color = colorB,
				thick,
				mark = none,
			]
			table
			[
				x expr = -2*pi*\coordindex/(0.05*2048),
				y expr = \thisrowno{2},
			]
			{\data};
			\addlegendentry{$n=1, m_{\mathrm{max}}=1000$};
			\addplot
			[
				color = colorC,
				thick,
				mark = none,
			]
			table
			[
				x expr = -2*pi*\coordindex/(0.05*2048),
				y expr = \thisrowno{3},
			]
			{\data};
			\addlegendentry{$n=2, m_{\mathrm{max}}=1000$};
			\pgfplotstabletranspose{\data}{../spectral_functions/neq/global_quench/L_64/1TDVP/spectral_function/U_m4p0_V_0p25_dV_m0p5_chi_max_500/single_particle/results/kw-0.1.rows-0-4.real};
			\addplot
			[
				color = colorA,
				thick,
				densely dotted,
				mark = none,
			]
			table
			[
				x expr = -2*pi*\coordindex/(0.05*2048),
				y expr = \thisrowno{1},
			]
			{\data};
			\addlegendentry{$n=0, m_{\mathrm{max}}=500$};
			\addplot
			[
				color = colorB,
				thick,
				densely dotted,
				mark = none,
			]
			table
			[
				x expr = -2*pi*\coordindex/(0.05*2048),
				y expr = \thisrowno{2},
			]
			{\data};
			\addlegendentry{$n=1, m_{\mathrm{max}}=500$};
			\addplot
			[
				color = colorC,
				thick,
				densely dotted,
				mark = none,
			]
			table
			[
				x expr = -2*pi*\coordindex/(0.05*2048),
				y expr = \thisrowno{3},
			]
			{\data};
			\addlegendentry{$n=2, m_{\mathrm{max}}=500$};
		\end{axis}
	\end{tikzpicture}
	\caption
	{%
		\label{fig:neq-spectral-functions-1particle:comparison}%
		Comparison of single-particle spectral functions $S_{\hat c}(q,\omega,15)$ after quenching $V=0.25 \rightarrow V=-0.25$ evaluated at fixed values of the wave vector $q_n = \frac{2\pi}{L-1}n$ and for maximal bond dimensions $m_{\mathrm{max}}=500,1000$.
	}
\end{figure}
\begin{figure}
	\centering
	\tikzsetnextfilename{compare_spectral_functions_eq_neq_quench_d}
	\begin{tikzpicture}
		\begin{axis}
		[
			grid	= both,
			xlabel	= {$\omega$},
			ylabel	= {$S_{\hat d}(q,\omega,15)$},
			xmax	= 2.0,
			restrict x to domain = -10:1.0,
			legend style = {font=\scriptsize, at={(0.99,0.99)}, anchor=north east},
		]																			
			\pgfplotstabletranspose{\data}{../spectral_functions/neq/global_quench/L_64/1TDVP/spectral_function/U_m4p0_V_0p25_dV_m0p5/two_particle/results/kw-0.1.rows-0-4.real};
			\addplot
			[
				color = colorA,
				thick,
				mark = none,
			]
			table
			[
				x expr = -2*pi*\coordindex/(0.05*2048),
				y expr = -\thisrowno{1},
			]
			{\data};
			\addlegendentry{$n=0, m_{\mathrm{max}}=1000$};
			\addplot
			[
				color = colorB,
				thick,
				mark = none,
			]
			table
			[
				x expr = -2*pi*\coordindex/(0.05*2048),
				y expr = -\thisrowno{2},
			]
			{\data};
			\addlegendentry{$n=1, m_{\mathrm{max}}=1000$};
			\addplot
			[
				color = colorC,
				thick,
				mark = none,
			]
			table
			[
				x expr = -2*pi*\coordindex/(0.05*2048),
				y expr = -\thisrowno{3},
			]
			{\data};
			\addlegendentry{$n=2, m_{\mathrm{max}}=1000$};													
			\pgfplotstabletranspose{\data}{../spectral_functions/neq/global_quench/L_64/1TDVP/spectral_function/U_m4p0_V_0p25_dV_m0p5_chi_max_500/two_particle/results/kw-0.1.rows-0-4.real};
			\addplot
			[
				color = colorA,
				thick,
				densely dotted,
				mark = none,
			]
			table
			[
				x expr = -2*pi*\coordindex/(0.05*2048),
				y expr = -\thisrowno{1},
			]
			{\data};
			\addlegendentry{$n=0, m_{\mathrm{max}}=500$};
			\addplot
			[
				color = colorB,
				thick,
				densely dotted,
				mark = none,
			]
			table
			[
				x expr = -2*pi*\coordindex/(0.05*2048),
				y expr = -\thisrowno{2},
			]
			{\data};
			\addlegendentry{$n=1, m_{\mathrm{max}}=500$};
			\addplot
			[
				color = colorC,
				thick,
				densely dotted,
				mark = none,
			]
			table
			[
				x expr = -2*pi*\coordindex/(0.05*2048),
				y expr = -\thisrowno{3},
			]
			{\data};
			\addlegendentry{$n=2, m_{\mathrm{max}}=500$};
		\end{axis}
	\end{tikzpicture}
	\caption
	{
		\label{fig:neq-spectral-functions-2particle:comparison}%
		Comparison of two-particle spectral functions $S_{\hat d}(q,\omega,15)$ after quenching $V=0.25 \rightarrow V=-0.25$ evaluated at fixed values of the wave vector $q_n = \frac{2\pi}{L-1}n$ and for maximal bond dimensions $m_{\mathrm{max}}=500,1000$.
	}
\end{figure}
In order to check the results for convergence we performed simulations with reduced maximal bond dimension $m_{\mathrm{max}} = 500$.
In \cref{fig:neq-neq-spectral-functions-2particle,fig:neq-neq-spectral-functions-1particle}, the calculated spectral functions of two- and single-particle excitations for those bond dimensions are shown.
The fidelity is very high, only small derivations around $q=0$ are visible, which we relate to an induced length scale due to the truncation errors, but are not relevant for the scope of our investigations.
Those derivations are shown in more detail in \cref{fig:neq-spectral-functions-1particle:comparison,fig:neq-spectral-functions-2particle:comparison} for modes near $q=0$.
As expected the spectral functions obtained with higher maximal bond dimension resolve more features.
The probably most interesting one can be found in the two-particle spectral functions \cref{fig:neq-spectral-functions-2particle:comparison}. 
Therein, an additional peak structure appears suggesting further fine features, which will be interesting to study in future work.

\bibliographystyle{prsty}
\bibliography{Literatur}